\documentclass[10pt,letterpaper]{article}

\usepackage[top=1in,left=1.5in,right=1in,footskip=0.75in]{geometry}
\usepackage{adjustbox}

\usepackage{amsmath,amssymb}

\usepackage{changepage}

\usepackage[utf8x]{inputenc}

\usepackage{textcomp,marvosym}

\usepackage{cite}

\usepackage{nameref}

\usepackage[final,nopatch=footnote]{microtype}
\DisableLigatures[f]{encoding = *, family = * }

\usepackage[table]{xcolor}

\usepackage{array}

\usepackage{graphicx} 
\usepackage{svg}
\usepackage{makecell}
\usepackage{amsmath}
\usepackage{amssymb}
\usepackage{url}
\usepackage{comment}
\usepackage{subfig}
\usepackage{placeins}
\usepackage[table]{xcolor}
\definecolor{mycolor}{RGB}{220,230,240} 

\newcolumntype{+}{!{\vrule width 2pt}}

\newlength\savedwidth

\usepackage{setspace} 

\raggedright
\usepackage[aboveskip=1pt,labelfont=bf,labelsep=period,justification=raggedright,singlelinecheck=off]{caption}

\makeatletter
\renewcommand{\@biblabel}[1]{\quad#1.}
\makeatother

\usepackage{lastpage,fancyhdr,graphicx}
\usepackage{epstopdf}
\fancyheadoffset[L]{2.25in}
\fancyfootoffset[L]{2.25in}
\DeclareMathOperator*{\argmaxB}{\arg\max} 

\usepackage[hidelinks]{hyperref}

\begin{document}
	
	\vspace*{0.2in}
	
	\begin{flushleft}
		{\LARGE
			\textbf\newline{Troika algorithm: approximate optimization for accurate clique partitioning and clustering of weighted networks} 
		}
		\newline
		\\
		Samin Aref\textsuperscript{1*} and
		Boris Ng\textsuperscript{1}
		\\
		\bigskip
		\textbf{1} Department of Mechanical and Industrial Engineering, University of Toronto, Toronto, Ontario, Canada
		\bigskip
		
		* Corresponding author's email: aref@mie.utoronto.ca
		
	\end{flushleft}
	\section*{Abstract}
	Clique partitioning is a fundamental network clustering task, with applications in a wide range of computational sciences. It involves identifying an optimal partition of the nodes for a real-valued weighted graph according to the edge weights. An optimal partition is one that maximizes the sum of within-cluster edge weights over all possible node partitions. This paper introduces a novel approximation algorithm named \textit{Troika} to solve this NP-hard problem in small to mid-sized networks for instances of theoretical and practical relevance. Troika uses a branch-and-cut scheme for branching on node triples to find a partition that is within a user-specified optimality gap tolerance. Troika offers advantages over alternative methods like integer programming solvers and heuristics for clique partitioning. Unlike existing heuristics, Troika returns solutions within a guaranteed proximity to global optimality. And our results indicate that Troika is faster than using the state-of-the-art integer programming solver Gurobi for most benchmark instances.
	Besides its advantages for solving the clique partitioning problem, we demonstrate the applications of Troika in community detection and portfolio analysis. Troika returns partitions with higher proximity to optimal compared to eight modularity-based community detection algorithms. When used on networks of correlations among stocks, Troika reveals the dynamic changes in the structure of portfolio networks including downturns from the 2008 financial crisis and the reaction to the COVID-19 pandemic. Our comprehensive results based on benchmarks from the literature and new real and random networks point to Troika as a reliable and accurate method for solving clique partitioning instances with up to 5000 edges on standard hardware. 

	\section{Introduction}
	
	Clustering is the unsupervised task of grouping objects based on their similarities. This task becomes \textit{network clustering} if the objects are modeled as nodes of a graph and their pairwise similarities are modeled as weighted edges. The \textit{clique partitioning} (CP) problem is a specific network clustering problem defined on undirected weighted graphs \cite{grotschel1989cutting} that have positive and negative real values as edge weights. Network models that have signed edges appear in a wide range of contexts \cite{aref_balance_2019} including gene regulatory networks in biology, spin glass models in physics, portfolio networks in finance, networks of international relations, and social ties of opposite nature in social network analysis. Besides its wide applicability, clique partitioning is a problem of crucial importance because it is computationally challenging \cite{Wakabayashi1986AggregationOB} despite its simple definition. For a weighted signed network as input, the CP problem is defined as clustering the nodes into a partition that maximizes the sum of within-cluster edge weights \cite{koshimura2022concise}. Clique partitioning is also important from a theoretical standpoint because other challenging optimization problems that are defined on unsigned networks (such as modularity-based community detection \cite{aref2023suboptimality}) reduce to it.
	
	
	The CP problem is NP-hard \cite{Wakabayashi1986AggregationOB}, so the quest for efficient approximation and exact algorithms has led to a variety of approaches including inexact approaches such as heuristics and meta-heuristics. These approaches are scalable to large networks, but do not provide any guarantee for solution quality. For small and mid-sized instances, there are also exact and approximation approaches including methods based on mathematical optimization models \cite{grotschel1989cutting}. 
	
	\subsection*{Related work}
	Grötschel and Wakabayashi formulated the CP problem as an integer linear program in 1989 \cite{grotschel1989cutting}. In the past 30 years, a myriad of inexact algorithms have been proposed for the CP problem relying on heuristics and meta-heuristics such as simulated annealing \cite{gao2022improving}, neighborhood search \cite{brimberg2017solving}, iterated tabu search \cite{palubeckis2014iterated, pacheco2023stepped}, and local search \cite{zhou2016three}. In 2021, Lu et al. \cite{lu2021hybrid} introduced a merge-divide memetic algorithm, which incorporates a merge-divide crossover operator along with simulated annealing-based local optimization and pool management. Their method uses a population-based approach to solve ``challenging'' CP instances, and is the first to have several complementary search components. According to their experiments, the algorithm has the ability to produce high-quality solutions, reporting improved best-known lower bounds for benchmark instances with 2000 nodes. 
	
	There are some applications of CP where optimization accuracy is not a concern because network instances are very large-scale. Named entity disambiguation is one such application from natural language processing which involves finding a cluster of candidates from a knowledge base which are likely to be the target of a name mentioned in a text \cite{belalta_graph_2024}. In a recent study, Belalta et al.\ proposed a method for this task which heuristically solves a large-scale CP problem under the hood \cite{belalta_graph_2024}.
	
	While heuristics and meta-heuristics are typically fast for small and mid-sized instances, their solutions have no guarantee of proximity to optimal solutions. Compared to them, there are fewer exact and approximation approaches proposed for solving the CP problem. In 2019, Simanchev et al. \cite{simanchev2019branch} introduced an exact branch-and-cut method tailored to two specific instances of the CP problem. They proposed a cutting-plane algorithm designed to explore facet inequalities, which was used to construct lower bounds, with special heuristics for searching for upper bounds. Their method is capable of finding optimal solutions for the two particular cases of the CP problem in a fair amount of time (within 3 hours) for networks up to 300 nodes. In 2023, Belyi et al.\ \cite{belyi2023subnetwork} estimated a tighter upper bound for the CP problem and combined it with branch-and-bound to obtain exact solutions. Their method, which relies on Integer Programming (IP), produced exact solutions faster than other existing methods at the time \cite{belyi2023subnetwork}. 
	
	Some exact and approximation approaches rely on theoretical results from the polyhedral analysis of the IP formulations of the CP problem \cite{letchford_separation_2024}. Letchford and S{\o}rensen proposed a polynomial-time separation algorithm for finding the valid inequalities for CP \cite{letchford_separation_2024}. In their more recent work, they proposed two families of new valid inequalities that can speed-up solving the IP models of CP \cite{letchford_new_2025}. Irmai et al.\ have recently generalized another family of valid inequalities for the CP problem \cite{irmai_chorded_2024}. For a detailed review of the literature on CP models and their performance, we refer the reader to \cite{du2022solving}. A detailed dichotomy of clique-based partitioning problems is provided in \cite{levin_clique_2022}.
	
	\subsection*{Our contributions}
	
	In this study, we propose a method for approximating the optimal solutions of the CP problem. Our proposed approximation algorithm, named Troika, delivers solutions with guaranteed proximity to the optimal solution. Troika solves integer programs by node triple branching. This type of branching was proven useful \cite{aref2022bayan} for dealing with constraints that involve triples. These constraints are called transitivity constraints and are a common challenge among integer programming models for clustering problems \cite{miyauchi2018exact,belyi2023subnetwork,aref2023suboptimality}.
	
	The main focus of this study is on developing an exact and approximation approach for solving the CP problem on small and mid-sized networks of practical relevance. Our proposed method pushes the practical limits on solving the CP problem on ordinary computers as demonstrated by comprehensive results on multiple benchmark datasets. While approximating an NP-hard optimization problem cannot possibly scale to large networks, the design of our method offers a key advantage over existing heuristics: it guarantees solution quality in the form of a user-specified optimality gap tolerance. In simpler terms, it allows the user to specify in advance their tolerance for the potential optimality gap of the partition. Compared to alternative exact methods, our method offers the advantage of having better time-quality speed-up: given the same time limit, it offers a partitions closer to the optimal; and given the same optimality gap tolerance, it converges faster. We conduct extensive experiments to demonstrate the applicability of our proposed method on CP instances and two other use cases. As a side contribution, we analytically demonstrate the connection that converts a modularity maximization \cite{brandes_modularity_2007} instance into a CP instance and show how Troika compares to eight modularity-based algorithms. 
	
	The technical background is outlines in Section \ref{s:math}. The Troika algorithm is explained in Section \ref{s:troika}. Section \ref{s:results-cp} provides comparative analysis results for Troika on five datasets demonstrating its practical advantages over the existing methods. Section \ref{s:results-community} discusses a use-case in community detection and demonstrates the advantages of Troika over eight modularity-based algorithms. Section \ref{s:results-portfolio} deals with the applicability of Troika for portfolio analysis. The main results are discussed in Section \ref{s:discuss}. Finally, materials and methods are provided in Section \ref{s:materials}.
	
	\section{Mathematical Background}
	\label{s:math}
	
	We represent the weighted graph $G$ with node set $V$, edge set $E$, and weight matrix $\textbf{W}$ as $G=(V,E,\textbf{W})$. Graph $G$ may have self-loops, but has at most one edge per each pair of nodes. Graph $G$ has $|V|=n$ nodes and $|E|=m$ undirected weighted edges. Its symmetric weight matrix (weighted adjacency matrix) $\textbf{W}=[w_{ij}]$ has real-valued entries $w_{ij}\in \mathbb{R}$. In some definitions of the CP problem, the graph $G$ is restricted to be a complete graph \cite{levin_clique_2022} and therefore the clusters are actual cliques. We study the more general version of the CP problem where the input graph the input graph $G$ is not necessarily complete $E=\{(i,j) \in V^2, i \leq j, w_{ij}\neq0\}$. The non-zero entry $w_{ij}\in \mathbb{R}$ indicates the weight of the undirected edge $(i,j) \in E$ between node $i$ and node $j$. Both positive and negative weights must exist in graph $G$ for CP to be a non-trivial problem. The degree of node $i$ is calculated by $d_i=\sum_j{w_{ij}}$. 
	
	The node set $V$ of the input graph $G$ can be partitioned into (any unspecified number $k$ of) disjoint clusters based on partition $P=\{V_1,V_2, \dots, V_k \}$ such that $ \bigcup_{1}^k V_i = V$ and $ V_i \cap V_j = \emptyset$. Given partition $P$, the relative cluster assignment of a pair of nodes $(i,j)$ is same (represented by $x_{ij}=0$) or different (represented by $x_{ij}=1$). The partition $P$ can therefore be alternatively represented as the symmetric binary matrix $\textbf{X}=[x_{ij}]$ which is interpreted as follows: The binary entry $x_{ij}$ indicates the relative cluster assignments of nodes $i$ and $j$. The diagonal entries $x_{ii}$ are 0's. Given partition $\textbf{X}$, edges with endpoints in the same cluster (different clusters) are called internal (external) edges.

	\subsection{Problem Statement}
	\label{ss:problem-statement}
	The clique partitioning problem \cite{grotschel1989cutting} for the graph $G=(V,E,\textbf{W})$ is defined below. Given graph $G$ and partition $\textbf{X}$, the \textit{weight} (the sum of within-cluster edge weights) of the partition $W_{(G,\textbf{X})}$ is computed according to Eq.\ \eqref{eq2}.
	\begin{equation}
		\label{eq2}
		W_{(G,\textbf{X})}= \sum \limits_{(i,j) \in E} w_{ij}(1-x_{ij})
	\end{equation}
	In the CP problem, we look for an \textit{optimal} partition: a partition $\textbf{X}^*_{(G)}$ whose weight is maximum over all possible partitions: $\textbf{X}^*_{(G)}=\argmaxB_{\textbf{X}}W_{(G,\textbf{X})}$. Any partition of $G$ that in not an optimal partition is a \textit{sub-optimal} partition.
	
	\subsection{Integer Programming Formulations}
	\label{ss:ip}
	The CP problem can be formulated as the Integer Programming (IP) model \cite{grotschel1989cutting} in Eq.\ \eqref{eq_cp_ip}.
	
	\begin{equation}
		\label{eq_cp_ip}
		\begin{split}
			\text{CP}(G): \quad &\max_{x_{ij}} W = \sum\limits_{(i,j) \in E} w_{ij}(1- x_{ij}) \\
			\text{s.t.} &\quad x_{ik}+x_{jk} \geq x_{ij} \quad \forall (i,j,k) \in T \\ 
			&\quad x_{jk}+x_{ij} \geq x_{ik} \quad \forall (i,j,k) \in T \\ 
			&\quad x_{ij}+x_{ik} \geq x_{jk} \quad \forall (i,j,k) \in T \\ 
			& \quad \quad x_{ij} \in \{0,1\} \quad \forall (i,j) \in E
		\end{split}
	\end{equation}
	
	In Eq.\ \eqref{eq_cp_ip}, the optimal objective function value equals the optimal weight (maximum within-cluster weight) for the input graph $G$. An optimal partition is characterized by the optimal values of the $x_{ij}$ variables. $T$ indicates the set of all unique node triples $T=\{(i,j,k)\in V^3 | 1\leq i< j< k \leq n\}$ for graph $G$. The 3 constraints defined for each triple in $T$ are called \textit{transitivity constraints}. They are a common challenge of network clustering problems formulated as integer programs \cite{miyauchi2018exact,belyi2023subnetwork,aref2023suboptimality}.
	
	The computational complexity of the classic formulation in Eq.\ \eqref{eq_cp_ip} becomes impracticable for large networks due to the voluminous number of $3|T|$ constraints, scaling as \(3\binom{n}{3}\) which is \(\mathcal{O}(n^3)\) \cite{miyauchi2015redundant}. The formulation in Eq.\ \eqref{eq_cp_ip} comprises numerous redundant constraints. The redundant constraints \cite{miyauchi2015redundant} are as follows: 
	
	$$x_{ik} + x_{jk} \geq x_{ij} \quad \forall\, 1 \leq i < j < k \leq n,\, w_{ij} < 0 \land w_{jk} < 0$$
	$$x_{ik} + x_{jk} \geq x_{ij} \quad \forall\, 1 \leq i < j < k \leq n,\, w_{ij} < 0 \land w_{ik} < 0$$
	$$x_{ij} + x_{ik} \geq x_{jk} \quad \forall\, 1 \leq i < j < k \leq n,\, w_{jk} < 0 \land w_{ik} < 0$$
	
	The classic formulation of the problem \cite{grotschel1989cutting} in Eq.\ \eqref{eq_cp_ip}, can be strengthened by using negative edges for removing the redundant constraints \cite{miyauchi2015redundant}. The redundancy of these constraints is due to the pressure from the maximization objective function. After removing the redundant constraints \cite{miyauchi2015redundant} and applying other simplifications \cite{miyauchi2018exact}, we arrive at the model RP$^*(G)$ as in Eq.\ \eqref{eq_cp_rp_ip}.
	
	\begin{equation}
		\label{eq_cp_rp_ip}
		\begin{split}
			\text{RP}^*(G): \quad &\max_{x_{ij}} W = \sum\limits_{(i,j) \in E} w_{ij}(1- x_{ij}) \\
			\text{s.t.} &\quad x_{ik}+x_{jk} \geq x_{ij} \quad \forall (i,j,k) \in T^k_+ \\ 
			&\quad x_{jk}+x_{ij} \geq x_{ik} \quad \forall (i,j,k) \in T^j_+ \\ 
			&\quad x_{ij}+x_{ik} \geq x_{jk} \quad \forall (i,j,k) \in T^i_+ \\ 
			& \quad \quad x_{ij} \in \{0,1\} \quad \forall (i,j) \in E
		\end{split}
	\end{equation}
	
	In the formulation $\text{RP}^*(G)$ which is proposed by \cite{miyauchi2018exact}, the set $T$ is replaced by the subsets $T^k_+,T^j_+,T^i_+$ which are defined as follows:
	\begin{equation}
		\label{eq_t_plus}
		\begin{split}
			T^k_+&=\{(i,j,k) \in T \ |\ w_{ik}>0 \ \lor \ w_{jk}>0\}\\
			T^j_+&=\{(i,j,k) \in T \ |\ w_{jk}>0 \ \lor \ w_{ij}>0\}\\
			T^i_+&=\{(i,j,k) \in T \ |\ w_{ij}>0 \ \lor \ w_{ik}>0\}.
		\end{split}
	\end{equation}
	
	The optimal solution from $\text{RP}^*(G)$ is required to go through a linear-time ($\mathcal{O}(m)$) post-processing step, called \(\textbf{pp}\) and described in \cite{miyauchi2018exact}, to ensure that an optimal solution has been obtained. This post-processing step ensures the solution found from solving the $\text{RP}^*(G)$ formulation does not violate the transitivity constraints in the classic formulation in Eq.\ \eqref{eq_cp_ip}. 
	Miyauchi et al.\ \cite{miyauchi2018exact} demonstrated that the $\text{RP}^*(G)$ formulation combined with the \(\textbf{pp}\) post-processing step is a more efficient approach for solving the CP compared to solving the classic formulation \cite{grotschel1989cutting} in Eq.\ \eqref{eq_cp_ip}. In recent years, several attempts have been made for obtaining more efficient IP formulations for CP \cite{koshimura2022concise}. Koshimura et al.\ have proposed two new IP formulations with fewer constraints that $\text{RP}^*(G)$. Like $\text{RP}^*(G)$, both new models require a post-processing to ensure that the optimal solution represents a feasible partition. Numerical results suggest that these two new formulations are only sometimes faster than $RP^*$ \cite{koshimura2022concise}. Therefore, we use $\text{RP}^*(G)$ as the base IP foundation on which we build the Troika algorithm.
	
	Despite the efficiency gain in using $\text{RP}^*(G)$, this strengthened formulation does not fully take advantage of the structural characteristics of the input graph. We address this shortcoming in formulation by using the graph structure for developing pre-processing steps (discussed in Section \ref{s:materials}).
	
	\section{The Troika algorithm}
	\label{s:troika}
	The development of Troika heavily relies on the lessons learned from the Bayan algorithm \cite{aref2022bayan}. Bayan was developed for another network clustering problem that has the same challenge of transitivity constraints \cite{brandes_modularity_2007}. Inspired by the key components of the Bayan algorithm, like branching on node triples, Troika approximates the optimal solution of the CP problem. The two most important technical components of Troika are discussed in the following two sections \ref{ss:b&c}--\ref{ss:termination}. Additional details about the Troika algorithm are provided in Section \ref{s:materials} including a flowchart in Fig. \ref{fig:flowchart}.
	
	\subsection{The branch and cut implementation}
	\label{ss:b&c}
	The feasible space of the $\text{RP}^*(G)$ formulation in Eq.\ \eqref{eq_cp_rp_ip} is defined by constraints on node triples. Consider $\mathcal{T}$ to be the union of the subsets $T^k_+,T^j_+,T^i_+$ defined in Eq. \eqref{eq_t_plus}. $\mathcal{T}$ is the relevant set of node triples over which the transitivity constraints from \cite{miyauchi2018exact} are defined. Given a node triple, $(i,j,k) \in \mathcal{T}$, the three transitivity constraints are equivalent the logical disjunction of Eq.\ \eqref{con:left_cut} and Eq.\ \eqref{con:right_cut}.
	\begin{equation}\label{con:left_cut}
		x_{ij} + x_{ik} + x_{jk} = 0
	\end{equation}
	\begin{equation}\label{con:right_cut}
		x_{ij} + x_{ik} + x_{jk} \geq 2
	\end{equation}
	
	Unlike Bayan, Troika starts by obtaining one lower bound and one upper bound before forming any IP models. The upper bound is obtained using the method proposed by Belyi et al.\ in \cite{belyi2023subnetwork}. The lower bound is obtained using the CP version of the Combo algorithm \cite{sobolevsky2014general} (from the Python library \href{https://github.com/Casyfill/pyCombo}{PyCombo} \cite{pycombo}). Combo is a heuristic network optimization algorithm that can be reconfigured to solve the CP problem. Solving the natural LP relaxation (resulted from dropping the integrality constraints) of the IP model $\text{RP}^*(G)$ \cite{miyauchi2018exact} provides an additional upper bound. The minimum of the two upper bounds is used as the tight upper bound for starting the branch-and-cut scheme. Note that we have used the Gurobi LP solver \cite{gurobi} for solving all the LP models involved within the Troika algorithm. 
	
	At the root node, if the two bounds differ more than the optimality gap tolerance (as explained in \ref{ss:termination}), the algorithm does not terminate. It selects a triple of nodes whose corresponding values (from the LP solution) violate both Eq.~\eqref{con:left_cut} and Eq.~\eqref{con:right_cut}. Adding each of the violated constraints Eq.~\eqref{con:left_cut} and Eq.~\eqref{con:right_cut} forms a cut to the root node problem and divides the problem into left and right sub-problems. Recursively, for each of the two sub-problems, Gurobi LP and Combo are used to obtain an upper and a lower bound. This recursive process creates a search tree where node triples are used for branching. 
	
	Within Troika and after branching on the node triple $(i,j,k)$, we use additional techniques to obtain the lower bound in the right and left branch respectively, to speed up the convergence. In the search for lower bound in the right branch, the pre-defined value $\delta$ is subtracted from the edge weights associated with nodes $i$, $j$, $k$. This adjustment can enhance the likelihood of identifying heuristic solutions that adhere to the constraint $ x_{ij} + x_{ik} + x_{jk} \geq 2$ on the right branch. The $\delta$ value is set to be the absolute value of the median of all edge weights within the graph. This choice of value can ensure that the edge weights associated with nodes $i$, $j$, $k$ get small enough to deter the Combo algorithm from grouping those triples together. This alteration does not ensure the compliance of the heuristic solution with the constraint in Eq.~\eqref{con:right_cut}; however, such compliance is not a prerequisite for Troika's convergence to optimality.
	
	Conversely, the constraint $x_{ij} + x_{ik} + x_{jk} = 0$ is added to the left sub-problem in the branching process. The associated nodes $i$, $j$, $k$ are grouped together and represented by the supernode $ijk$. This supernode gets connected to all neighbours of $i$, $j$, $k$. The edges between the three nodes are conserved as a weighted self-loop on the supernode $ijk$. This ensures the Combo algorithm groups the node triple together in the returned partition, adhering to the constraint in the left sub-problem. The branching process is a key component among several other components of the Troika algorithm which are discussed in Section \ref{s:materials}. A schematic representation of the Troika algorithm is provided as a flowchart in Figure \ref{fig:flowchart}.
	
	Exploration of search tree continues through branching on new triples whose LP solution violates both Eq.~\eqref{con:left_cut} and Eq.~\eqref{con:right_cut}. After completing the computations of all Branch and Bound (B\&B) nodes at a given level of the search tree, Troika determines the \textit{incumbent} and the \textit{best bound}. The incumbent is chosen as the higher value between the best heuristic solution and the best integer solution discovered during the search. The highest upper bound value from the level is recorded as the best bound. During the branching process, a B\&B node is considered \textit{fathomed} under three circumstances: when the LP solution turns integral; when the LP becomes infeasible; or when the LP objective function value falls below the current incumbent. Under these conditions, further branching from the B\&B node is halted, and it is subsequently closed. 
	
	\subsection{Search termination criteria}
	\label{ss:termination}
	Troika is designed with two search termination criteria that grant users the flexibility to choose between computational efficiency and the precision of solutions. These criteria include \textit{optimality gap tolerance} and \textit{solve time limit}. 
	
	During the search process, Troika aims for the convergence of the best bound and incumbent to identify globally optimal solutions. This convergence is indicative of the exactness of the solution as per the branch-and-cut method, demonstrating the algorithm's capability to deliver globally optimal results under a stringent criteria for the optimality gap tolerance.
	
	Alternatively, one can use a larger optimality gap tolerance to obtain an approximate solution efficiently. 
	The \textit{optimality gap}, denoted by $g$, is the percentage difference between the current incumbent, $i$, and the best bound, $b$, according to the equation $g=(b-i)/b$. The optimality gap tolerance is a user-specified threshold for the acceptably low optimality gap to terminate the search. Using an optimality gap tolerance of $0< 1-\alpha < 1$ makes Troika an \textit{$\alpha$-approximation} algorithm for CP and terminates it once the solution gets within the $1-\alpha$ proximity of the optimal value.
	
	Furthermore, the criterion of solve time limit defines a maximum duration for the search process. This criterion is particularly suitable for scenarios where timely results are desirable. Additional technical details about the Troika algorithm are provided in Section \ref{s:materials} including a flowchart in Fig. \ref{fig:flowchart}.

	\section{Results on solution quality and time}
	\label{s:results-cp}
	In this section, we compare three methods: (1) Troika, (2) the Combo heuristic algorithm for CP \cite{sobolevsky2014general}, and (3) the $\text{RP}^*(G)$ formulation \cite{miyauchi2018exact}, solved using the commercial IP solver Gurobi \cite{gurobi} followed by the \(\textbf{pp}\) post-processing step. As a shorthand, we refer to the latter method as Gurobi IP. Gurobi is considered to be among the fastest mathematical solvers for solving IP problems \cite{Miltenberger2024}, setting a challenging baseline for Troika to be compared against.
	
	To ensure a thorough and unbiased evaluation, we use a diverse set of five datasets and use the average and standard deviation of three runs for each method and instance. Our five comparative analyses in Sections \ref{abr_section}--\ref{ss:ba} show the performance of each of the three methods for solving the CP problem in terms of solve time and solution quality. The primary metrics for assessing performance include (1) \textit{solve time} - the time taken by each method to produce its final partition on each instance and (2) the \textit{Extent of Sub-optimality} (EOS) for the partition produced by each method for each instance. We define and use the EOS for method/algorithm $A$ on graph $G$ as $EOS_{(G,A)}=1-O_{(G,\textbf{X}_A)}/O^*_G$. In this equation, $O_{(G,\textbf{X}_A)}$ is the objective function value corresponding to the partition $\textbf{X}_A$ returned by method $A$ for graph $G$. $O^*_G$ denotes the globally maximum objective function value for graph $G$.
	
	A recent study by S{\o}rensen and Letchford \cite{sorensen2023cp} has consolidated known and new challenging instances of the CP problem, addressing the difficulty of locating benchmark instances scattered across literature. They classify the CP instances into 3 distinct classes: ``easy", ``non-trivial" and ``challenging" \cite{sorensen2023cp}. For ``easy" instances, their standard LP relaxation yields at least one optimal solution that is integral. Such instances were solved at the root node by the branch-and-cut implementation in \cite{sorensen2023cp}, either due to the LP solution being an integer solution or because their heuristic's lower bound matches the LP's upper bound. Instances that do not qualify as easy but are solvable by the algorithm in \cite{sorensen2023cp} within an hour are considered ``non-trivial". Lastly, ``challenging" instances are defined as those that cannot be solved by the algorithm in \cite{sorensen2023cp} within an hour time, making them particularly complex. S{\o}rensen and Letchford \cite{sorensen2023cp} have shared their CP instances and the optimal solutions that were available in a public GitHub repository \url{https://github.com/MMSorensen/CP-Lib}. 
	
	In Sections \ref{abr_section}--\ref{ss:clusedit}, we use instances from four datasets of \cite{sorensen2023cp} for which the optimal solutions are known. In Section \ref{ss:ba}, we generate a dataset of homogeneous synthetic networks and obtain its optimal partitions. The synthetic networks allow us to compare the three methods under varying time restrictions.
	
	
	All computational experiments were conducted using Python 3.10 on a MacBook computer equipped with an Apple M1 Pro and 16GB of RAM, operating under macOS 14. All experiments in Sections \ref{abr_section}--\ref{ss:clusedit} are run with a time limit of 10 minutes per instance. Unlike Combo, Troika and Gurobi IP take optimality gap tolerance and time limit as optional user inputs. All experiments of Section \ref{s:results-cp} are run with an optimality gap tolerance of $0.05$ used for Troika and Gurobi IP to put them on an equal footing that is also reasonable for comparison with Combo. We have empirically observed that using the \textit{start separate} flag in Combo is crucial for obtaining high quality partitions and therefore configured Combo with it for all experiments in Section \ref{s:results-cp}. 
	
	\subsection{Comparisons on ABR benchmarks}
	\label{abr_section}
	
	Some early works on the CP problem were based on modeling and solving CP instances in the context of qualitative data analysis \cite{grotschel1989cutting, Wakabayashi1986AggregationOB}. This context of the CP problem is known as ``Aggregation of Binary Relations into an equivalence relation" (\textit{ABR}). One instance of the problem is denoted through $z$ ``objects", each possessing $q$ qualitative ``attributes". These objects, along with their attribute values, can be organized into a matrix, with each element representing the value of attribute $v$ for object $i$ \cite{sorensen2023cp}. For every pair of objects and for each attribute $v$, the following binary constant is defined:
	
	\[r^v_{ij} = {\left\{ \begin{array}{ll} 1, &{} \text {if attribute $v$ has the same value for objects {i} and {j}} \\ 0, &{} \text {otherwise}. \end{array}\right. }\]
	
	The similarities between objects \(i\) and \(j\) regarding the $q$ attributes are then quantified through the edge weight \(w_{ij} = 2 \sum _{v = 1}^q r^v_{ij} - q\) \cite{grotschel1989cutting}. In the case of missing $r^v_{ij}$ values, some adjustments (using the approach from \cite{lorena2019qualitative}) were made to create the weighted graphs as discussed in \cite{sorensen2023cp}.
	
	The 26 ABR test instances that we use from \cite{sorensen2023cp} consist of real-life use cases of the ABR from the literature \cite{brusco2009clustering, lorena2019qualitative, dorndorf1994fast}. They feature node counts $n$ ranging between 30 and 797, with edge counts $m$ ranging from 381 to 306,915. Most instances are deemed as ``easy" and with four classified as ``non-trivial" and one as ``challenging" \cite{sorensen2023cp}. Next, we present the performance results for each of the three methods on the 26 ABR instances.

\begin{figure}[!ht]
		\centering
		\subfloat[EOS on ABR benchmarks]{\includegraphics[width=0.5\textwidth]{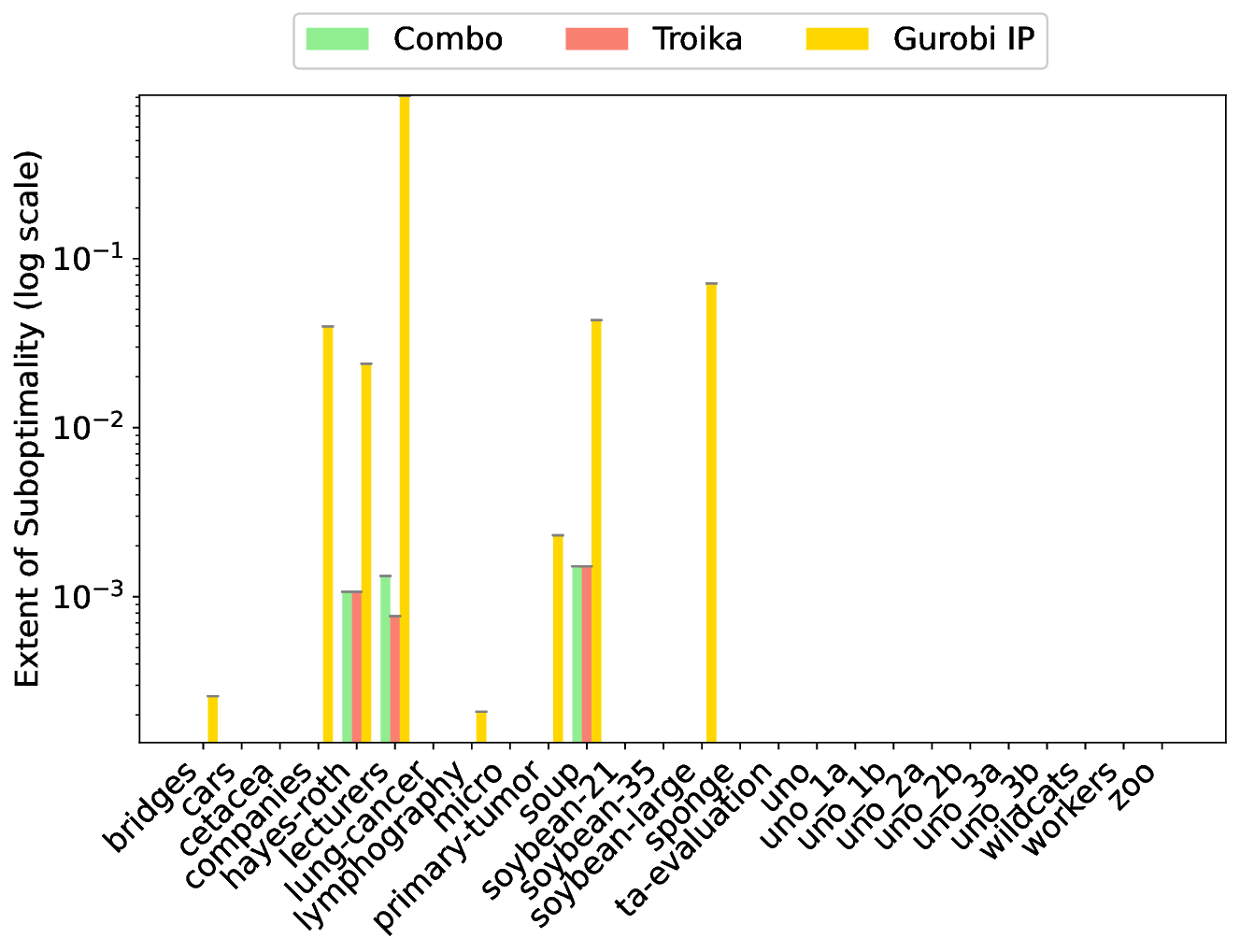}\label{fig:abr_eos}}
		\subfloat[Solve time on ABR benchmarks]{\includegraphics[width=0.5\textwidth]{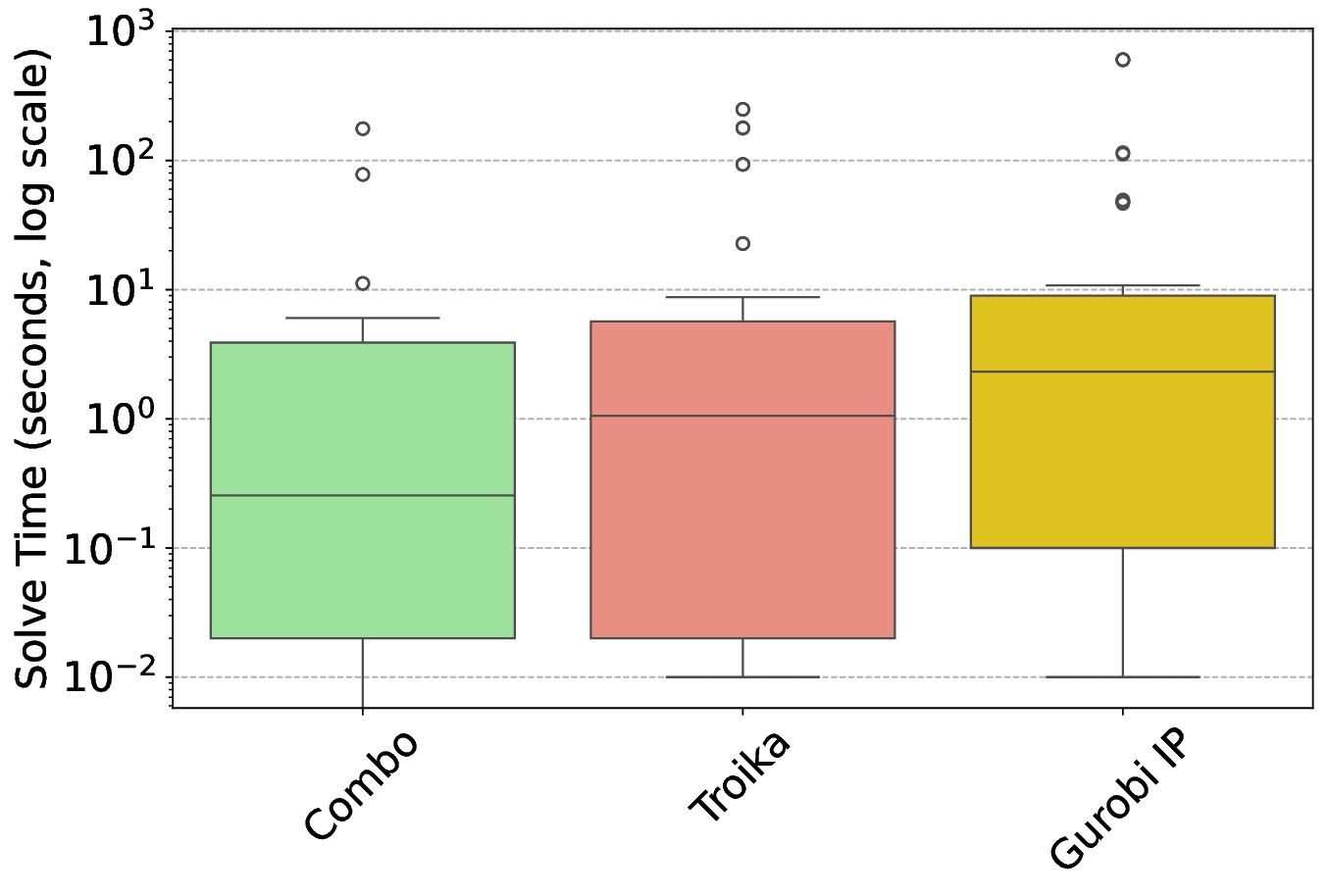}\label{fig:abr_solve_time}}
		\caption{Two comparative performance measures for the three method Troika, Combo, and Gurobi IP on the ABR benchmark dataset: (a) Extent of sub-optimality, (b) solve time. }
		\label{fig:abr}
	\end{figure}

	Fig.\ \ref{fig:abr} provides a comparison between the three methods based on EOS and solve time. Fig.\ \ref{fig:abr_eos} shows that most ABR instances are solved to global optimality by these methods because EOS is zero for them. For the few instances where EOS is not zero, the difference in solution quality between Troika and Gurobi IP is substantial. Fig.\ \ref{fig:abr_solve_time} indicates that Combo expectedly has the lowest median solve time on the ABR dataset followed by Troika and Gurobi IP respectively. Troika and Combo arrive at optimal solutions in 23 out of 26 ABR instances. Gurobi IP reached the globally optimal solutions in only 18 instances though. 
	
	More detailed results on objective values and solve times are provided in Table~\ref{table:abr_table} (in the appendices). An important distinction is observed in the ``lecturers'' instance, where Gurobi IP reaches the time limit and returns a substantially inferior partition compared to Troika and Combo. Additional experiments showed that Gurobi IP fails to converge for the ``lecturers'' within a four-hour limit. As shown in Fig. \ref{fig:abr_eos}, the optimal partition of this instance is unattainable by Troika and Combo as well, while they produce partitions with orders-of-magnitude lower EOS compared to Gurobi IP. Besides Troika producing higher quality solutions, the time columns of Table~\ref{table:abr_table} (in the appendices) demonstrate that Troika is faster than the Gurobi IP in 23 out of 26 ABR instances. 
	
	The quality of partitions produced by Combo and Troika is comparable for the ABR instances. Combo is particularly faster that Troika for the ``hayes-roth'', ``lecturers'', and ``soup'' instances because the extra computations of Troika (to ensure the optimality gap tolerance is met) take substantial time. Among these 26 instances, the mean solves times for Troika, Gurobi IP, and Combo are 22.53, 56.59, and 11.37 seconds, respectively, illustrating that Troika, on average, executes over 2.51 times faster than the Gurobi IP and around 1.98 times slower than Combo, on average.
	
	\subsection{Comparisons on equicut benchmarks}
	\label{ss:equicut}
	Some benchmark instances of \cite{sorensen2023cp} are derived from the \textit{equicut} problem. The ``equicut'' or ``equipartition'' problem is similar to the CP problem, but has the additional constraint that the partition must be a partition into two clusters of equal or almost equal size. S{\o}rensen and Letchford used the instances from the equicut literature that had negative edges and defined CP instances based on them by removing the cluster count and cluster size restrictions \cite{sorensen2023cp}. 
	
\begin{figure}[!ht]
		\centering
		\subfloat[EOS on Equicut benchmarks]{\includegraphics[width=0.5\textwidth]{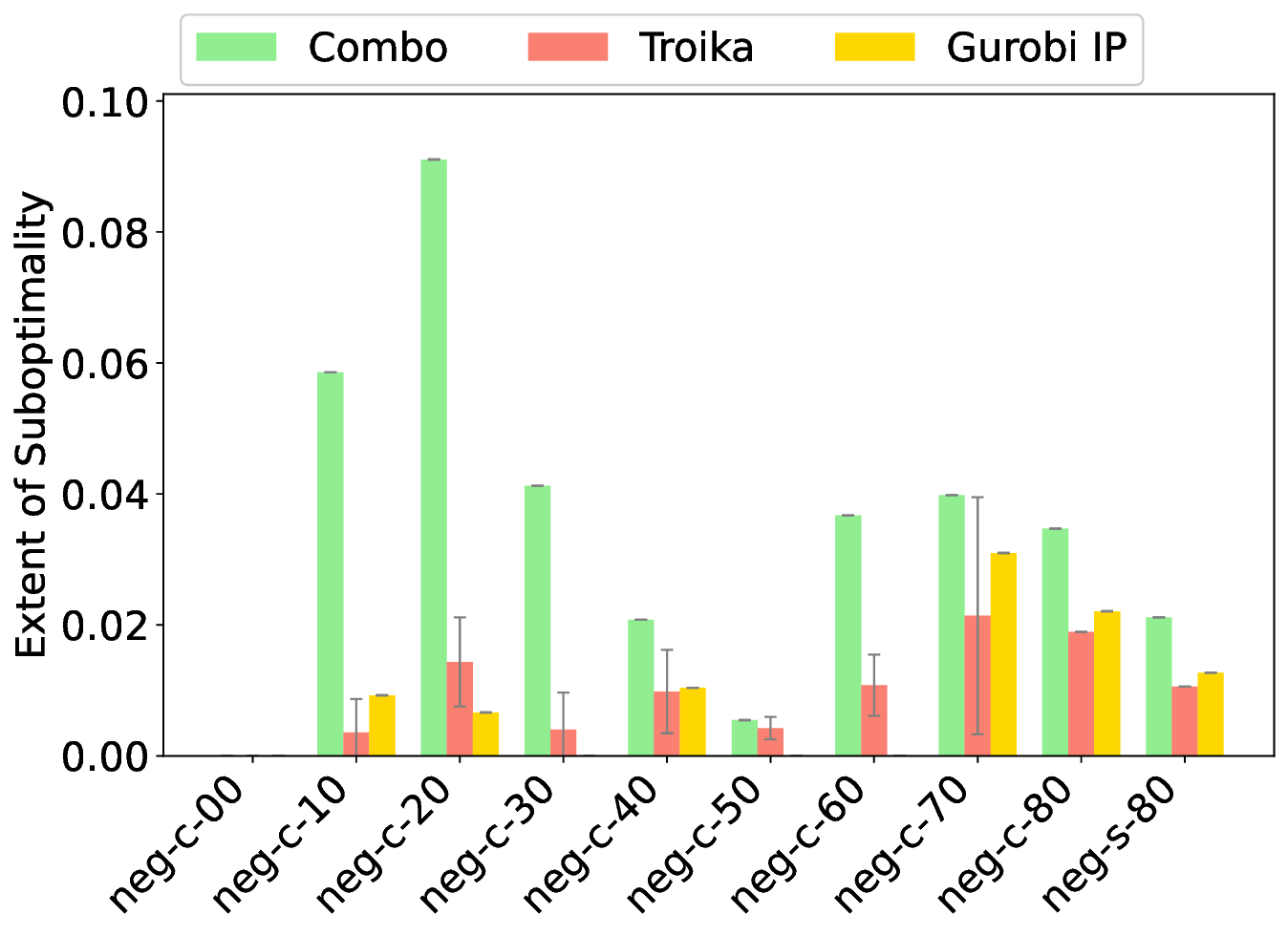}\label{fig:equicut_eos}}
		\subfloat[Solve time on Equicut benchmarks]{\includegraphics[width=0.5\textwidth]{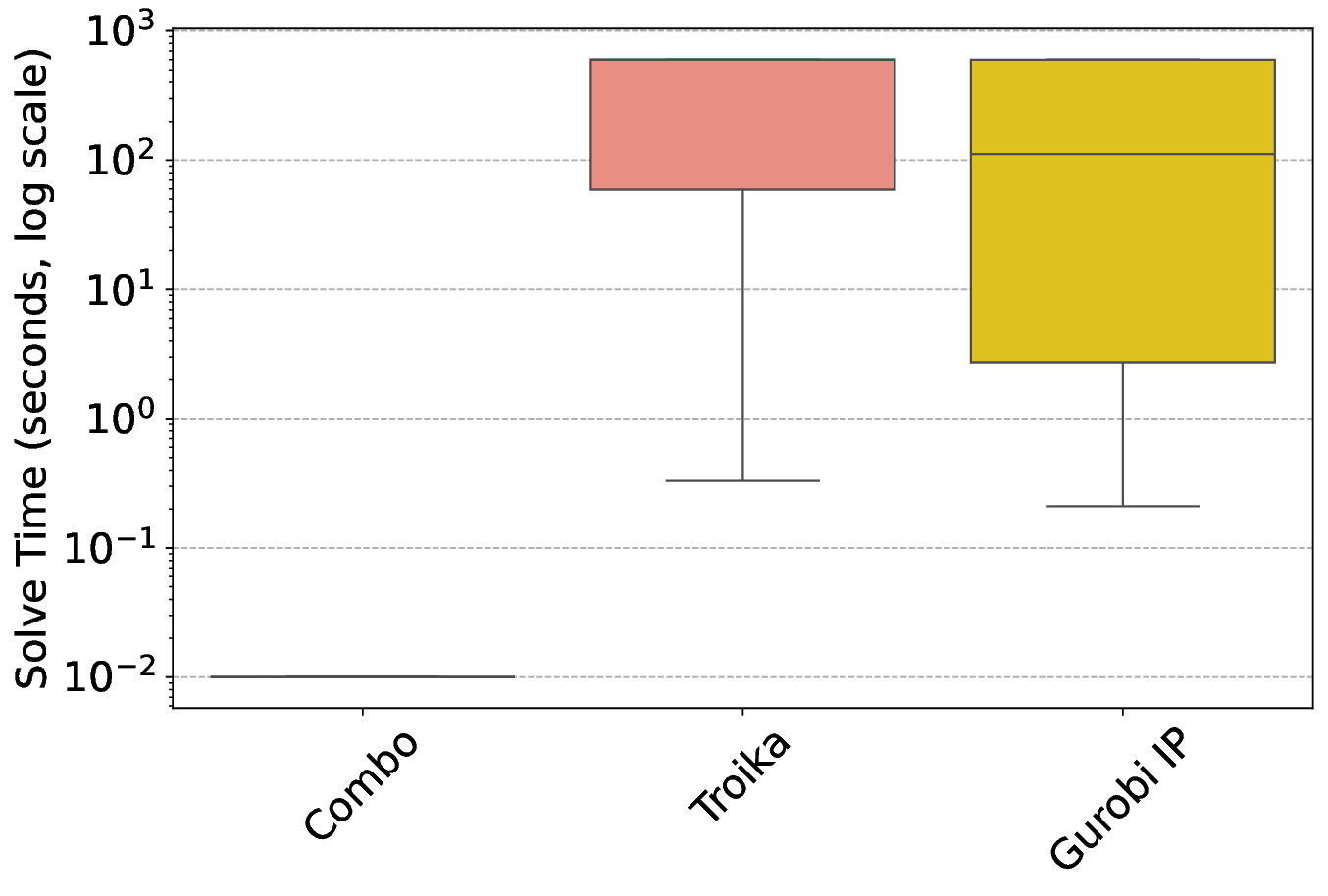}\label{fig:equicut_solve_time}}
		\caption{Two comparative performance measures for the three method Troika, Combo, and Gurobi IP on the Equicut benchmark dataset: (a) Extent of sub-optimality, (b) solve time. }
		\label{fig:equicut}
	\end{figure}

	We solve ten challenging and non-trivial equicut benchmark instances using the three methods. They have 50 nodes, except the last instance which has 60 nodes. Fig. \ref{fig:equicut} shows EOS and solve time for each method on the ten equicut benchmark instances. Error bars in Fig. \ref{fig:equicut_eos} show one standard deviation for EOS values obtained over three runs for each algorithm. Figure \ref{fig:equicut_eos} shows that the partitions of Combo for nine out of ten equicut instances get improved by the extra work that Troika does. On five instances, Troika has a better (lower) average EOS compared to Gurobi IP. Figure \ref{fig:equicut_solve_time} shows that satisfying the optimality gap tolerance of 0.05 on these instances requires extra work from Troika or Gurobi IP that is substantially more time-consuming than obtaining a single heuristic solution from Combo.

	\subsection{Comparisons on correlation benchmarks}
	\label{ss:correlation}
	A set of new benchmark instances proposed in \cite{sorensen2023cp} are called \textit{correlation} benchmarks. These benchmarks are produced based on two steps: (1) creating a matrix with uniformly random entries from the unit interval, (2) defining the weighted edge $(i,j)$ to be the correlation coefficient between columns $i$ and $j$ of the matrix. 
	
\begin{figure}[!ht]
		\centering
		\subfloat[EOS on Correlation benchmarks]{\includegraphics[width=0.5\textwidth]{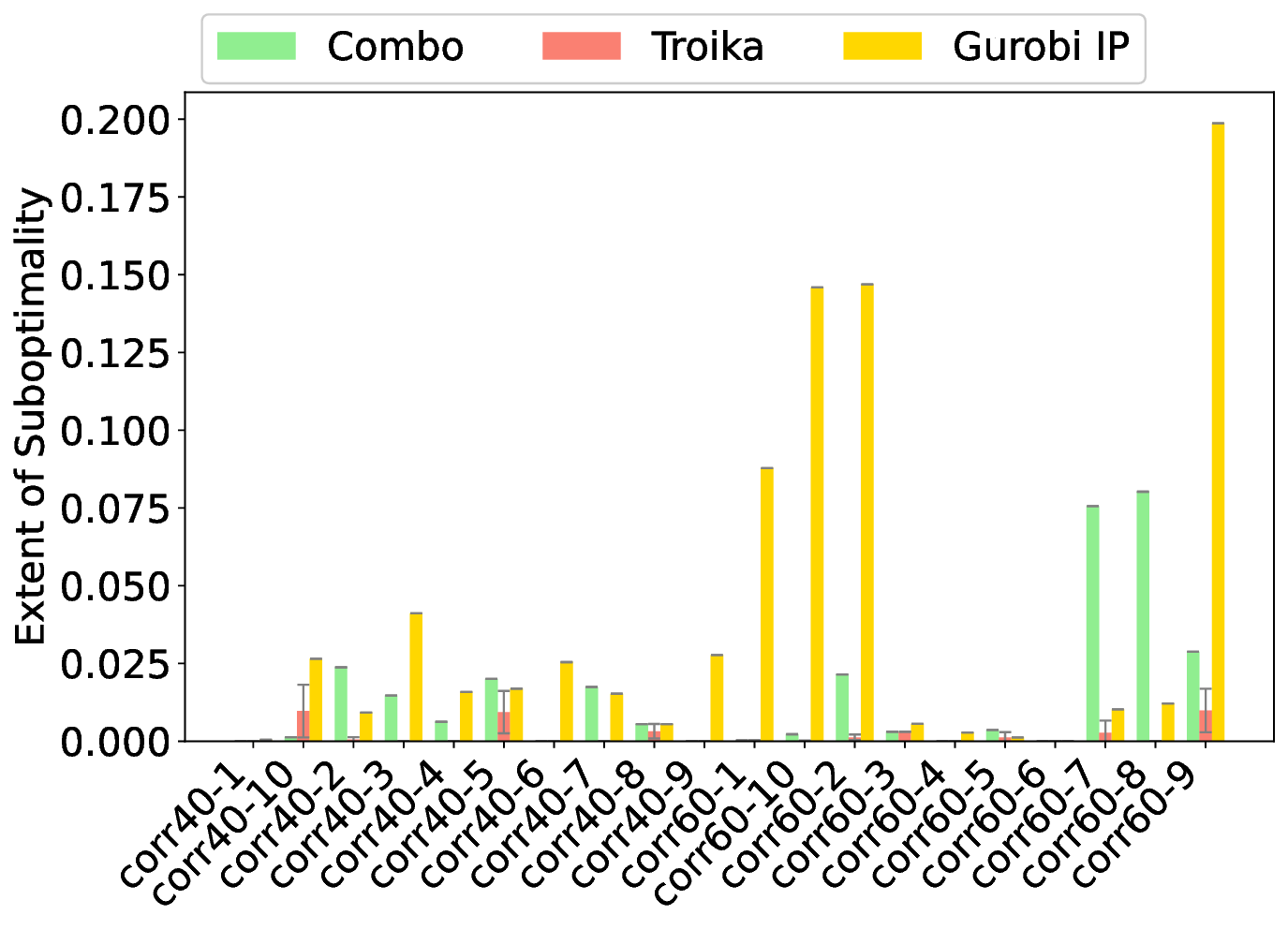}\label{fig:correlation_eos}}
		\subfloat[Solve time on Correlation benchmarks]{\includegraphics[width=0.5\textwidth]{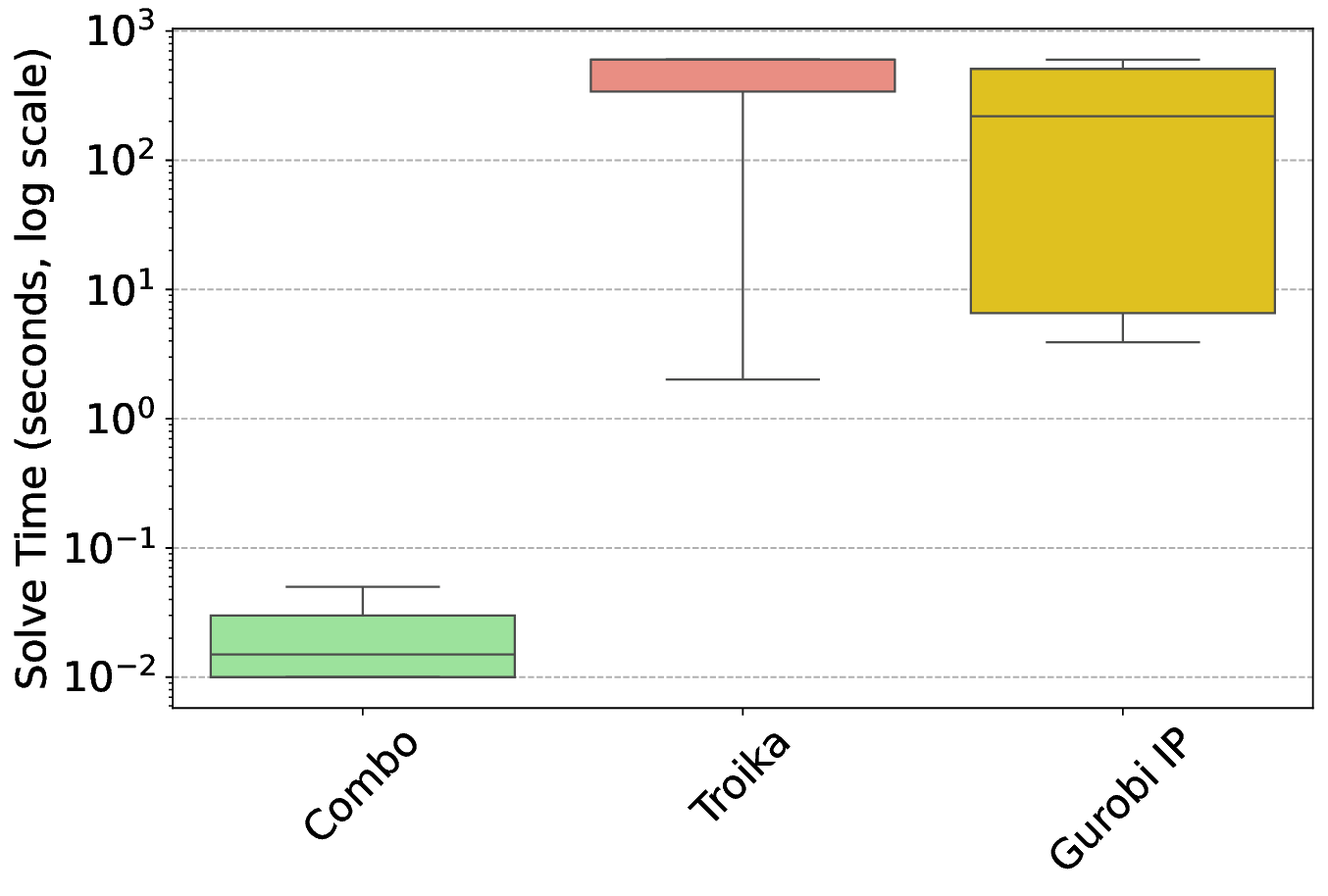}\label{fig:correlation_solve_time}}
		\caption{Two comparative performance measures for the three method Troika, Combo, and Gurobi IP on the Correlation benchmark dataset: (a) Extent of sub-optimality, (b) solve time. }
		\label{fig:correlation}
	\end{figure}

	We solve 20 challenging and non-trivial correlation instances using the three methods. The instance name denotes the number of nodes. Fig. \ref{fig:correlation_eos} shows that the partitions from Troika have much lower EOS compared to the two other methods on almost all these 20 instances. Figure \ref{fig:correlation_solve_time} shows Troika median solve time to be higher than that of Gurobi IP on these instances; this is justifiable by the higher quality solutions that Troika produces compared to Gurobi IP parametrized with the same time limit and optimality gap tolerance.
	
	\subsection{Comparisons on clusedit benchmarks}
	\label{ss:clusedit}
	Another set of CP benchmarks from \cite{sorensen2023cp} are defined based on the ``cluster editing" (\textit{clusedit}) problem. This problem, also known as the ``correlation clustering" problem \cite{aref_modeling_2020}, is defined on a signed graph $G=(V,E^-,E^+)$. The edges in the set $E^-$ all have the weight of -1 (are negative edges) and the edges in the set $E^+$ all have weight of +1 (are positive edges). The correlation clustering problem is the task of finding a partition of nodes into any number of clusters to minimize the total count of intra-cluster negative edges and inter-cluster positive edges. To convert instances of this problem to CP instances, S{\o}rensen and Letchford have defined the task of maximizing the sum of within-cluster edge weights for clusedit instances that have 20\% to 60\% negative edges.
	
\begin{figure}[!ht]
		\centering
		\subfloat[EOS on Clusedit benchmarks]{\includegraphics[width=0.5\textwidth]{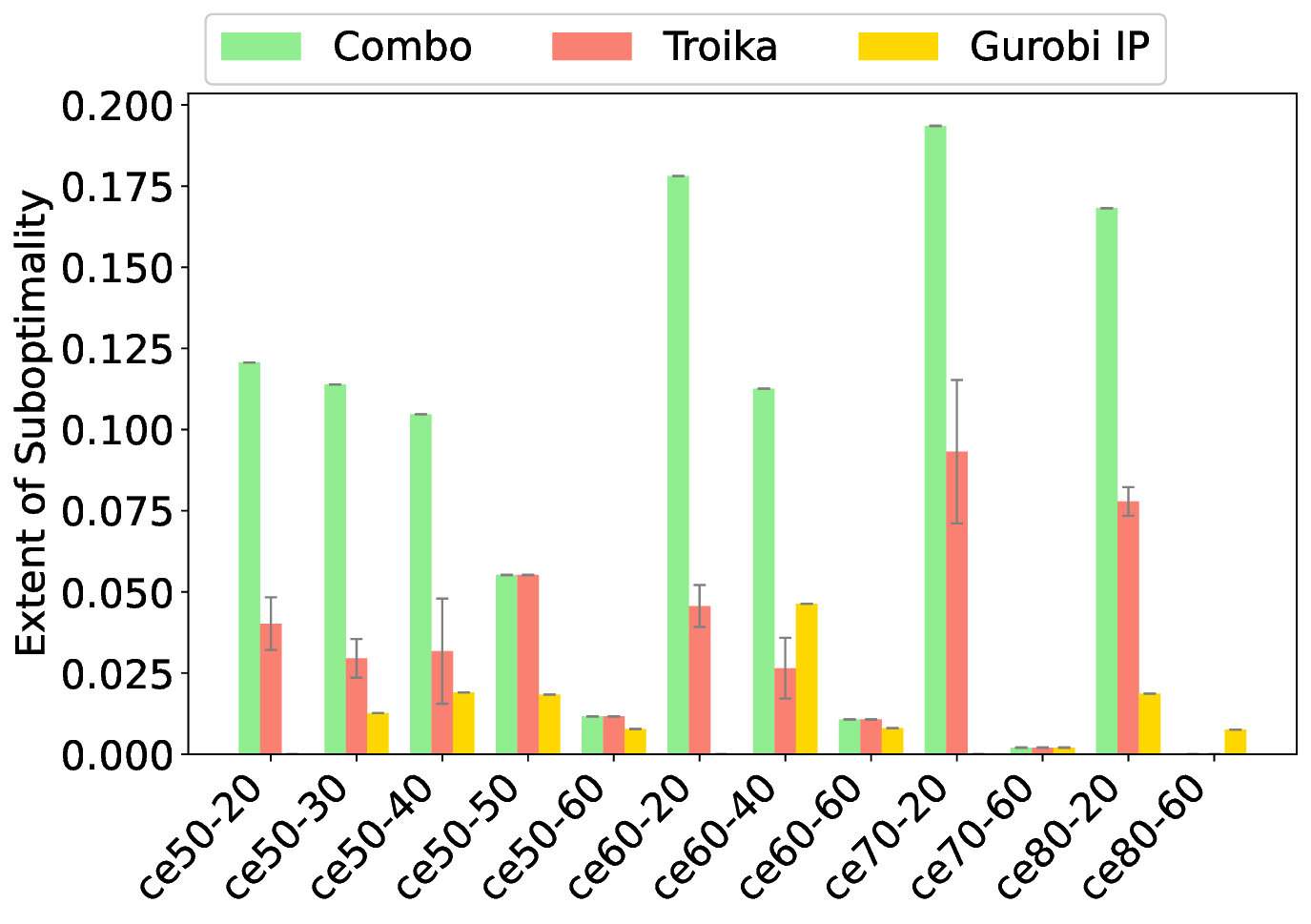}\label{fig:clusedit_eos}} 
		\subfloat[Solve time on Clusedit benchmarks]{\includegraphics[width=0.5\textwidth]{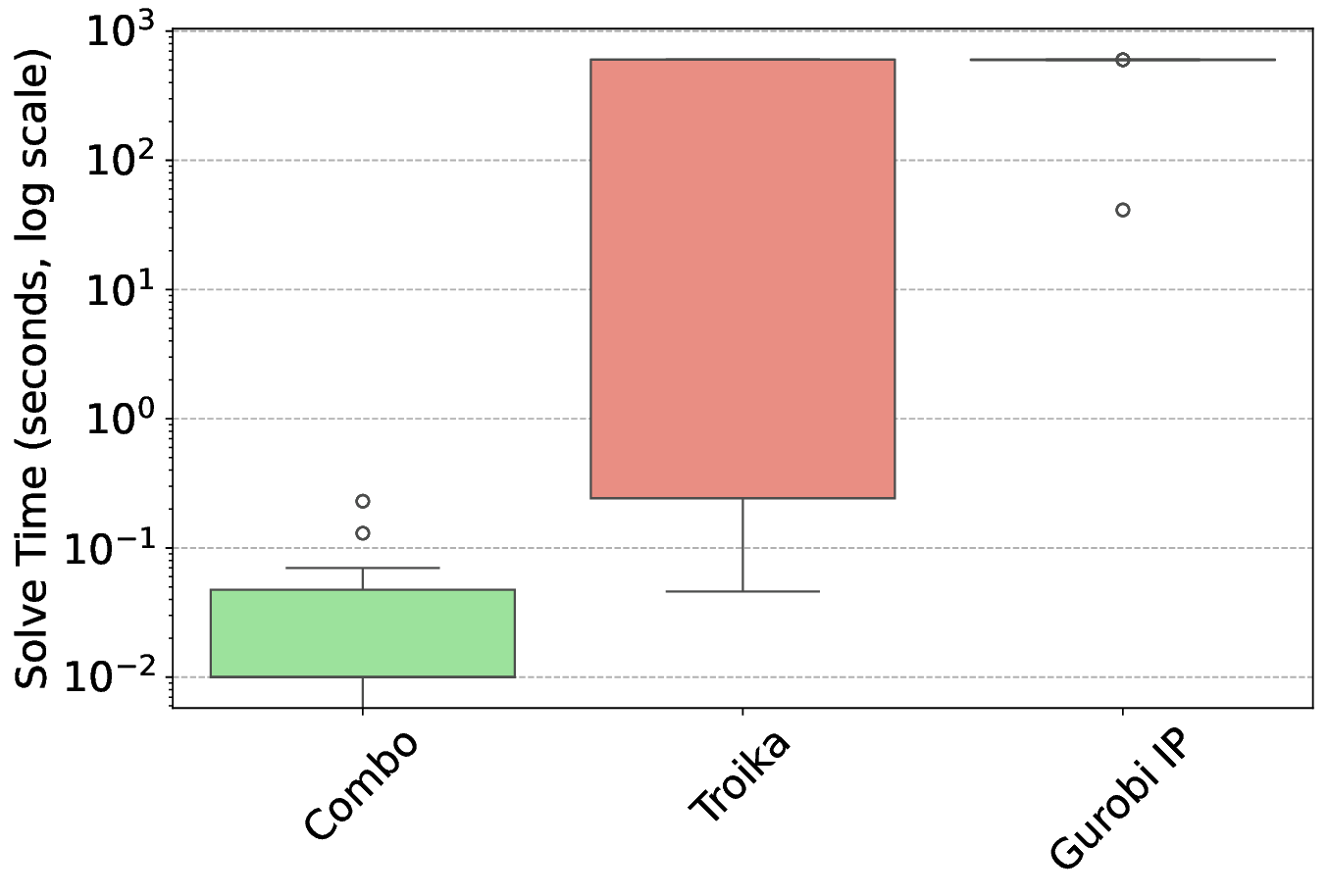}\label{fig:clusedit_solve_time}}
		\caption{Two comparative performance measures for the three method Troika, Combo, and Gurobi IP on the Clusedit benchmark dataset: (a) Extent of sub-optimality, (b) solve time. }
		\label{fig:clusedit}
	\end{figure}

	We solve 12 challenging and non-trivial instances of these clusedit benchmarks using the three methods. The instance name denotes the number of nodes followed by the fraction of negative edges. Fig. \ref{fig:clusedit_eos} demonstrates that Troika substantially improves the partitions from Combo, but rarely have better EOS compared to the partitions from Gurobi IP. Figure \ref{fig:clusedit_solve_time} shows Troika to have some relative advantages in terms of solve time compared to Gurobi IP on the clusedit benchmarks.
	
	The results provided in Figs.\ \ref{fig:equicut}--\ref{fig:clusedit} show that for non-trivial and challenging instances across four benchmark datasets, Troika consistently returns partitions with objective function values that are closer to the globally optimal solutions compared to the Combo heuristic. This result is expected given the design of the Troika algorithm which relies on Combo for lower bounds and does some extra work for improving those partitions.
	Note that the descriptive comparisons provided in Figs.\ \ref{fig:equicut}--\ref{fig:clusedit} are not all statistically significant because some performance differences between these algorithms are marginal as shown in the error bars in Figs. \ref{fig:equicut}--\ref{fig:clusedit}. While running Combo is considerably faster than Troika, the extra time that Troika spends leads to obtaining higher quality partitions as illustrated in Figs.\ \ref{fig:equicut}--\ref{fig:clusedit}. While the results showed that generally Troika has some advantage in solution quality and/or time compared to Gurobi IP; there are some instances on which Gurobi IP outperforms Troika.
	
	\subsection{Comparisons on Barab\'{a}si Albert Graphs}
	\label{ss:ba}
	
	Besides instances from four datasets of \cite{sorensen2023cp}, we create random weighted graphs based on the Barab\'{a}si-Albert (BA) graph generation process \cite{barabasi1999emergence}. We use them for evaluating Troika on graphs that mirror real network structures. 
	The Barab\'{a}si Albert process grows the network through preferential attachment \cite{barabasi1999emergence}, which mirrors the heterogeneous connectivity inherent in some real-world networks, such as social, technological, and biological systems \cite{zadorozhnyi2012structural}. The BA graphs are reflective of certain real-world contexts where preferential attachment is a reasonably realistic model despite its simplicity.

	\begin{figure}[!ht]
		\centering
		\subfloat[EOS measured at 1 second]{\includegraphics[width=0.5\textwidth]{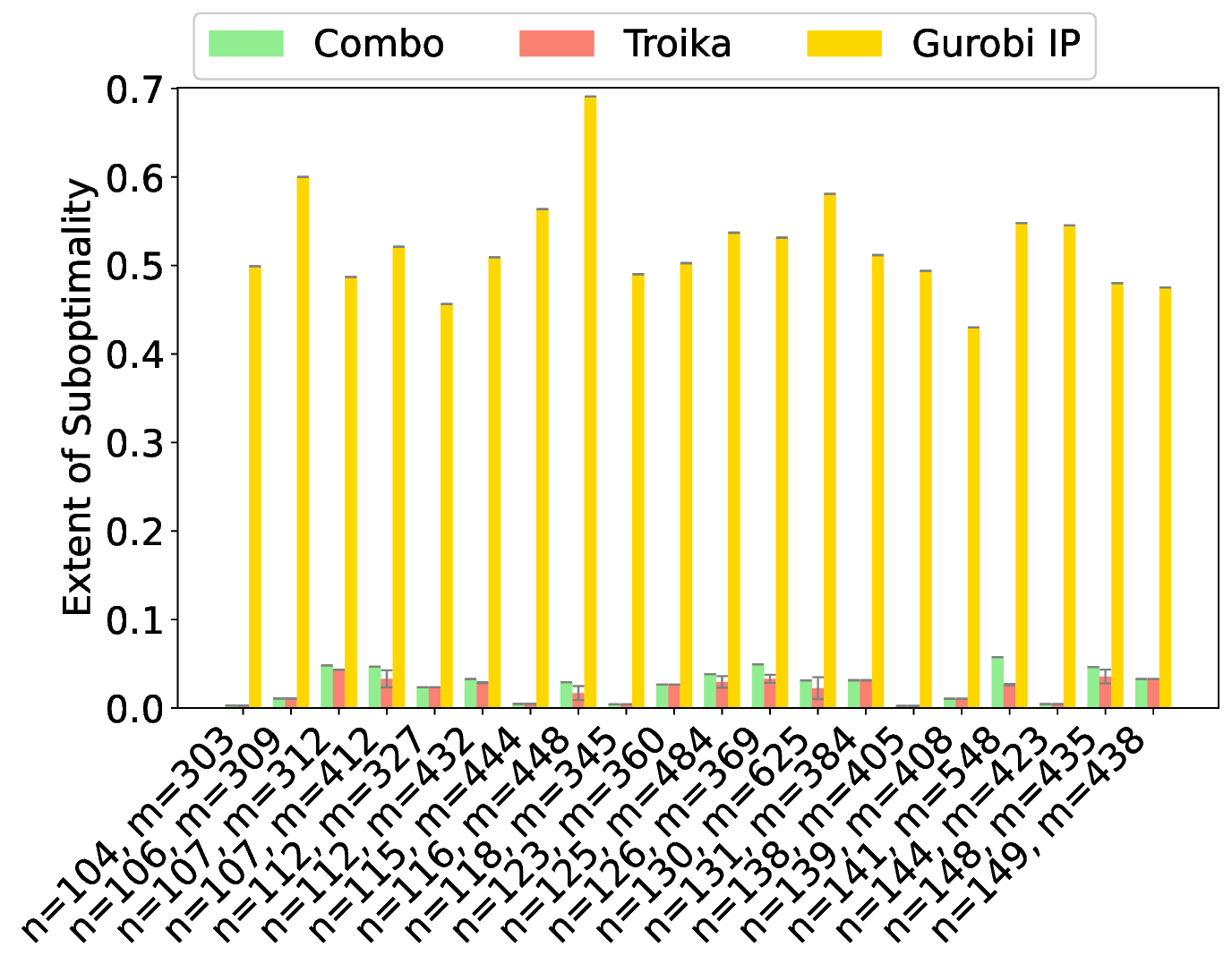}}
		\subfloat[Solve time capped at 1 second]{\includegraphics[width=0.5\textwidth]{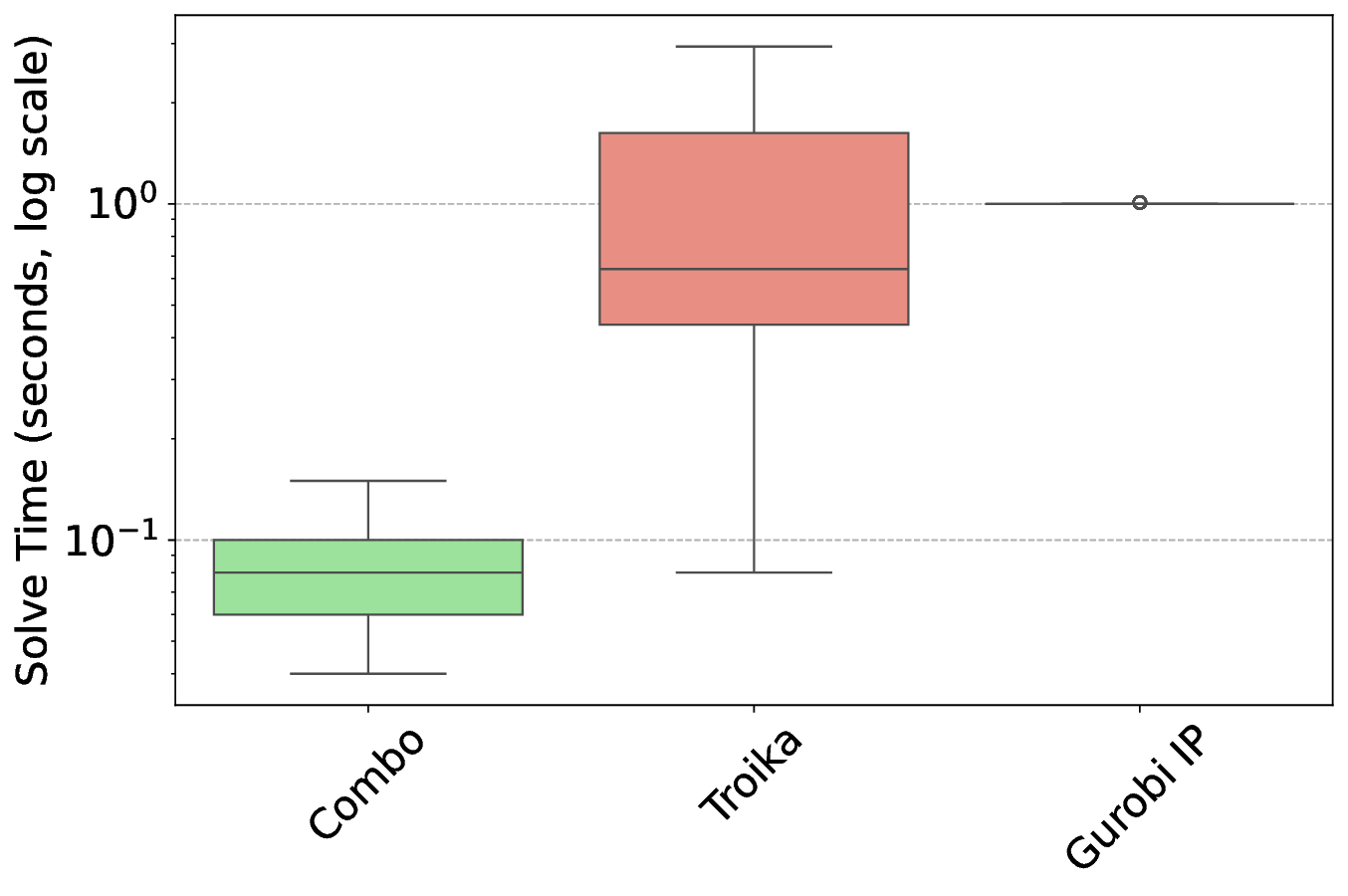}}\\
		\subfloat[EOS measured at 10 seconds]{\includegraphics[width=0.5\textwidth]{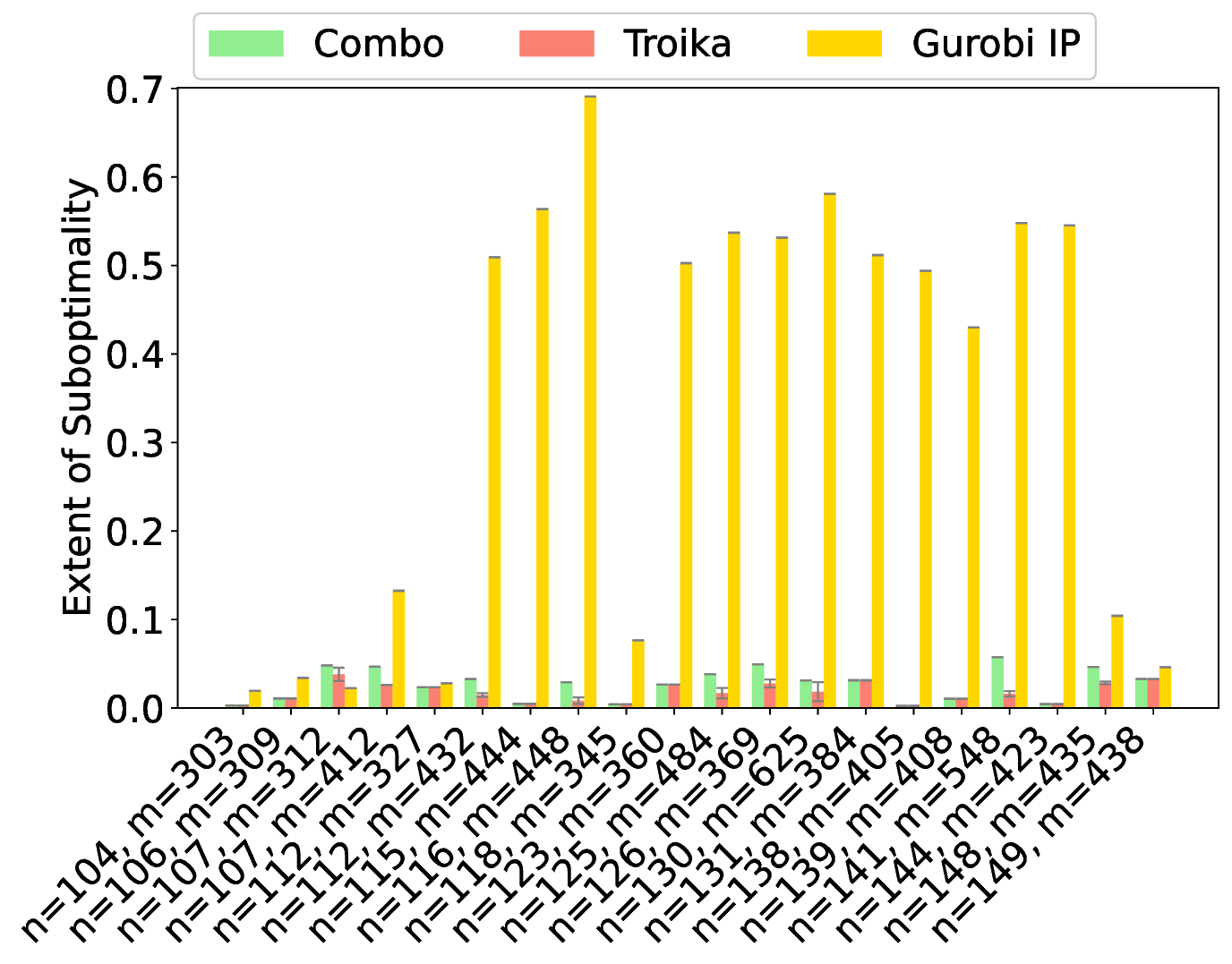}}
		\subfloat[Solve time capped at 10 seconds]{\includegraphics[width=0.5\textwidth]{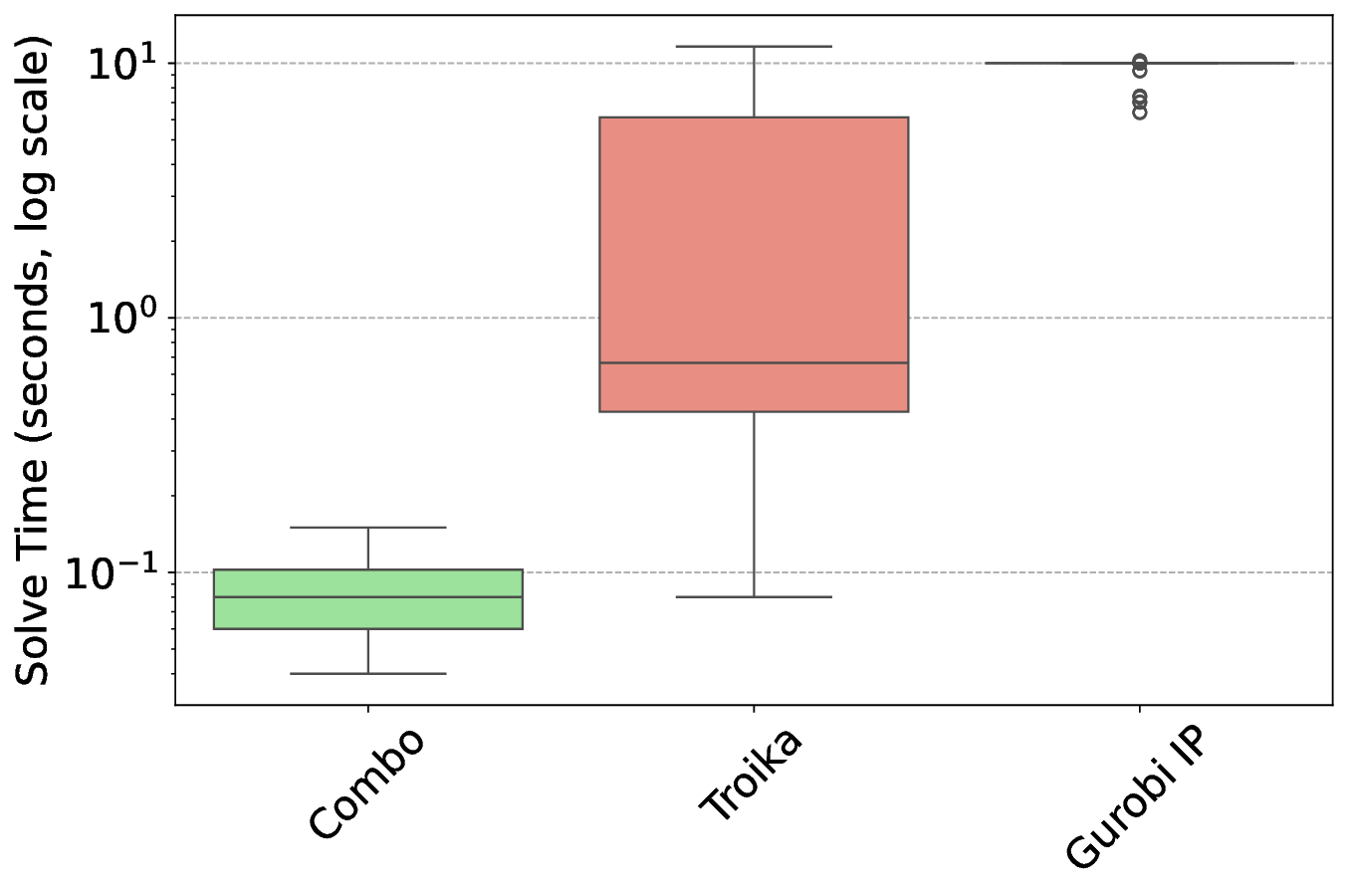}}\\
		\subfloat[EOS measured at 60 seconds]{\includegraphics[width=0.5\textwidth]{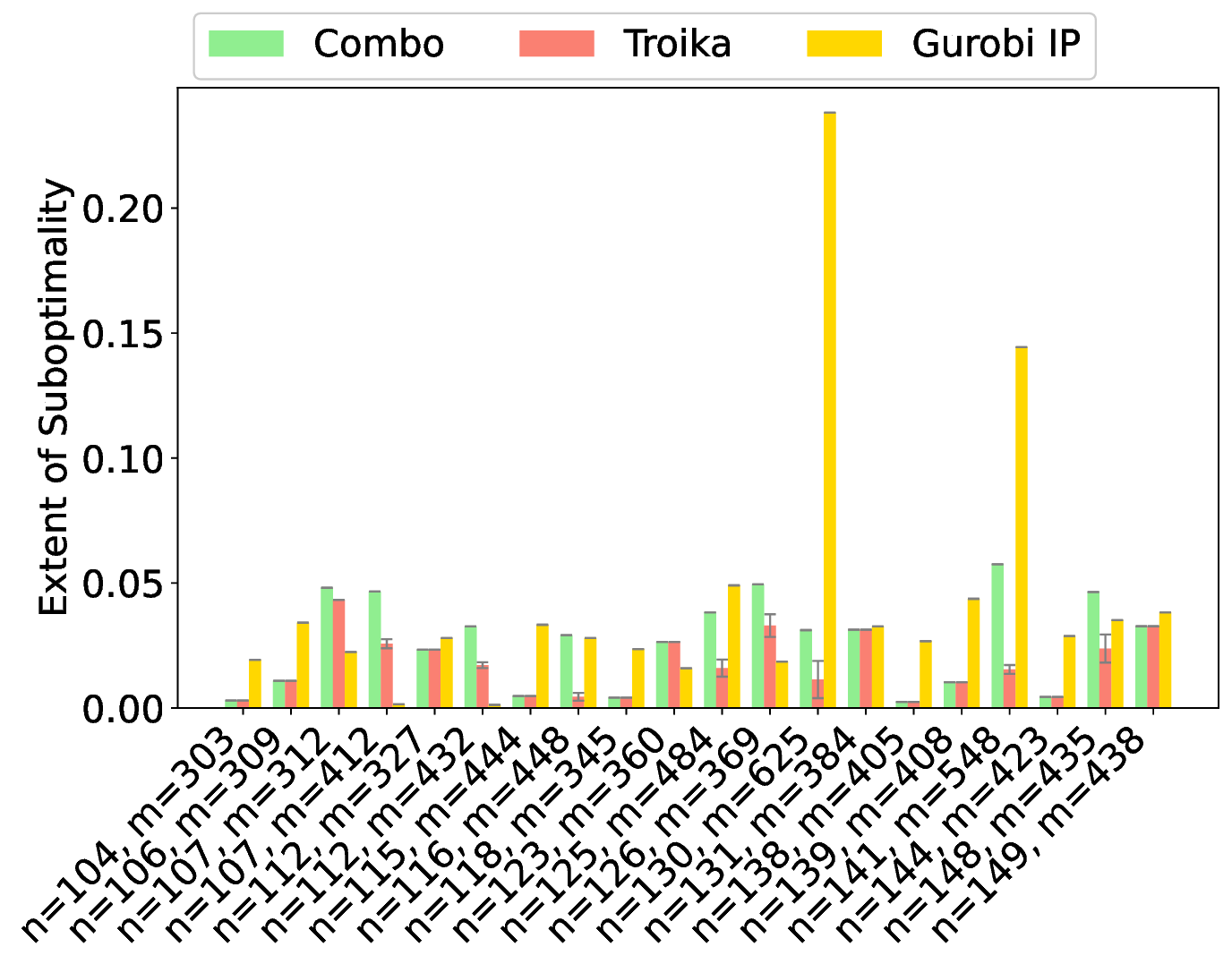}}
		\subfloat[Solve time capped at 60 seconds]{\includegraphics[width=0.5\textwidth]{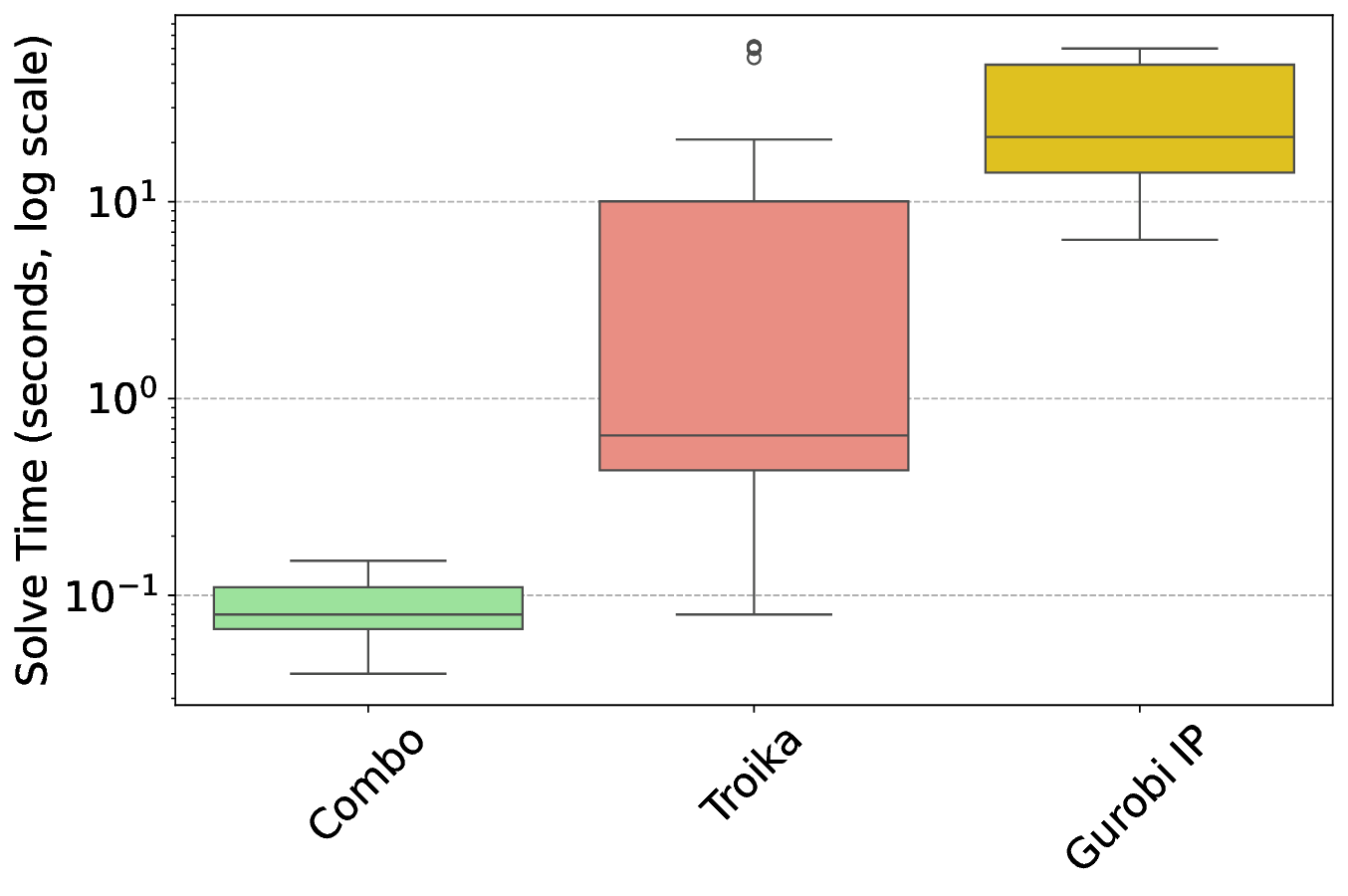}}
		\caption{Extent of sub-optimality and solve time for the three method Troika, Combo, and Gurobi IP on the BA instances.}
		\label{fig:ba}
	\end{figure}
	
	We generate 20 BA graphs to test all the three methods under varying time restrictions. These graphs are generated with the number of nodes \(n\) sampled from the discrete uniform distribution of $[100,150]$. For the number of edges to attach between a new node and existing nodes, the discrete uniform distribution of $[3,6]$ was used. Finally, each edge was assigned a random weight from the discrete uniform distribution of $[-10, 10]$. The data for these 20 BA networks are available in a FigShare repository \cite{Aref2025dataTroika}. Before running our comparison between the three methods, we obtain the globally optimal partitions of these instances by solving the $RP^*$ formulation using the Gurobi solver parametrized with an optimality gap tolerance of $1e-5$. Then, we use the optimal values for calculating EOS for the three methods. Unlike Sections \ref{abr_section}--\ref{ss:clusedit}, we define three experiment settings for assessing the three methods on these BA instances. The three experiment settings correspond to the time limits of 1, 10, and 60 seconds consistently used for the methods.

	In Figure \ref{fig:ba}, we observe that Troika consistently improves solutions from Combo on these 20 BA instances. We also see that this improvement increases as the time limit increases from 1 to 10 and then to 60 seconds. On a few instances, Troika's partition is same as Combo's partition; this is because the optimality gap of Combo's partition is below $0.05$ and therefore Troika terminates without further attempting to improve the partition. Note that the top whisker for the boxplot of Troika shows that its solve time exceeds the time limit in some cases. This marginal breach of the time limit is due to the graceful termination of Troika which is only possible after each branching step is complete (see the long chain of a branching process in Fig.\ \ref{fig:flowchart}).
	
	Across the three time limit settings, Troika's EOS is lower than Combo's in around half of the BA instances. This suggests that Troika potential to improve upon Combo's partitions makes a difference even if the user provides a split second of extra time to Troika. Figure \ref{fig:ba} shows that the median solve time of Troika is lower than that of Gurobi IP across the three time limit settings. Taken together with the consistently lower EOS values of Troika in Figure \ref{fig:ba}, we see that Troika outperforms Gurobi IP in both time and solution quality on almost all BA instances. Note that for very few BA instances, Gurobi IP terminates faster than Troika, but typically with a partition that has higher EOS.

	The comprehensive results provided in Figs.\ \ref{fig:abr}-\ref{fig:ba} show the practical advantages of Troika over two existing alternatives for obtaining approximate solutions for the CP problem. In Section \ref{s:results-community}, we investigate the applicability of Troika for another use case: community detection.
	
	\FloatBarrier
	
	\section{Applicability of Troika in community detection}
	\label{s:results-community}
	
	Descriptive community detection (CD) is the data-driven task of clustering nodes of an input (unsigned) graph into groups (communities) \cite{schaub2017many,aref2025pygenone}. Among a wide range of methods for community detection \cite{aref2025pygenone}, optimization-based algorithms are common approaches for CD \cite{aref2023analyzing}. They aim to optimize a network-level objective function, such as modularity \cite{newman_modularity_2006}, across all possible partitions of the input graph. We use the modularity objective function as an example to make an explicit connection between the CP problem and optimization-based community detection.
	
	
	\subsection{The modularity maximization problem}
	
	Given the NP-hard nature \cite{brandes_modularity_2007} of the modularity maximization problem, most modularity-based community detection algorithms are heuristics. We discuss solving the CP problem by the Troika algorithm as an indirect method for community detection through approximating maximum modularity. 
	
	The modularity function is directly extendable to graphs that have nonnegative edge weights. Therefore, we define the Modularity Maximization (MM) problem \cite{brandes_modularity_2007,aref2023suboptimality} for the simple (unsigned but generally weighted) graph $G=(V,E,\textbf{W})$ whose edge weights are nonnegative ($w_{ij}\geq0$). The modularity matrix of graph $G$ is represented by $\textbf{B}=[b_{ij}]$ whose entries are $b_{ij} = w_{ij}-{\gamma d_{i}d_{j}}/{\sum_{(i,j) \in V^2}{w_{ij}}}$. The resolution parameter for the modularity function \cite{lancichinetti_limits_2011} is denoted by $\gamma$. Without loss of generality, we use $\gamma=1$.
	
	
	Given partition $\textbf{X}$ (defined earlier in Section \ref{s:math}), for a pair of nodes $(i,j)$, their cluster assignment is same (represented by $x_{ij}=0$) or different (represented by $x_{ij}=1$). Given weighted graph $G$ and partition $\textbf{X}$, the \textit{modularity} of the partition $Q_{(G,\textbf{X})}$ is computed according to Eq.\ \eqref{eq0}.
	\begin{equation}
		\label{eq0}
		Q_{(G,\textbf{X})}= \frac{1}{\sum_{(i,j) \in V^2}{w_{ij}}} \sum \limits_{(i,j) \in V^2} b_{ij}(1-x_{ij})
	\end{equation}
	In the modularity maximization problem, we look for a partition $\textbf{X}^*$ whose modularity is maximum over all possible partitions: $\textbf{X}^*_{(G)}=\argmaxB_{\textbf{X}}Q_{(G,\textbf{X})}$
	
	\subsection{Converting an MM instance into a CP instance}
	The feasibility space of MM and CP problems is the same: all possible partitions of the input graph nodes into non-overlapping clusters. 
	In CP, only the edges ($w_{ij}\neq0$) contribute to the objective function if they become internal edges ($x_{ij}=0$). In MM, every pair of nodes (including pair of nodes without an edge) that have $b_{ij} \neq 0$ contributes to the objective function if they are assigned to the same cluster ($x_{ij}=0$). Therefore, as long as the input graph is not a complete graph, an IP instance of MM on the graph $G=(V,E,\textbf{W})$ has more decision variables compared to a CP instance of the same graph. MM requires a decision variable for node pairs with $b_{ij} \neq 0$ compared to CP which requires a variable for node pairs with $w_{ij}\neq0$. Therefore, it is expected that solving an instance of MM on the non-complete graph $G$ will be harder compared to solving an instance of the CP problem on a non-complete graph with the same number of nodes.
	
	
	Solving the MM problem on graph $G={(V,E,\textbf{W})}$ that has a modularity matrix $\textbf{B}$ is equivalent to solving the CP problem for graph $G'$ whose weight matrix equals $\textbf{B}$. Therefore, an instance of the MM problem on graph $G$ can be converted into an instance of the CP problem for graph $G'$ through using the weight matrix $\textbf{B}=[b_{ij}]$ based on the formula $b_{ij} = w_{ij}-{d_{i}d_{j}}/{\sum_{(i,j) \in V^2}{w_{ij}}}$. This implies that algorithms for solving the CP problem are also capable of solving the modularity maximization problem.
	
	\subsection{Baselines, data, and measures}
	
	We compare Troika with eight modularity maximization algorithms (baselines). These eight algorithms are (1) Clauset-Newman-Moore (CNM) \cite{clauset_finding_2004}, (2) Louvain \cite{blondel_fast_2008}, (3) Belief \cite{zhang2014}, (4) Paris \cite{paris_2018}, (5) Leiden \cite{traag_louvain_2019}, (6) EdMot \cite{edmot_2019}, (7) Combo for MM \cite{sobolevsky2014general}, and (8) graph neural network (GNN) \cite{sobolevsky2022gnn}. For algorithms (1-6), we use their Python implementation from the Community Discovery library (\textit{CDlib}) version 0.2.6 \cite{rossetti2019cdlib}. To access (7) Combo for MM, we use the Python library \href{https://github.com/Casyfill/pyCombo}{PyCombo} \cite{pycombo}. For the GNN algorithm, we use its Python implementation (the GNN-100 variation) from its \href{https://github.com/Alexander-Belyi/GNNS}{public GitHub repository} referenced in \cite{sobolevsky2022gnn}.
	
	For this comparison, we consider 100 networks comprising 53 real networks from a wide range of contexts, and 47 structurally diverse synthetic networks. The data for all these 100 networks are available in a FigShare repository \cite{Aref2025dataTroika}. The 47 synthetic networks are produced based on the graph generation models known as Lancichinetti-Fortunato-Radicchi (LFR) \cite{lancichinetti_benchmark_2008} and the Artificial Benchmark for Community Detection (ABCD) \cite{kaminski_artificial_2021}. They include 20 LFR networks with small mixing parameter values $\mu \in \{0.01, 0.1\}$ and 27 ABCD networks with small mixing parameter values $\xi \in \{0.01, 0.1, 0.3\}$. The choice of using small mixing parameters ensures that these 47 synthetic networks have modular structures.

	Our extent of sub-optimality measure, $EOS_{(G,A)}=1-O_{(G,\textbf{X}_A)}/O^*_G$, is applicable in the context of modularity maximization as well. We use it to compare Troika to the eight MM algorithms based on solution quality. In cases where an algorithm returns a partition with a non-positive modularity value, we set $EOS=1$ to facilitate easier interpretation of proximity to optimality based on non-negative EOS values. 
	
	\begin{figure}[!ht]
		\centering
		\includegraphics[width=0.7\linewidth]{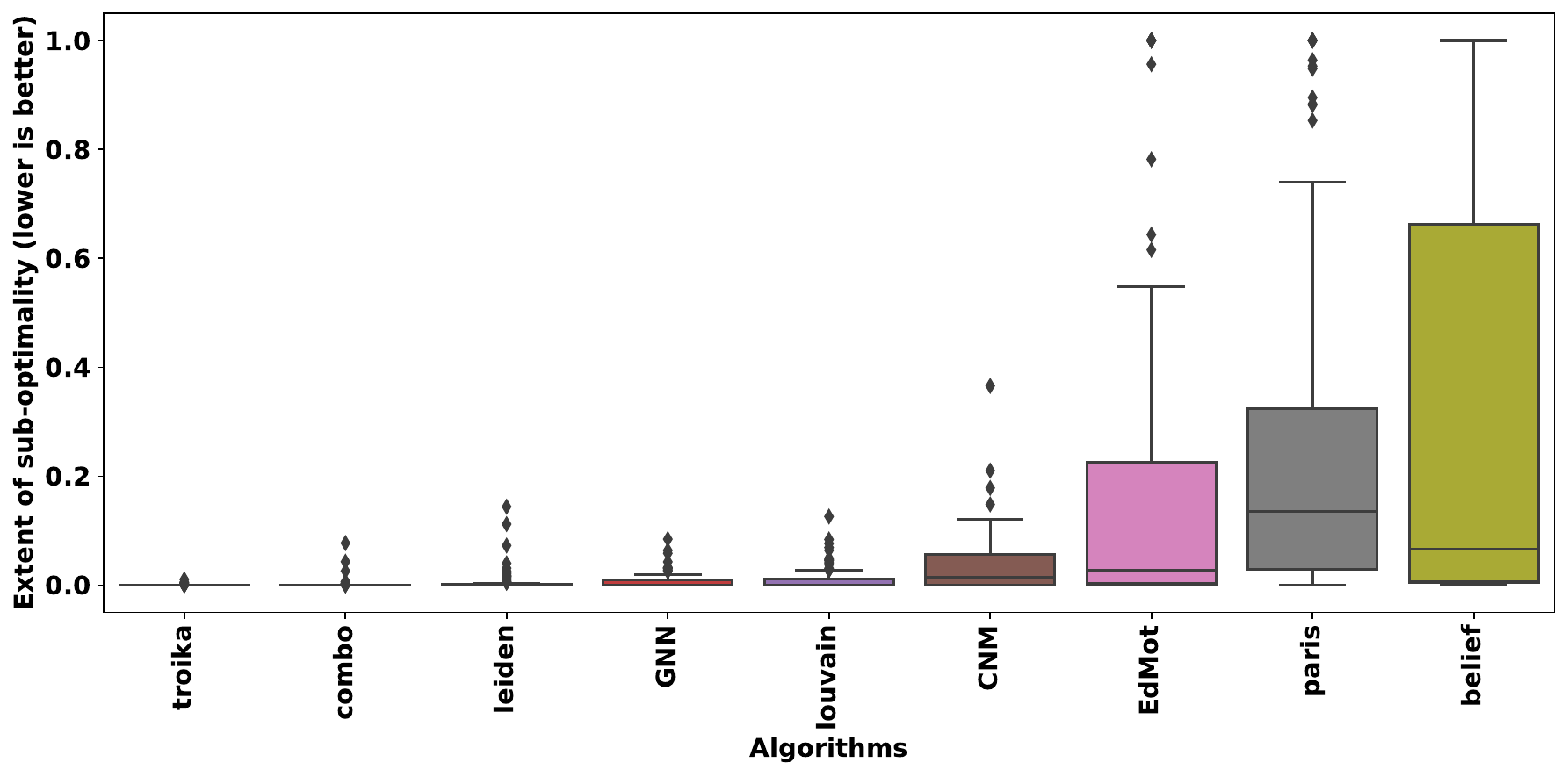}
		\caption{The boxplot of each algorithm illustrates the extent of sub-optimality for their partitions on the 100 real and synthetic networks.}
		\label{fig:sub-optimality}
	\end{figure}
	
	\subsection{Results on Troika for modularity maximization}
	
	Fig. \ref{fig:sub-optimality} shows the EOS for the partitions produced by nine algorithms including Troika, all of which aiming to maximize modularity. Among the nine algorithms, four algorithms have the best median performance indicated by a median EOS of zero: Troika, Combo (for MM), Leiden, and GNN. Among these algorithms, Troika has the lowest mean EOS. On these 100 networks, Troika returns globally optimal solutions for 88 networks. The median and mean of EOS for all algorithms are provided in Table \ref{tab:success} (in the appendices). The results in Fig. \ref{fig:sub-optimality} demonstrates that Troika can be reliably used to partition unsigned networks by approximating maximum modularity. In Section \ref{s:results-portfolio}, we move on to demonstrating a different use case of Troika: portfolio analysis.

	\section{Applicability of Troika in portfolio analysis}
	\label{s:results-portfolio}

	In portfolio analysis and management, gaining insights from the correlations between financial assets is paramount. Portfolio analysis is challenging due to the dynamic and networked nature of financial portfolios, where asset correlations play a crucial role in diversification-based risk reduction (e.g. hedging) and maximizing returns. In this section, we focus on using Troika's solutions for the CP problem to provide temporal and portfolio-level interpretations of stock return correlations within a well known market index from 2000 to 2024. The Standard and Poor's 500 market index (S\&P 500 for short) tracks the stock performance of 500 largest companies in the US stock exchanges. 
	
	\subsection{Creating weighted networks from correlations of returns}
	
	For each year between 2000 and 2024, we create a weighted network from the correlation matrix of the constituents (individual stocks within the S\&P 500). Almost all pairs of stocks have a non-zero correlation coefficient. Therefore, using all correlation coefficients as edge weights produces a complete graph without an particular structure. We create edges from correlations that are statistically significantly high (in absolute terms) using critical points of a normal distribution -2 and +2 (for the conventional significance threshold of 0.05). However, correlations data are often not normally distributed. Therefore, a transformation is required before using standard normal theory \cite{fisher1921probable}. The \textit{Fisher transformation} of correlation coefficients yields a variable that is approximately normally distributed \cite{fisher1921probable}. The Fisher transformation takes the correlation coefficient $r_{ij}$ corresponding to the stock $i$ and $j$ and returns $z_{ij}=0.5 \ln(\frac{1+r}{1-r})$. After the transformation, the distribution of transformed values $z_{ij}$ for each year can be used to decide which pairs have strong positive or strong negative correlations. We create an edge $(i,j)$ with the weight of $r_{ij}$ for each transformed variable $z_{ij}$ that differs from the mean $z$ value by more than 2 standard deviations. The data for these 25 financial networks are available in a FigShare repository \cite{Aref2025dataTroika}.
	
	
	\subsection{Troika's results on portfolio networks}
	
	In these networks, individual stocks are represented as nodes, and strong positive or strong negative correlations between them as weighted edges. Using the partitions returned by Troika on each network, we evaluate the changes in the clusters inferred from correlations of returns within the S\&P 500 index over 24 years.
	
	\begin{figure}[!ht]
		\centering
		\subfloat[2000]{\includegraphics[trim={0 4.5cm 0 4.5cm},clip,width=0.2\textwidth]{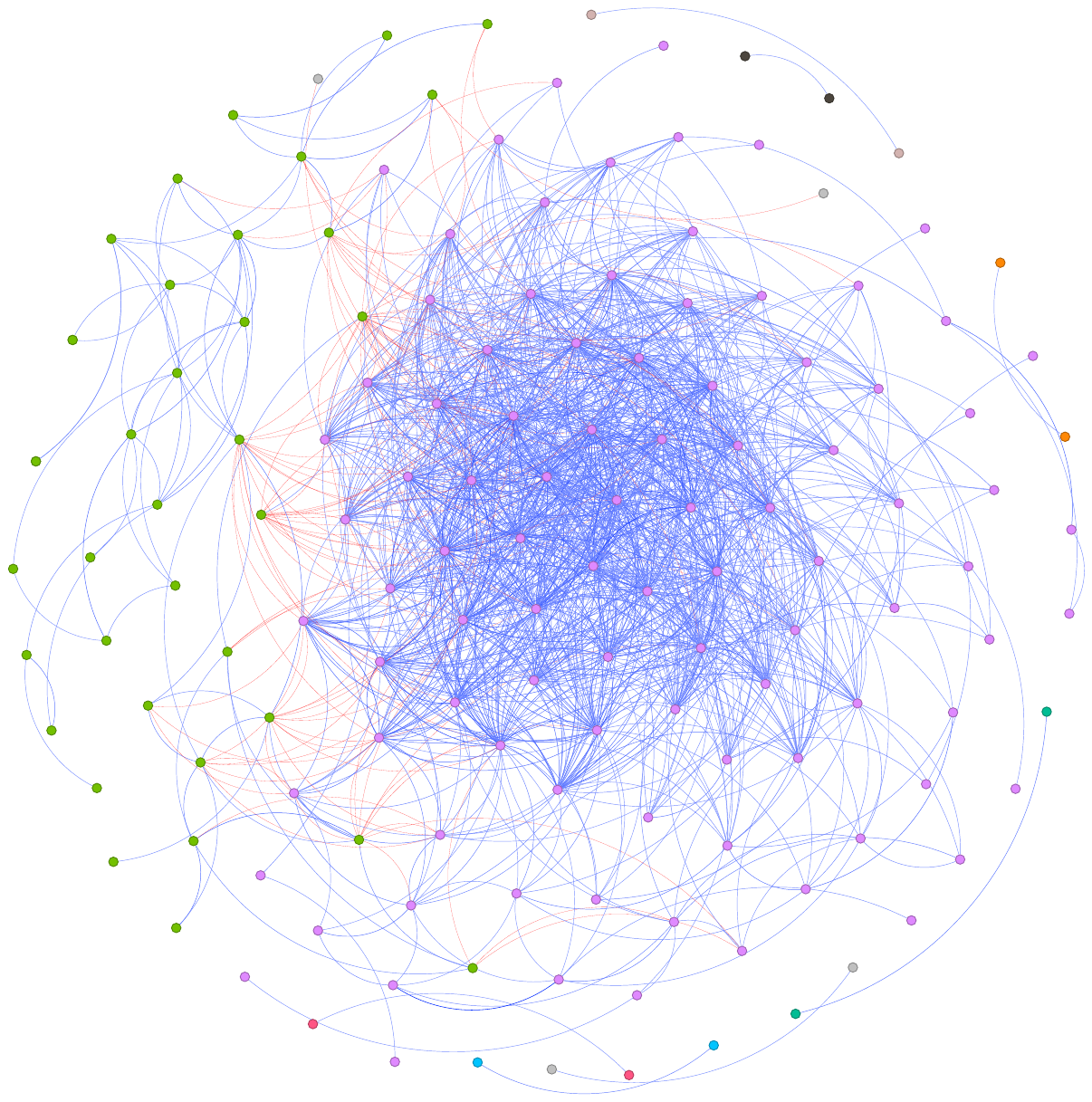}\label{fig:sub1}}
		\subfloat[2001]{\includegraphics[trim={0 4.5cm 0 4.5cm},clip,width=0.2\textwidth]{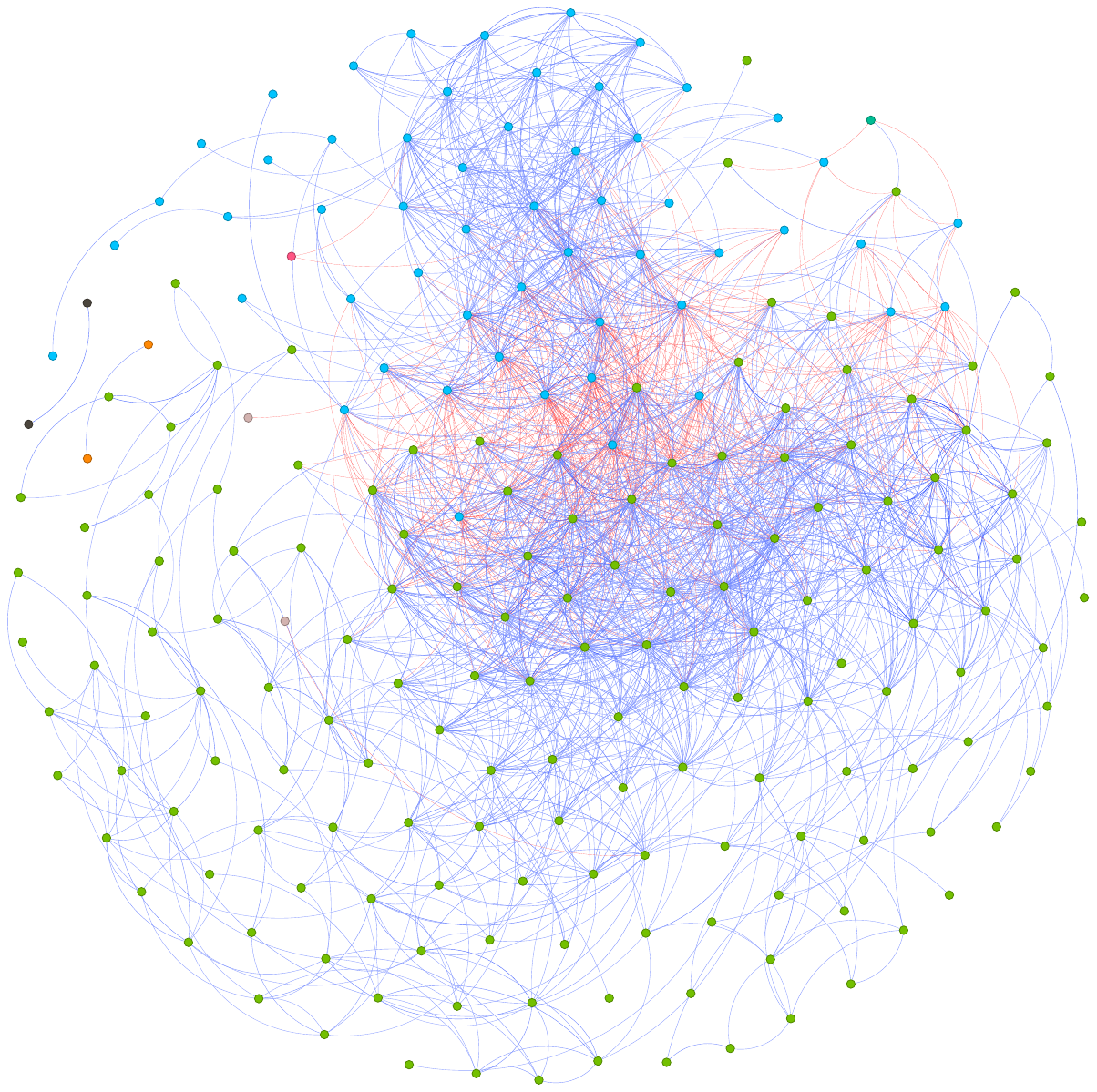}\label{fig:sub2}}
		\subfloat[2002]{\includegraphics[trim={0 4.5cm 0 4.5cm},clip,width=0.2\textwidth]{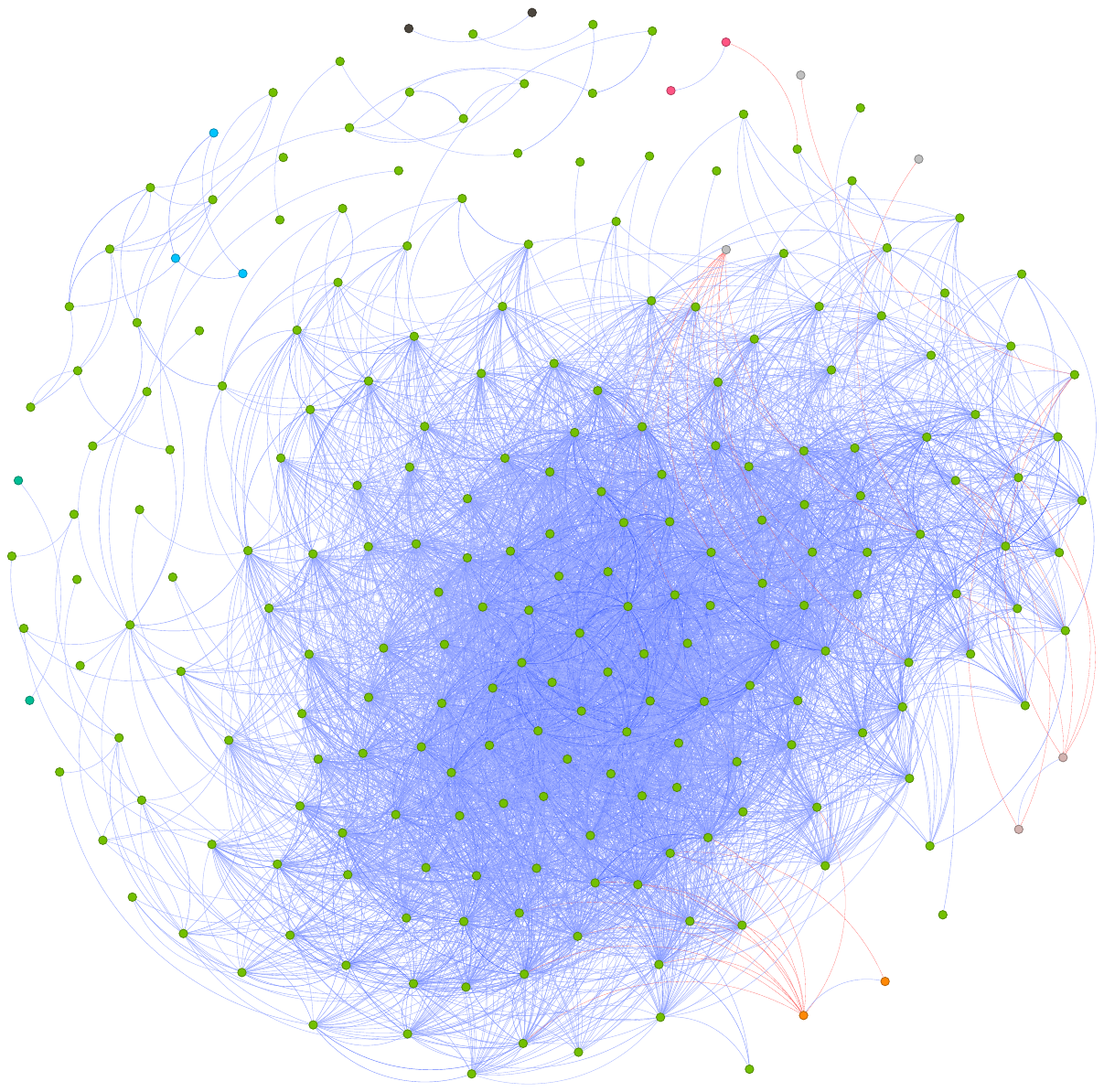}\label{fig:sub3}}
		\subfloat[2003]{\includegraphics[trim={0 4.5cm 0 4.5cm},clip,width=0.2\textwidth]{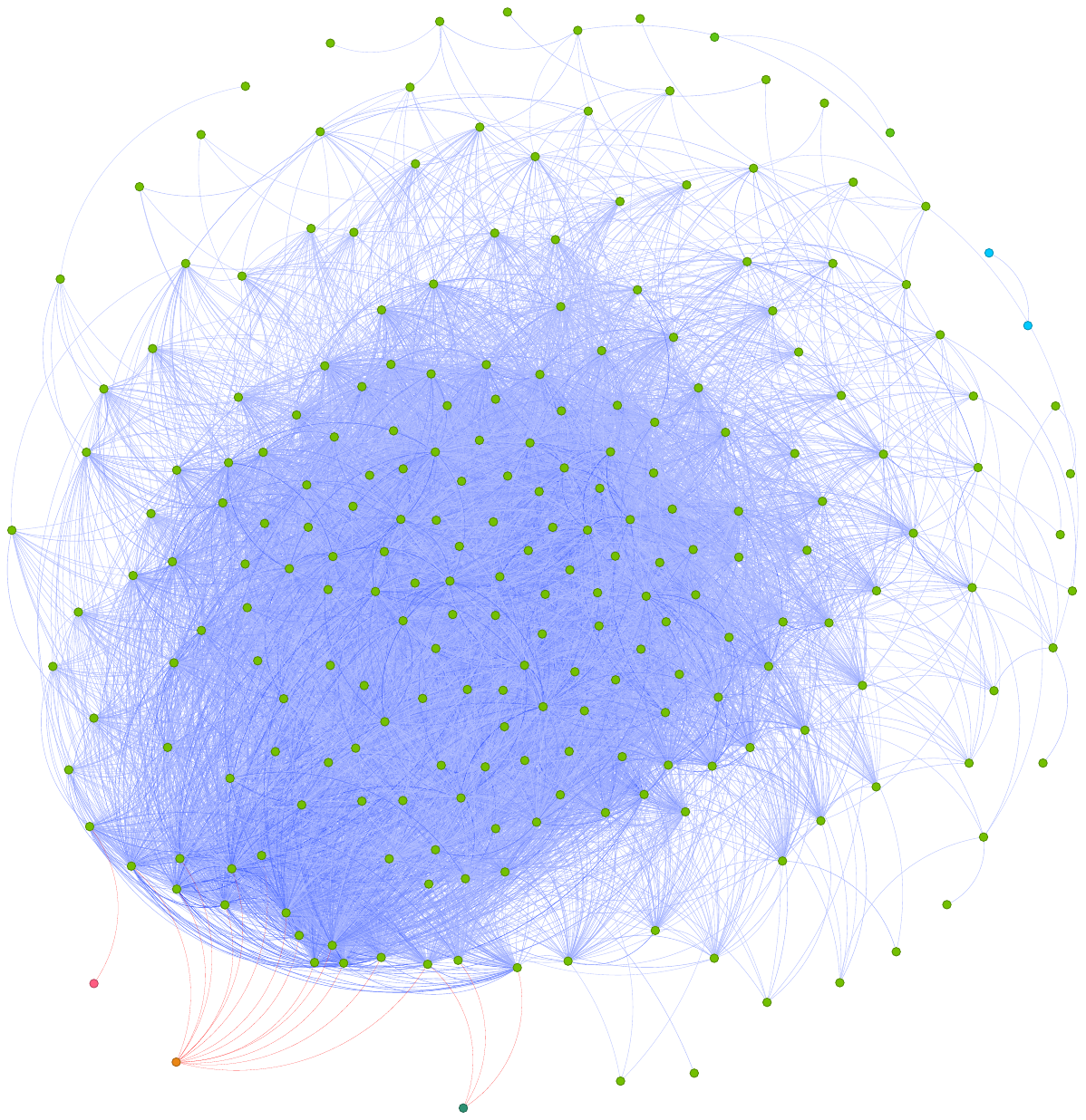}\label{fig:sub4}}
		\subfloat[2004]{\includegraphics[trim={0 4.5cm 0 4.5cm},clip,width=0.2\textwidth]{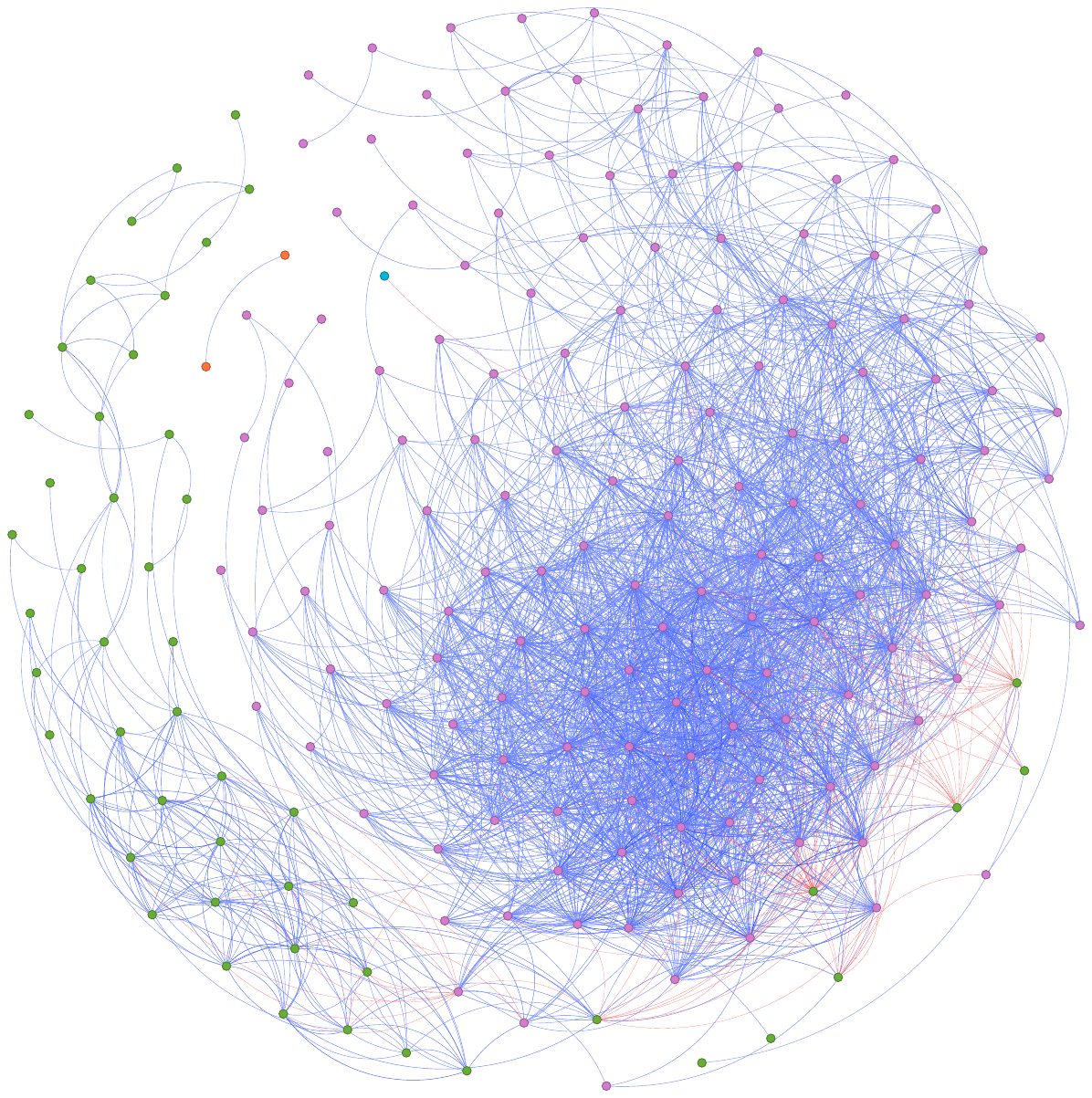}\label{fig:sub5}}\\
		\subfloat[2005]{\includegraphics[trim={0 4.5cm 0 4.5cm},clip,width=0.2\textwidth]{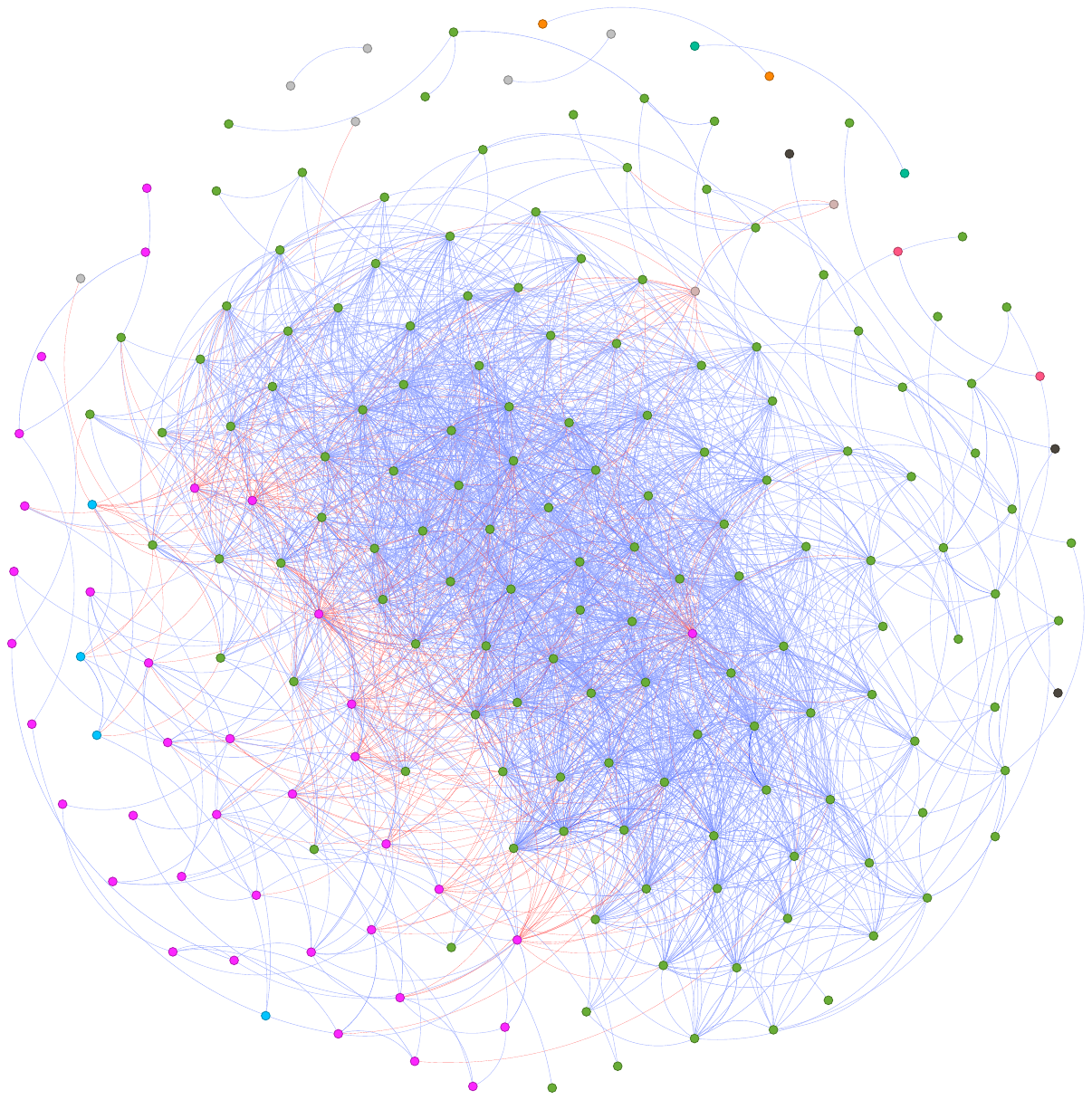}\label{fig:sub6}}
		\subfloat[2006]{\includegraphics[trim={0 4.5cm 0 4.5cm},clip,width=0.2\textwidth]{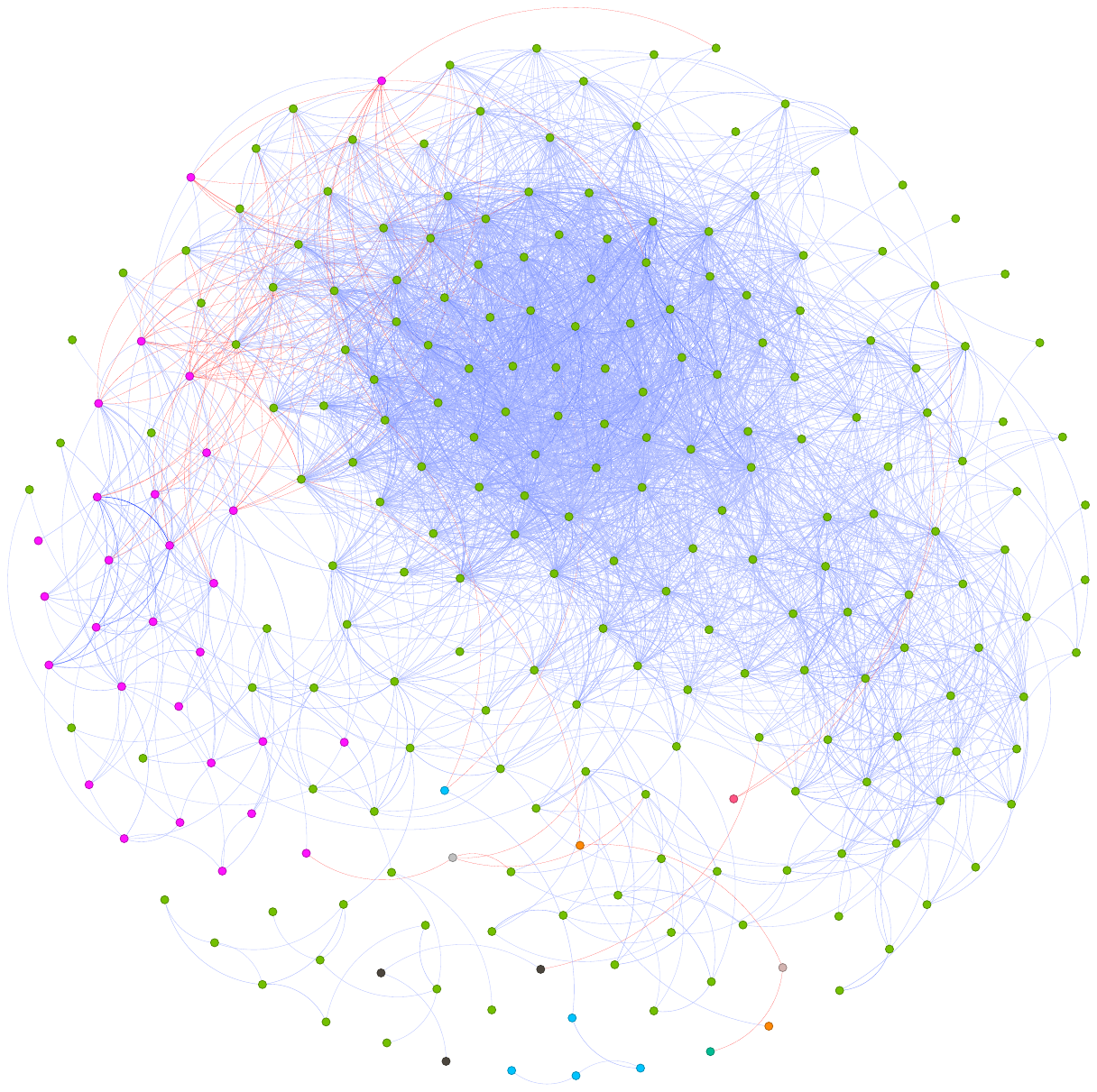}\label{fig:sub7}}
		\subfloat[2007]{\includegraphics[trim={0 4.5cm 0 4.5cm},clip,width=0.2\textwidth]{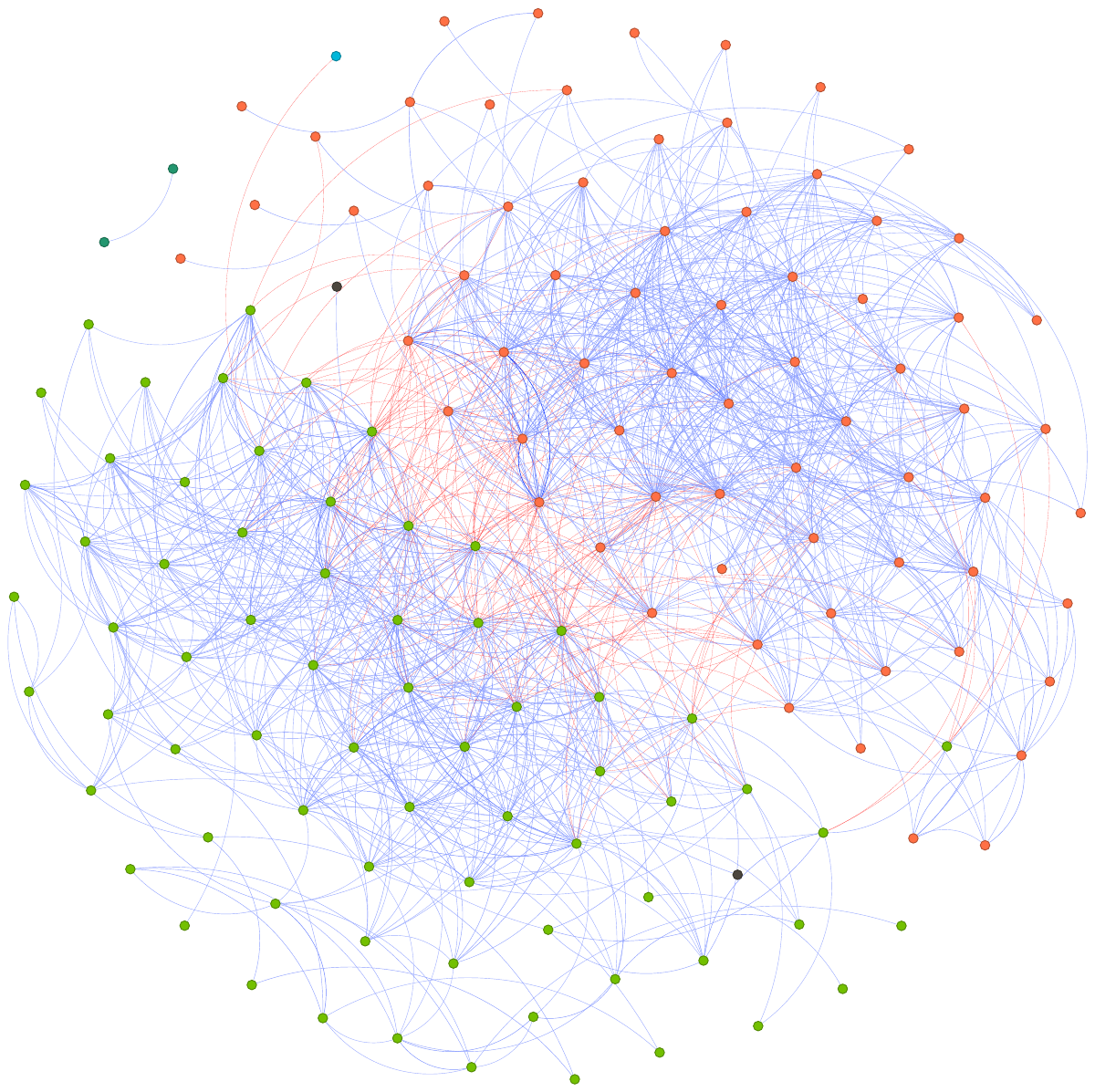}\label{fig:sub8}}
		\subfloat[2008]{\includegraphics[trim={0 4.5cm 0 4.5cm},clip,width=0.2\textwidth]{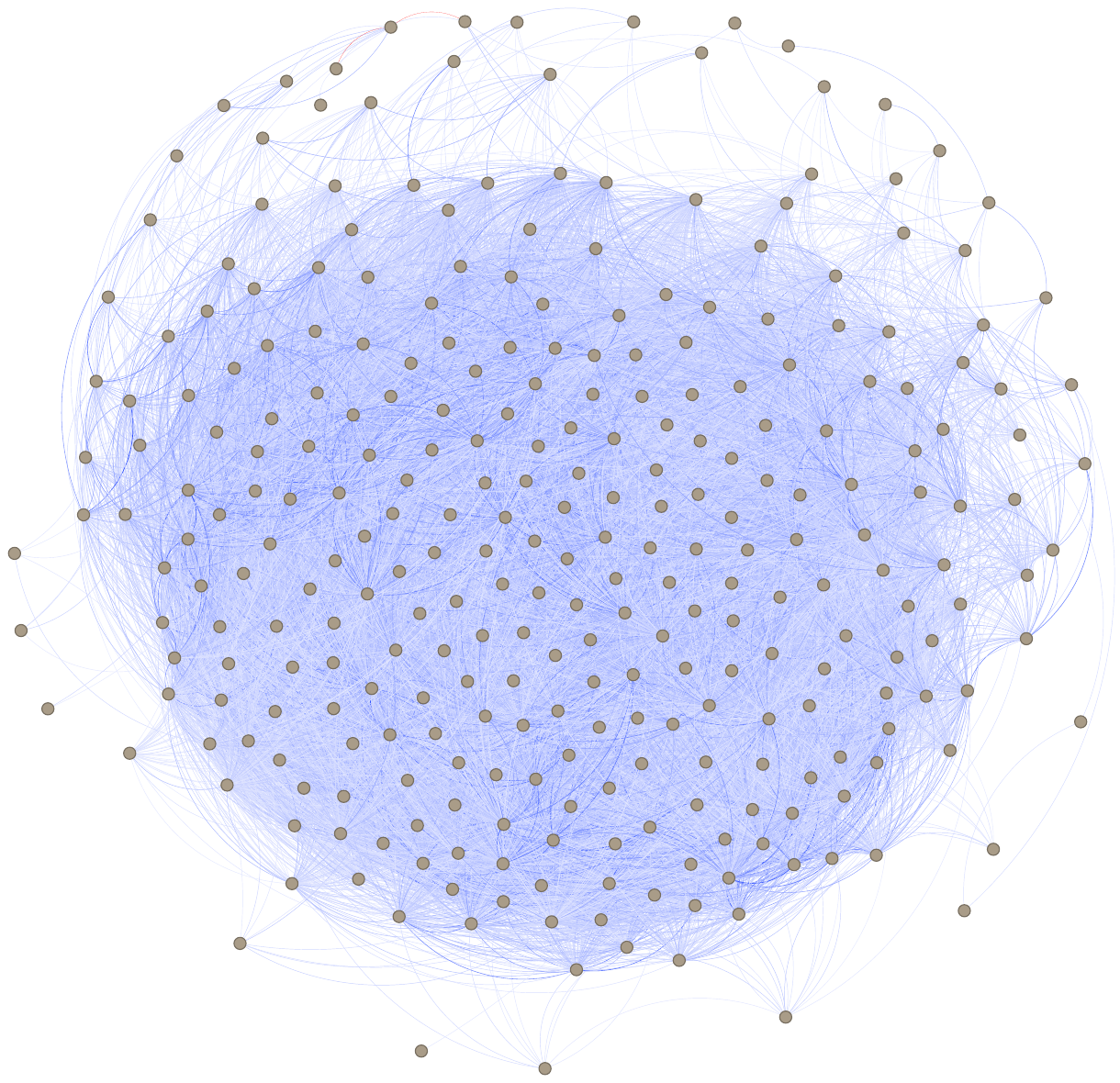}\label{fig:sub9}}
		\subfloat[2009]{\includegraphics[trim={0 4.5cm 0 4.5cm},clip,width=0.2\textwidth]{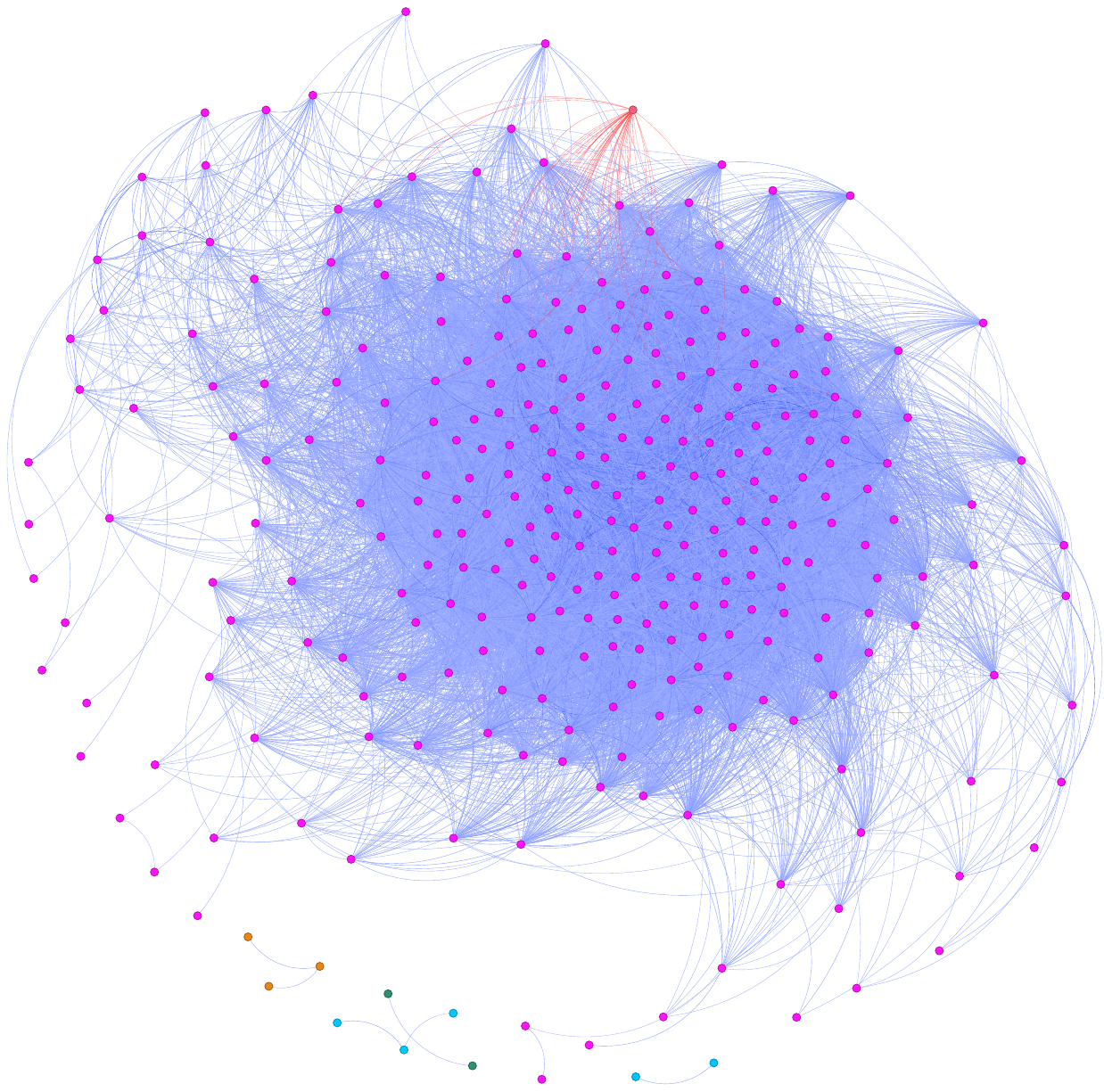}\label{fig:sub10}}\\
		\subfloat[2010]{\includegraphics[trim={0 4.5cm 0 4.5cm},clip,width=0.2\textwidth]{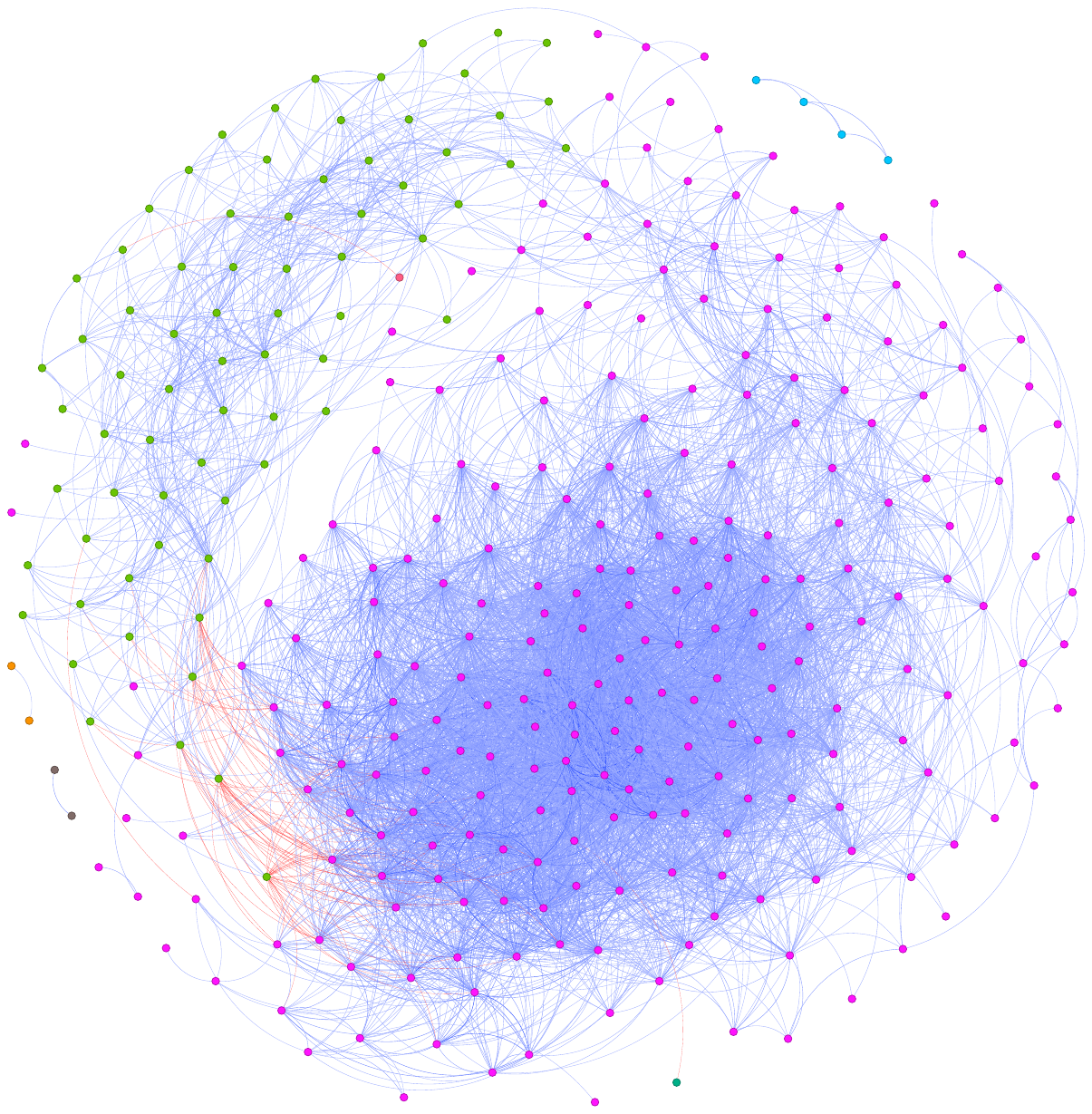}\label{fig:sub11}}
		\subfloat[2011]{\includegraphics[trim={0 4.5cm 0 4.5cm},clip,width=0.2\textwidth]{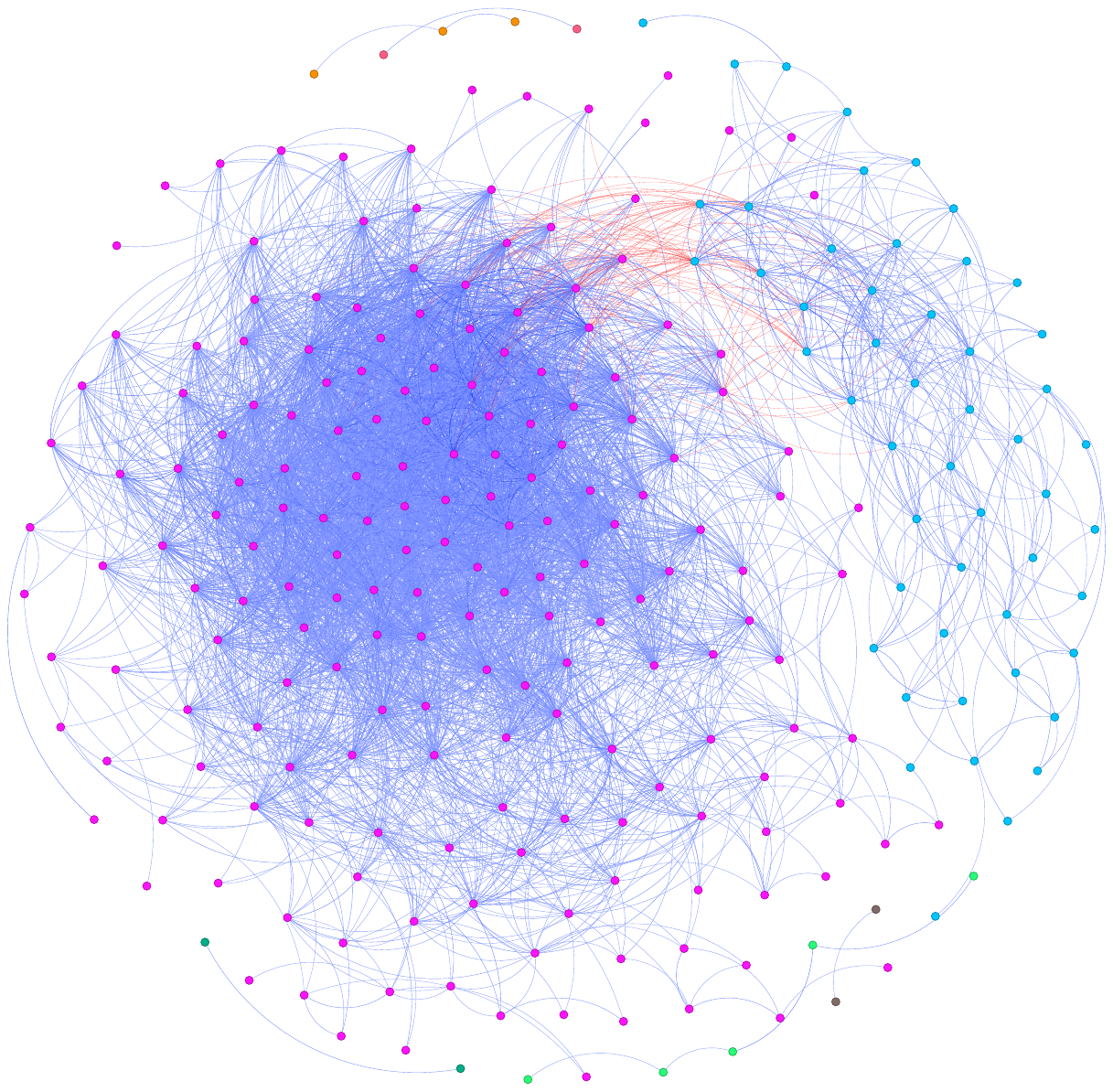}\label{fig:sub12}}
		\subfloat[2012]{\includegraphics[trim={0 4.5cm 0 4.5cm},clip,width=0.2\textwidth]{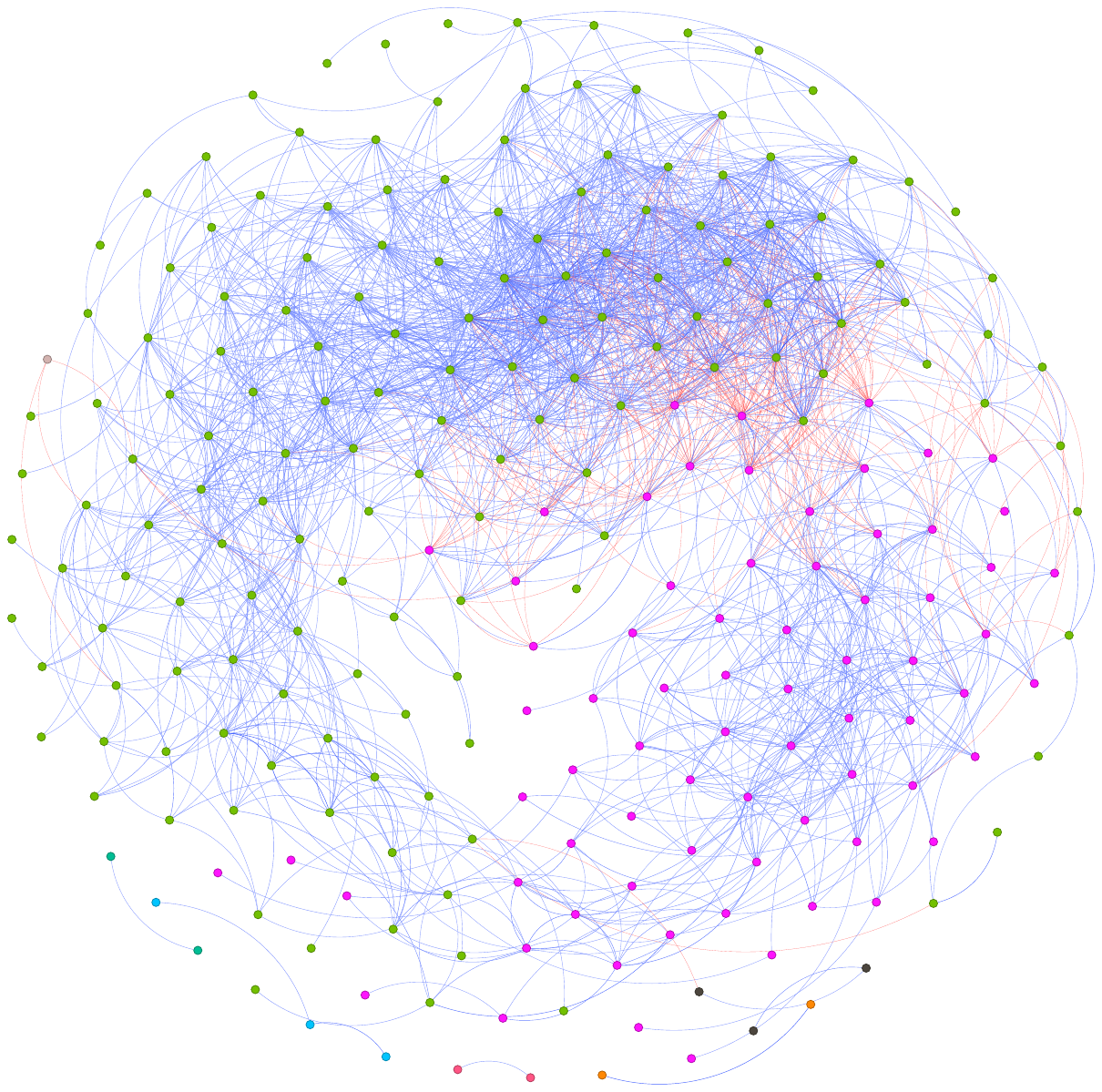}\label{fig:sub13}}
		\subfloat[2013]{\includegraphics[trim={0 4.5cm 0 4.5cm},clip,width=0.2\textwidth]{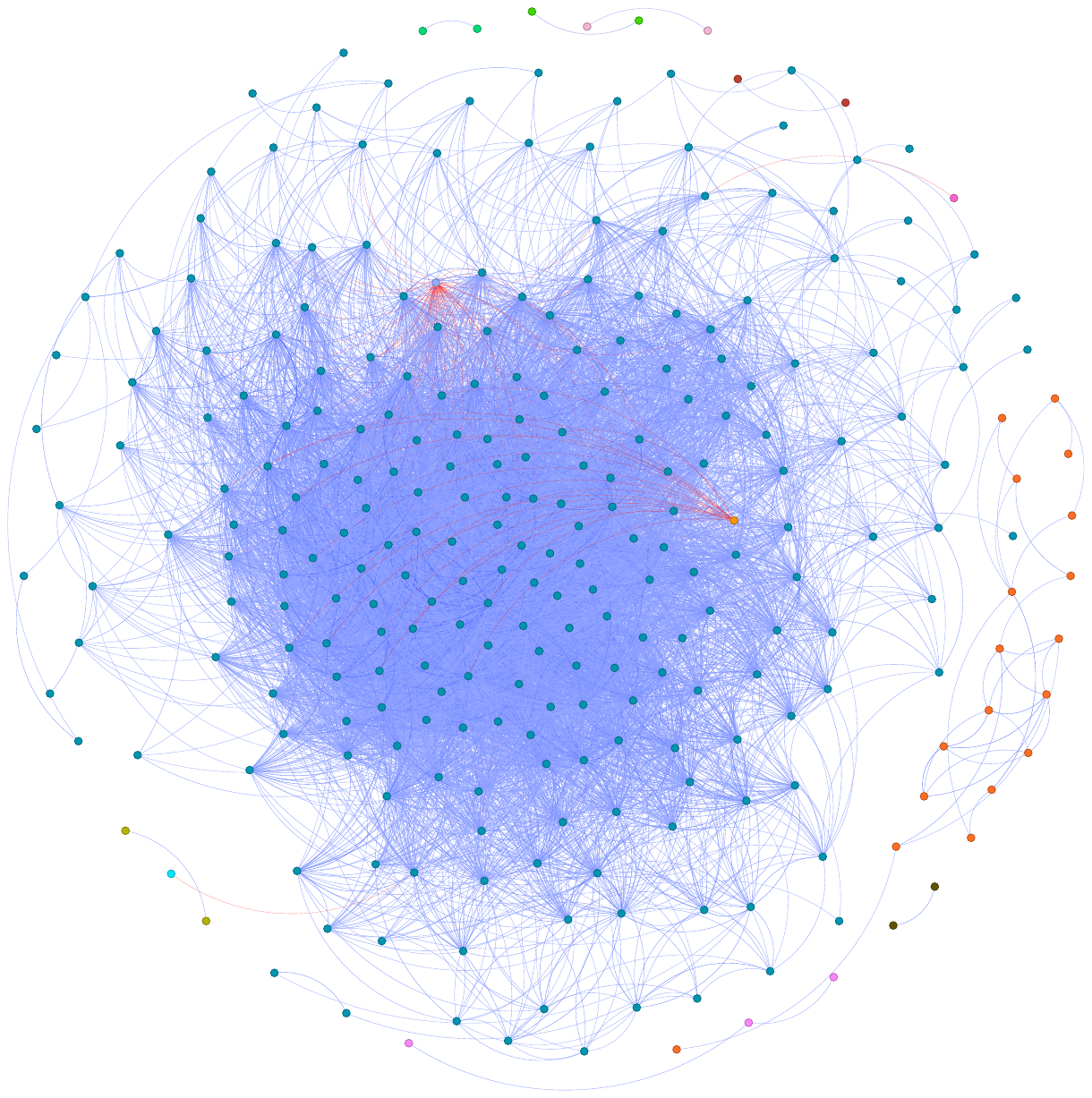}\label{fig:sub14}}
		\subfloat[2014]{\includegraphics[trim={0 4.5cm 0 4.5cm},clip,width=0.2\textwidth]{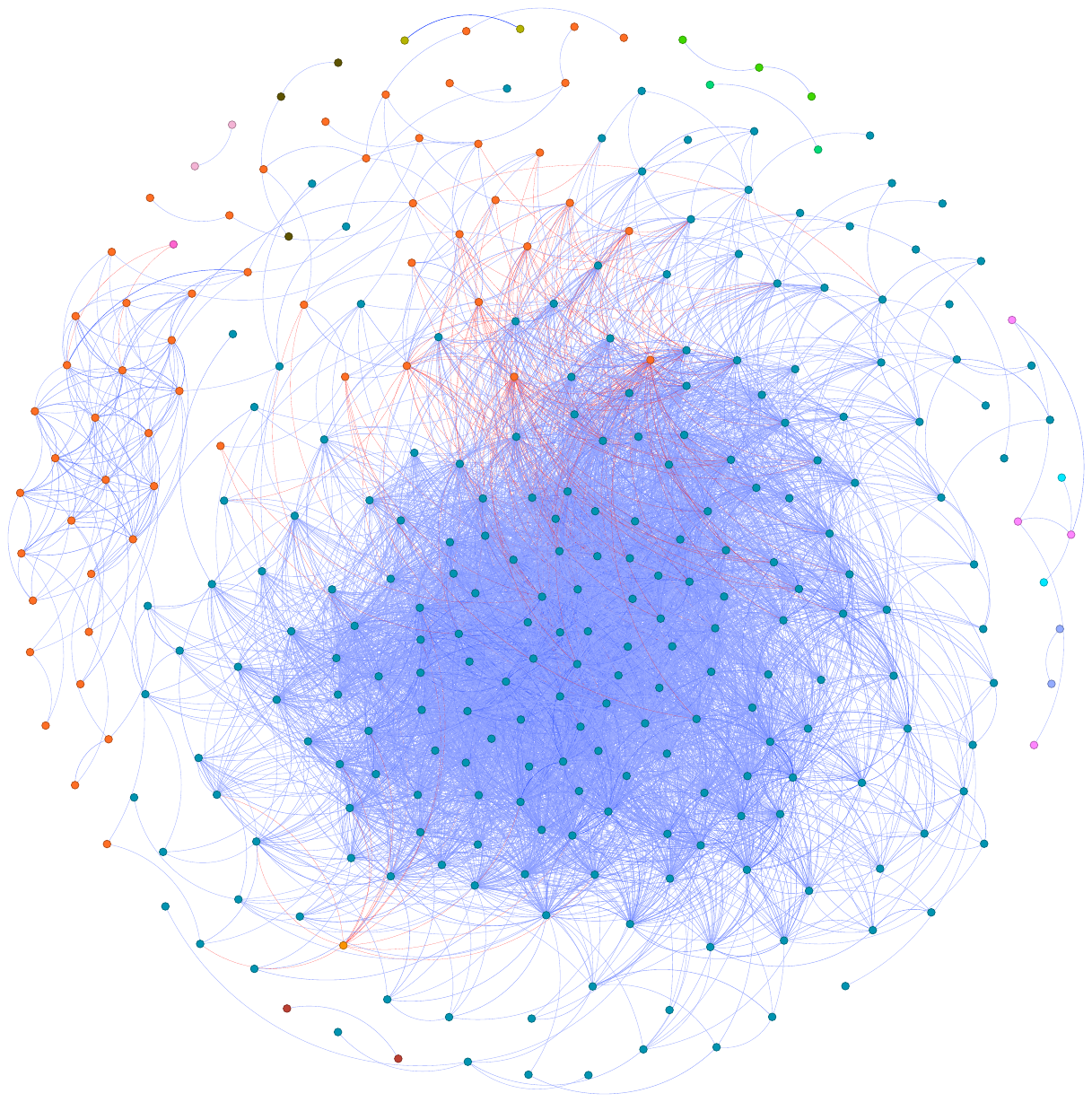}\label{fig:sub15}}\\
		\subfloat[2015]{\includegraphics[trim={0 4.5cm 0 4.5cm},clip,width=0.2\textwidth]{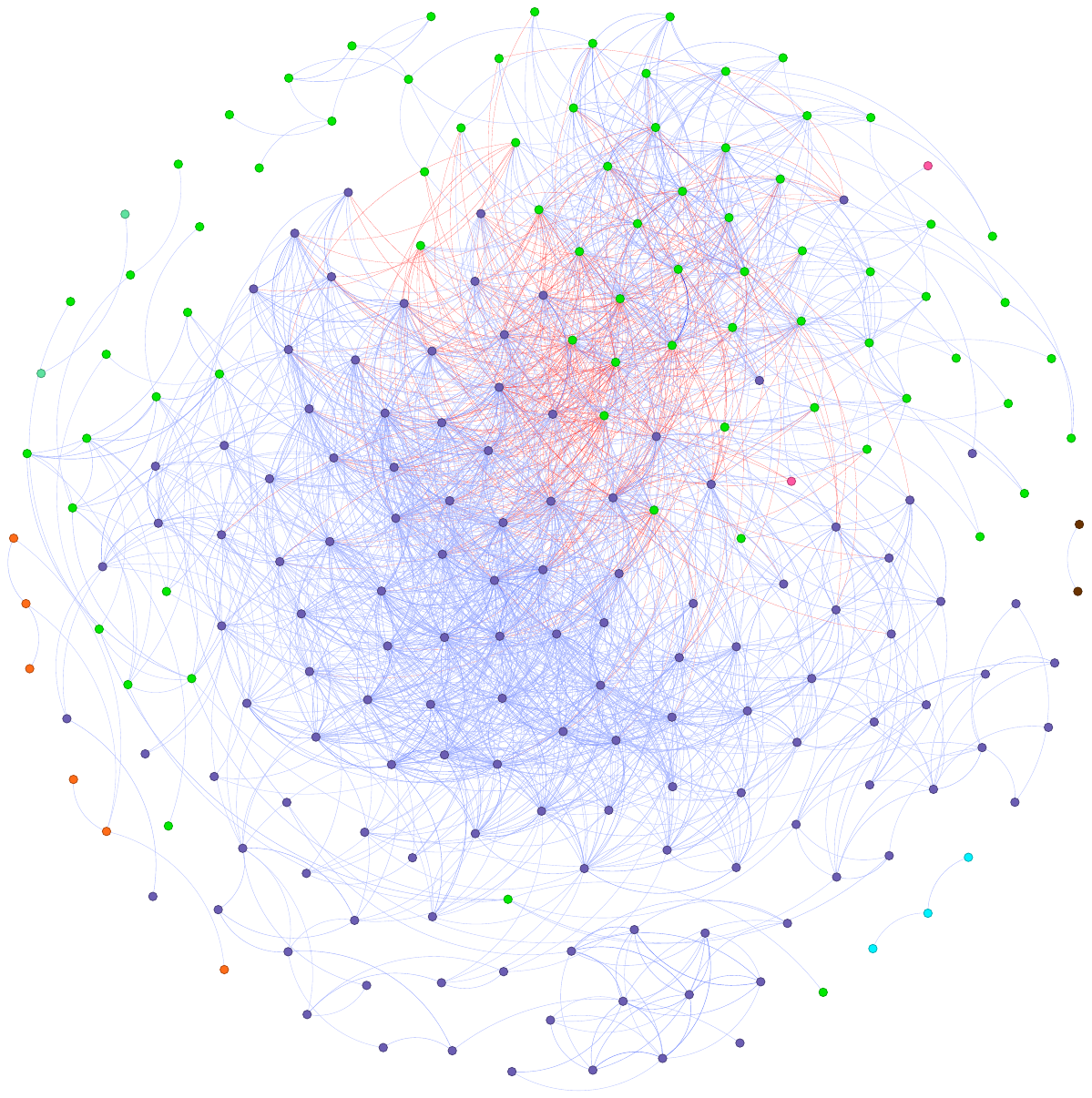}\label{fig:sub16}}
		\subfloat[2016]{\includegraphics[trim={0 4.5cm 0 4.5cm},clip,width=0.2\textwidth]{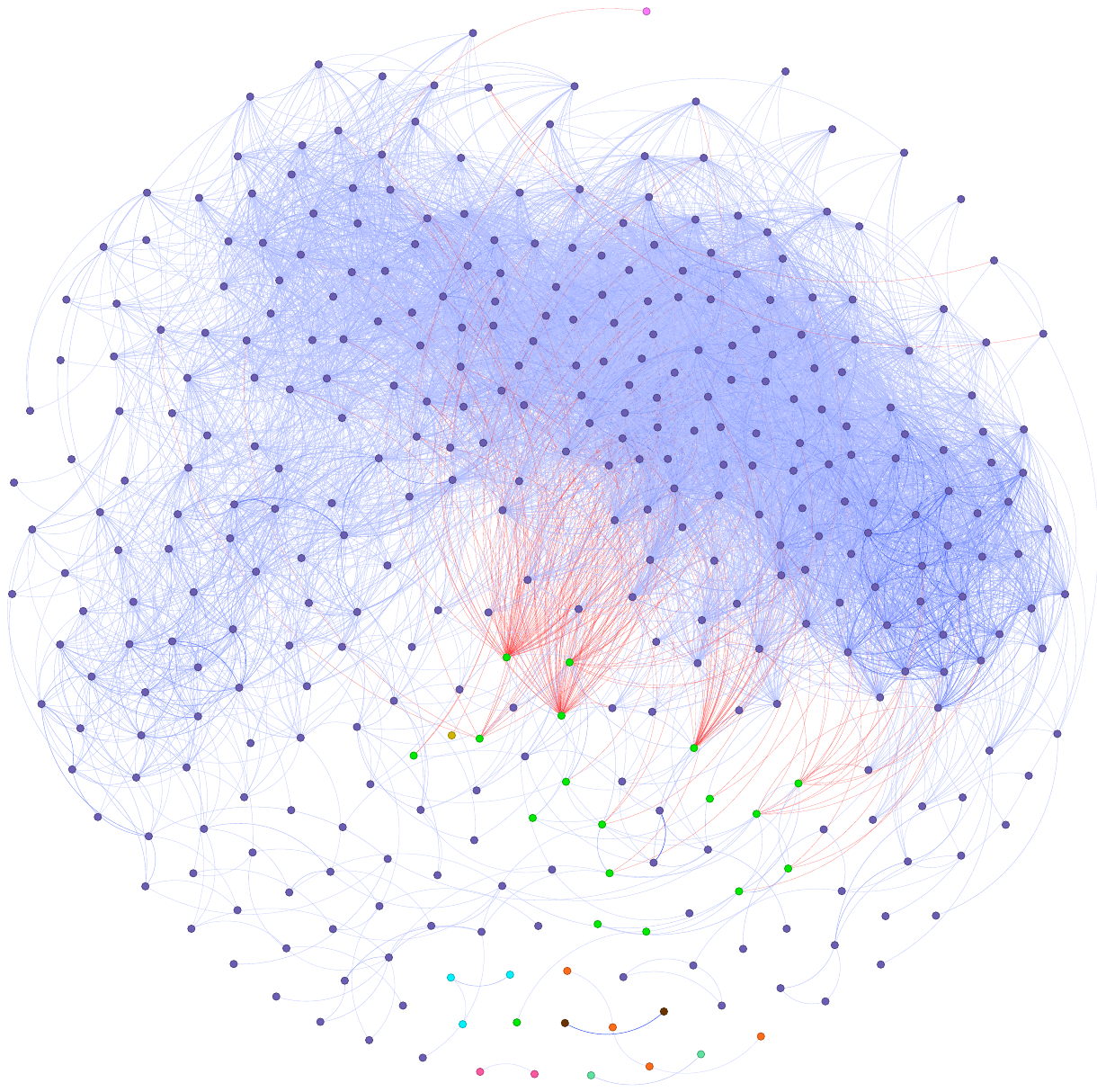}\label{fig:sub17}}
		\subfloat[2017]{\includegraphics[trim={0 4.5cm 0 4.5cm},clip,width=0.2\textwidth]{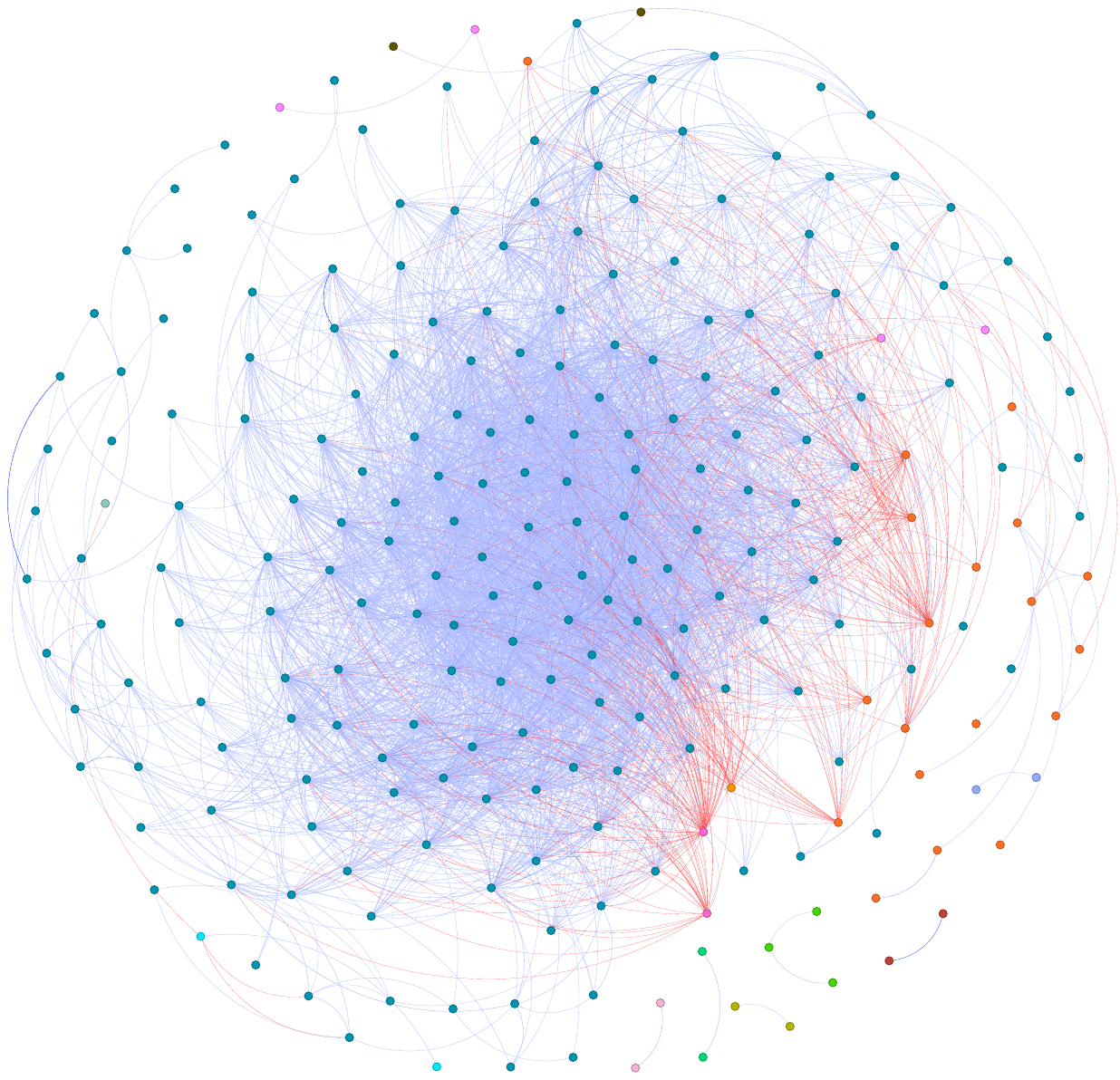}\label{fig:sub18}}
		\subfloat[2018]{\includegraphics[trim={0 4.5cm 0 4.5cm},clip,width=0.2\textwidth]{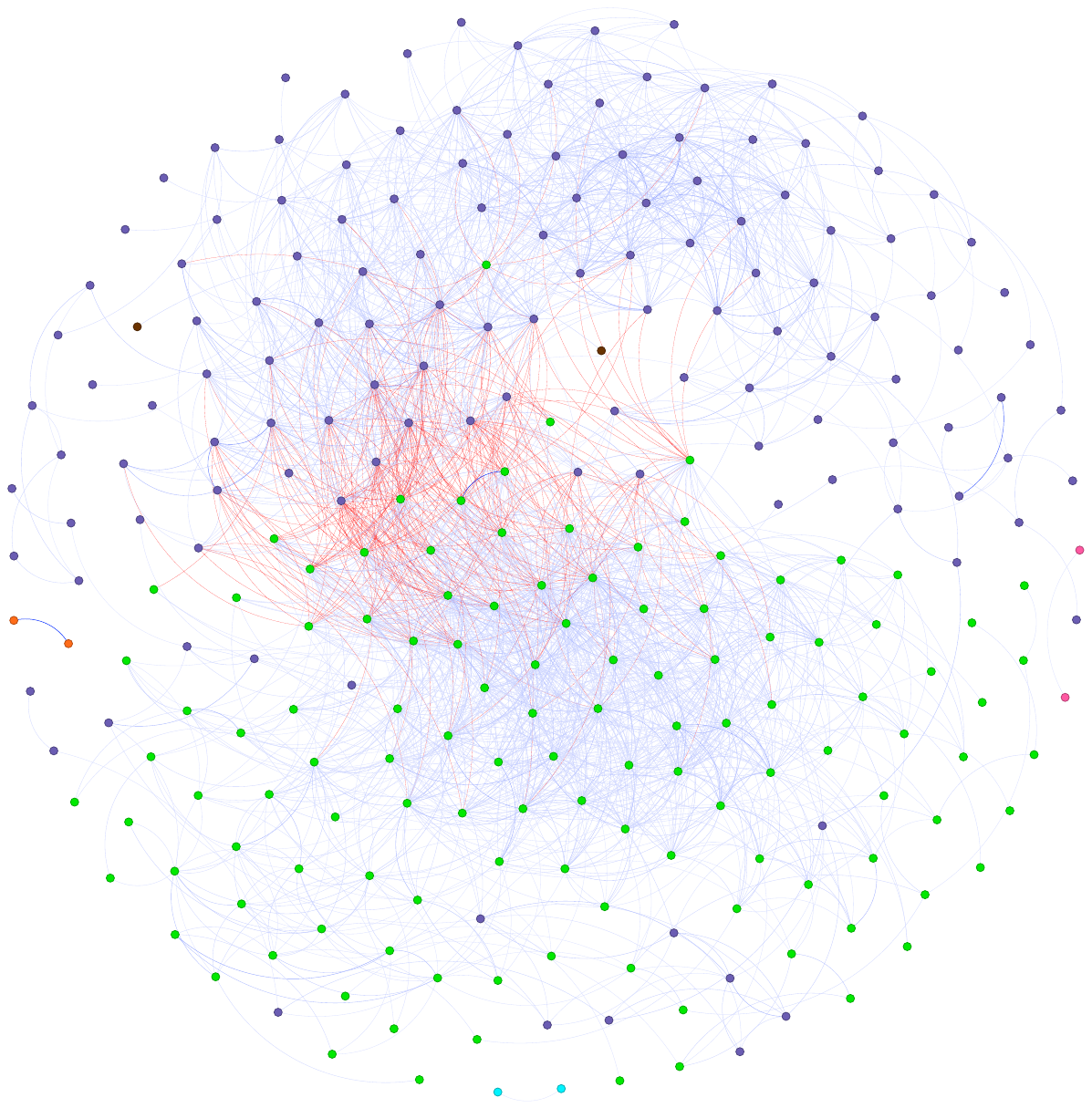}\label{fig:sub19}}
		\subfloat[2019]{\includegraphics[trim={0 4.5cm 0 4.5cm},clip,width=0.2\textwidth]{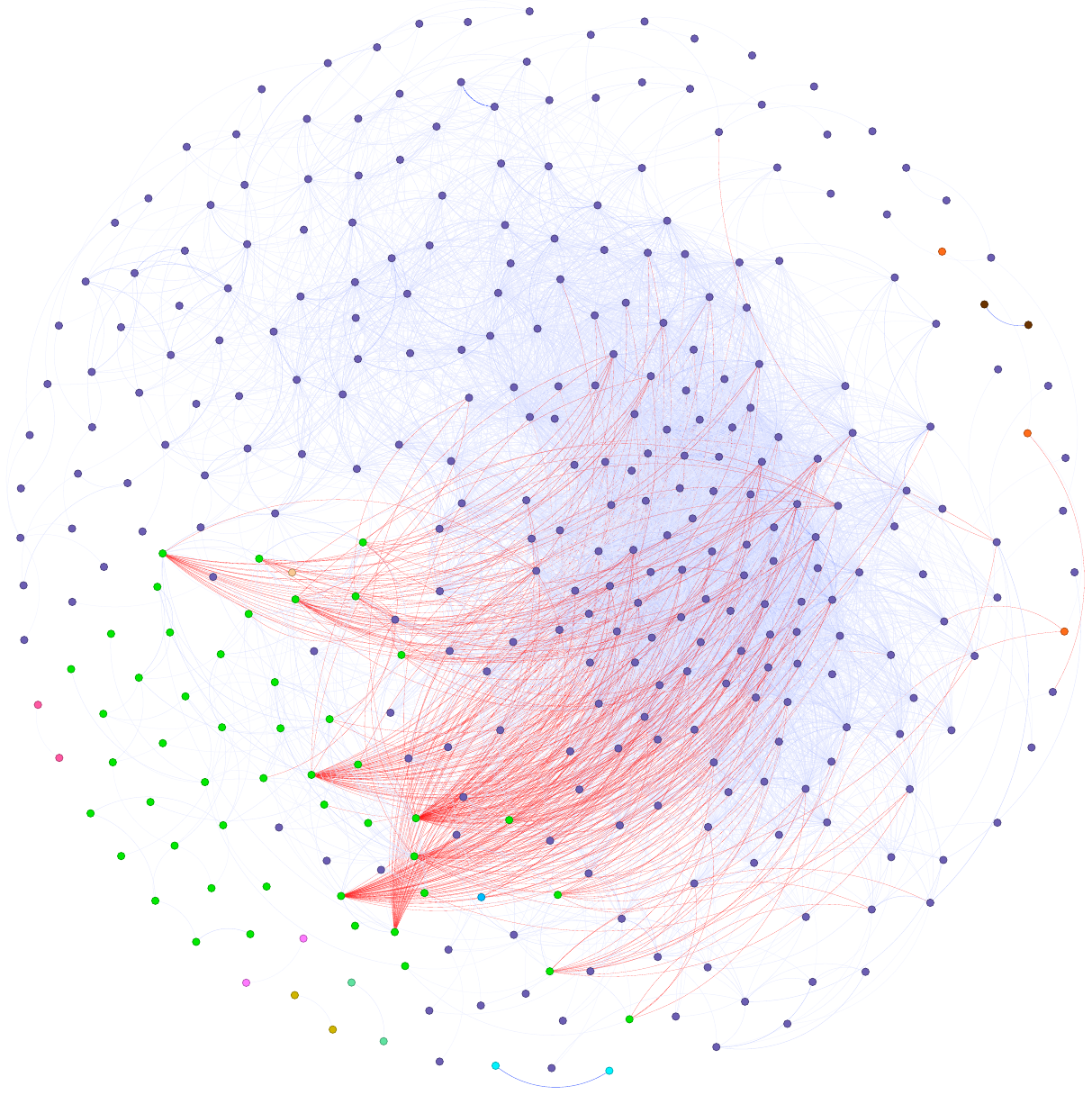}\label{fig:sub20}}\\
		\subfloat[2020]{\includegraphics[trim={0 4.5cm 0 4.5cm},clip,width=0.2\textwidth]{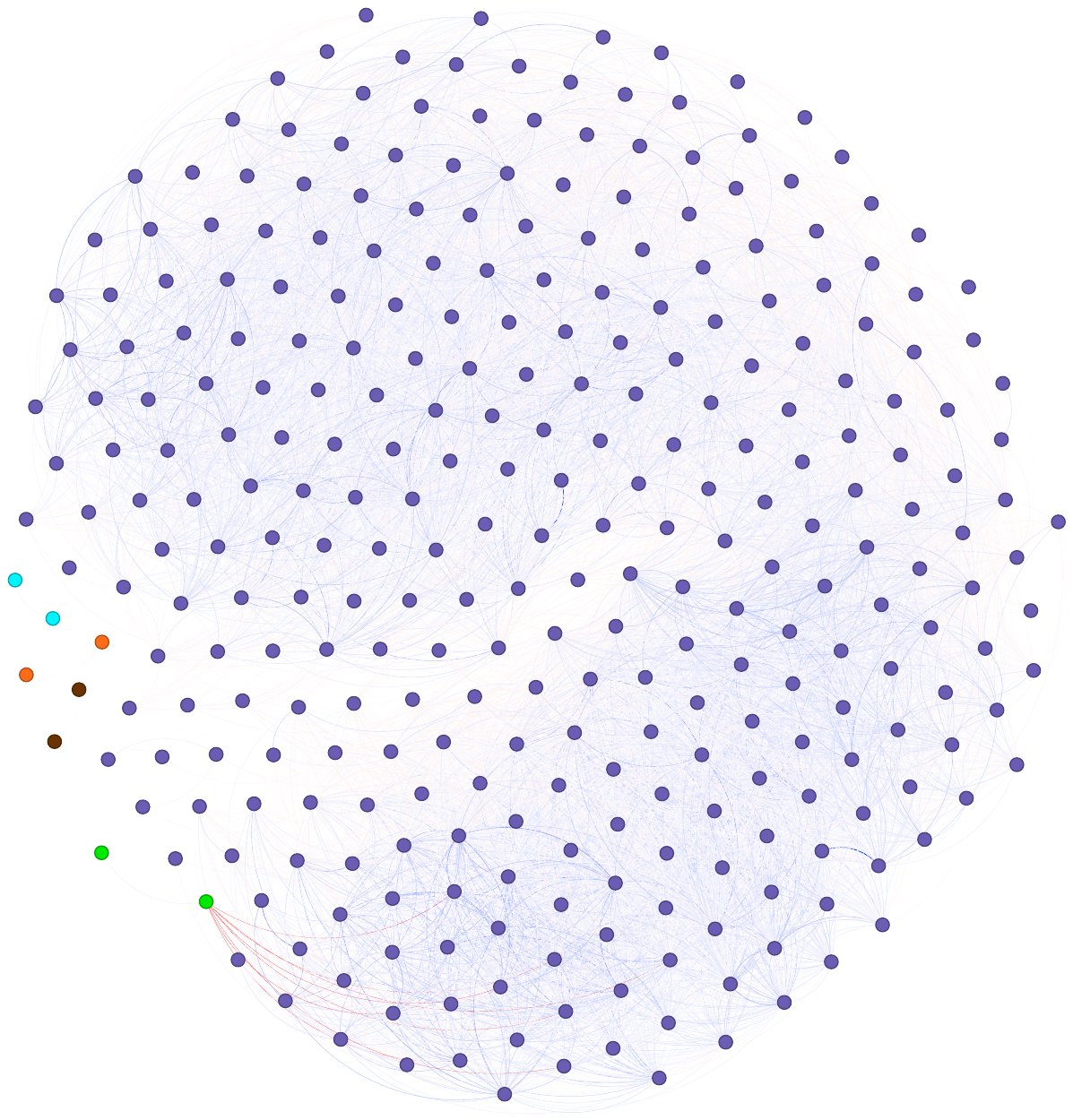}\label{fig:sub21}}
		\subfloat[2021]{\includegraphics[trim={0 4.5cm 0 4.5cm},clip,width=0.2\textwidth]{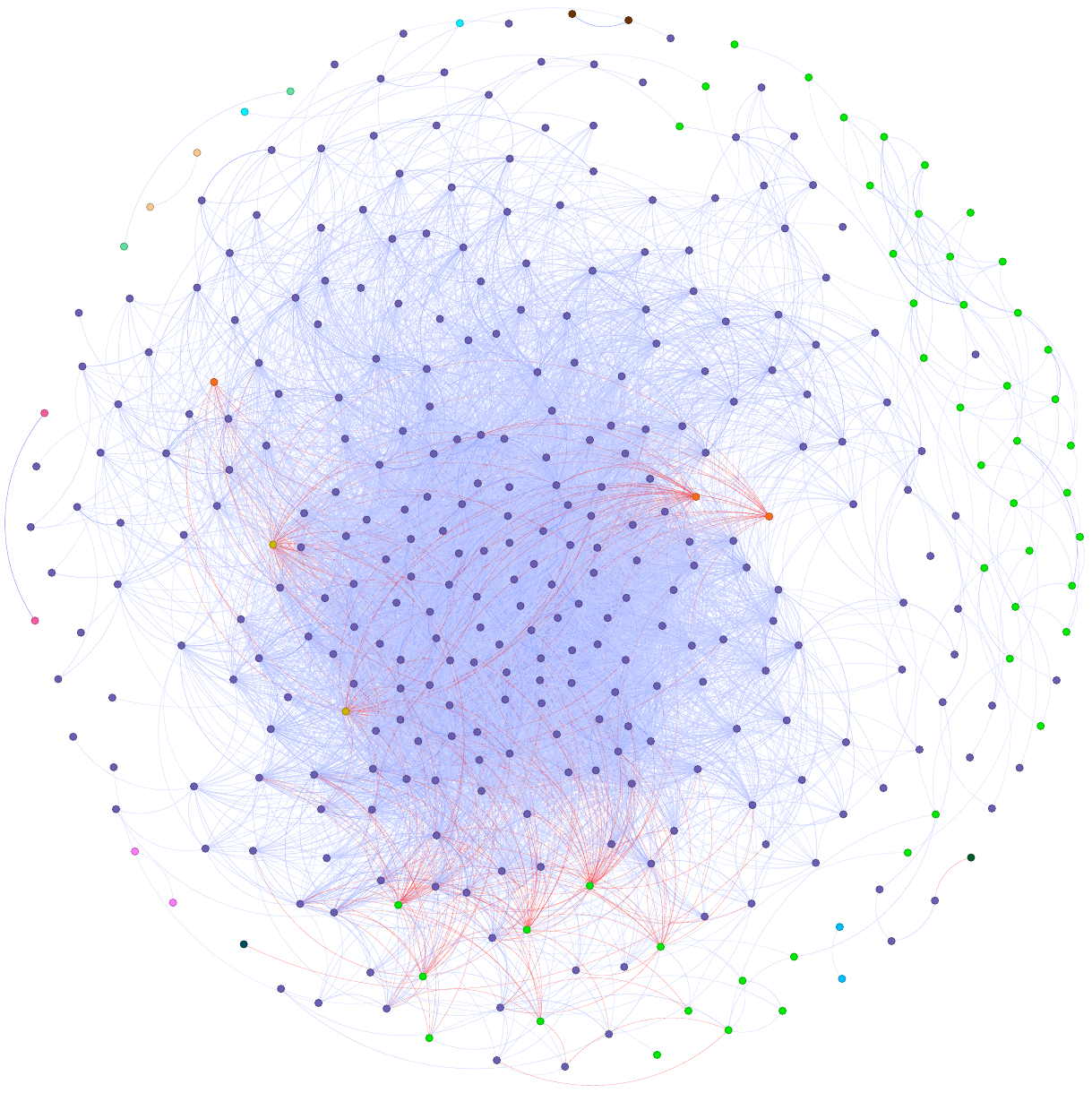}\label{fig:sub22}}
		\subfloat[2022]{\includegraphics[trim={0 4.5cm 0 4.5cm},clip,width=0.2\textwidth]{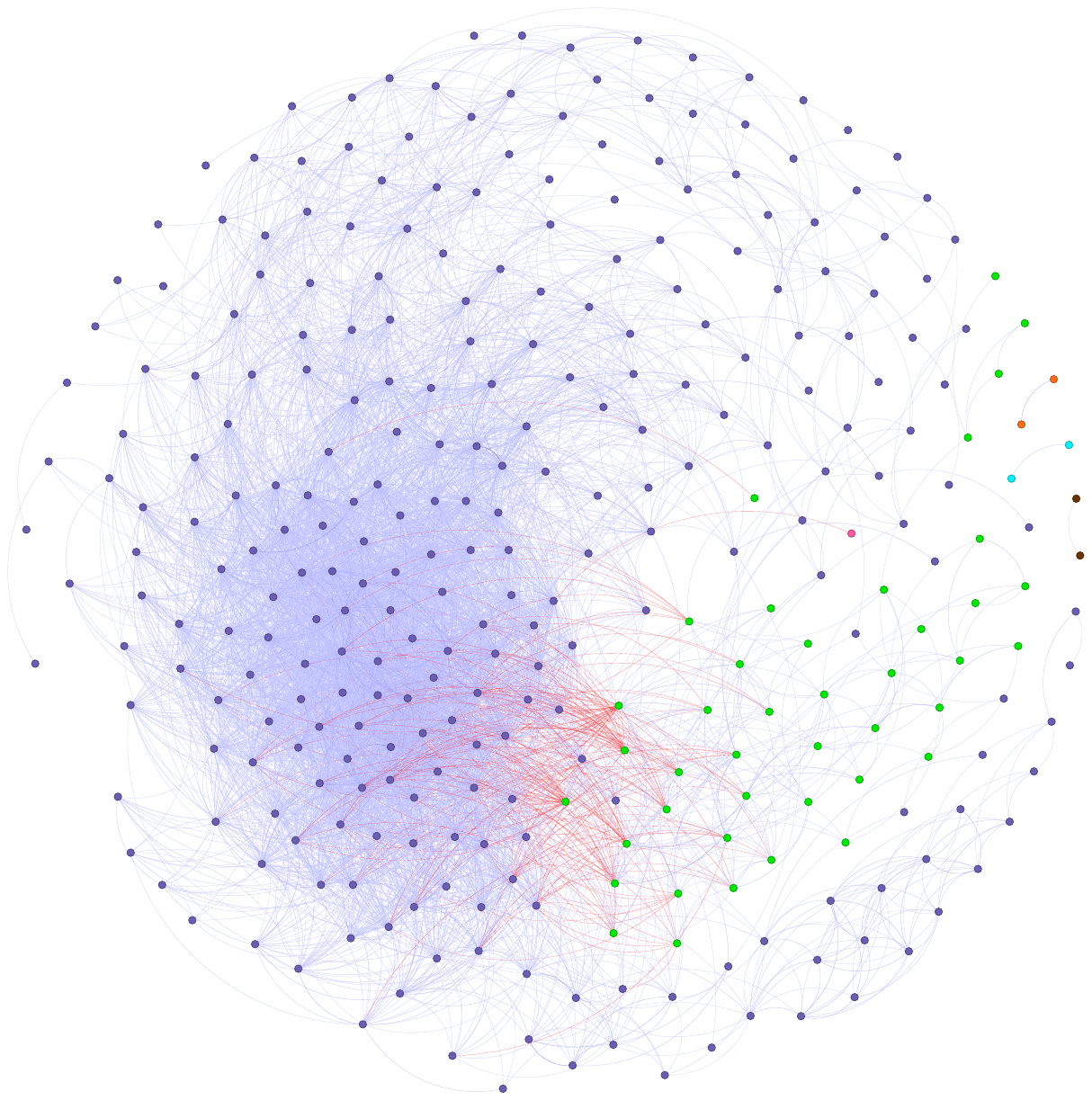}\label{fig:sub23}}
		\subfloat[2023]{\includegraphics[trim={0 4.5cm 0 4.5cm},clip,width=0.2\textwidth]{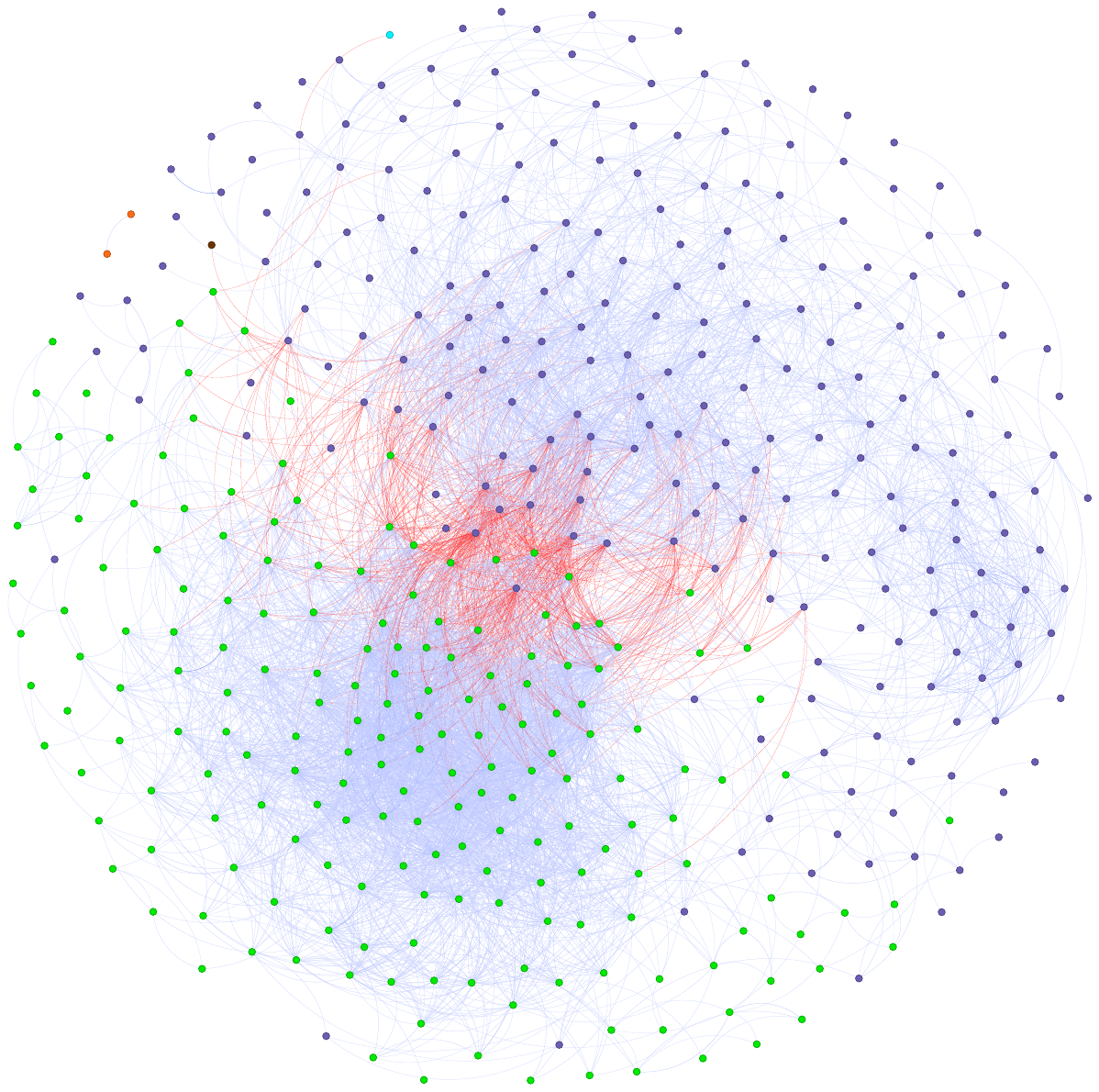}\label{fig:sub24}}
		\subfloat[2024]{\includegraphics[trim={0 4.5cm 0 4.5cm},clip,width=0.2\textwidth]{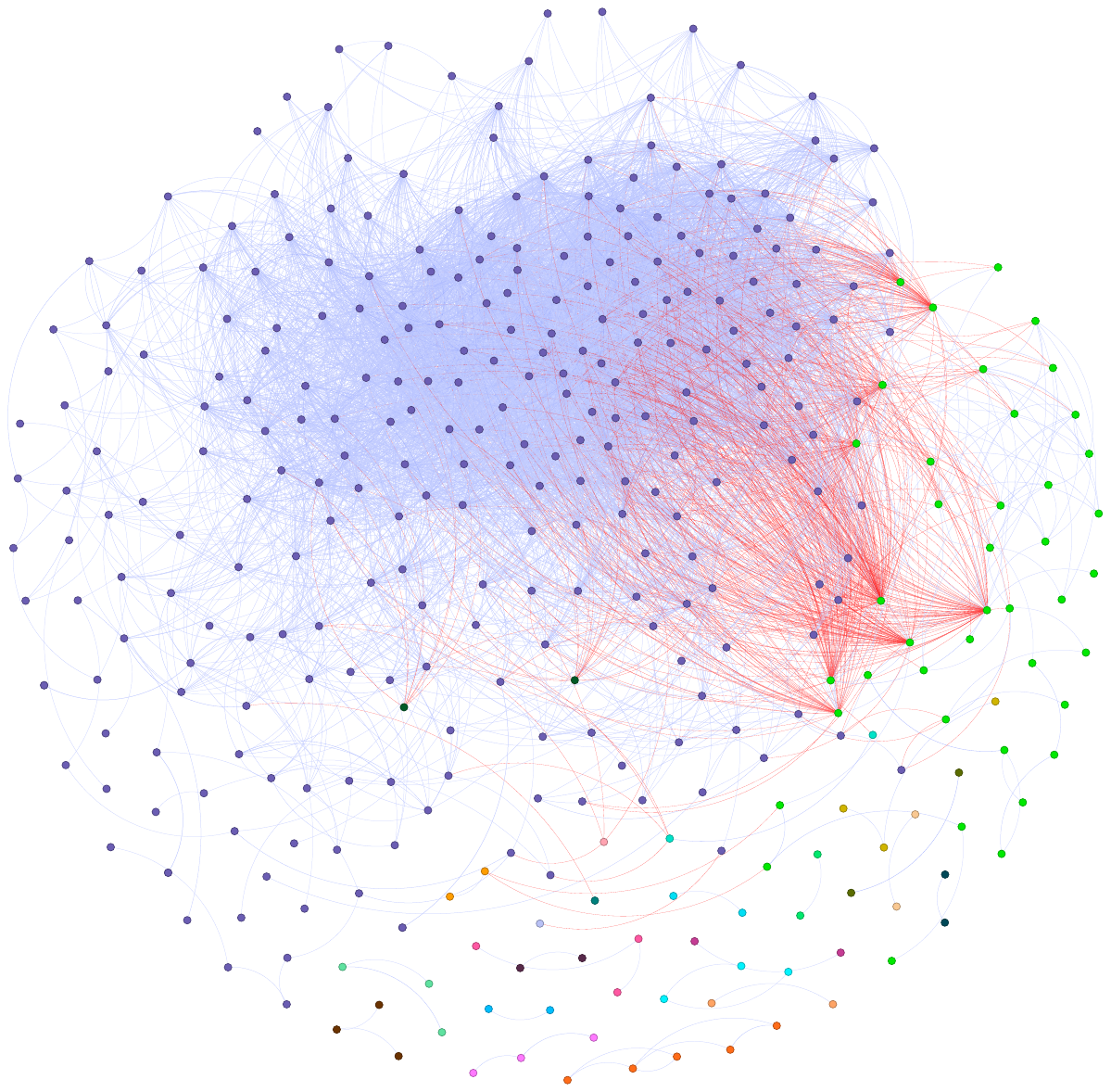}\label{fig:sub25}}
		\caption{The respective partitions returned by Troika for the networks of \textit{S\&P 500} market index from 2000 to 2024. Red and blue edge colours represent negative and positive correlations respectively. Different node colours represent the major clusters. Magnify the high-resolution figure on screen for the details.}
		\label{fig:s_and_p_500_viz_1}
	\end{figure}

	The analysis of the partitions produced by Troika on these networks reveals the dynamics in the structure of a major part of the US stock market over time. Table \ref{tab:sp500} (in the appendix) provides detailed results on the cluster size and the number of clusters for the partitions of the S\&P 500 networks obtained by the Troika algorithm. The optimal partition for each network is visualized in Fig. \ref{fig:s_and_p_500_viz_1} where clusters are shown using different node colours. The optimal partition for the year 2000 has 11 clusters (of positively correlated stocks) and an average cluster size of 13.54 nodes per cluster. In the three years leading to the 2008 financial crisis, we observe the number of partitions monotonically decreasing from 12 to 1 while the average cluster size monotonically increases from 16.3 to 294. In the network for year 2008, all edge weights are positive (from the consistently negative returns for most stocks). This makes the optimal partition become the trivial solution of all nodes belonging to one cluster. Such an unusual partition structure is only observed for the year 2008 when the most severe economic crisis of the contemporary era impacted the US stock market. The results in Fig. \ref{fig:s_and_p_500_viz_1} show that Troika can handle these portfolio networks and its solutions can reveal patterns from financial correlation data modeled as networks. 
	
	Moving on to the more recent time, the partitions also show a structural shift during the COVID-19 pandemic. In the year 2019 (and before the pandemic started), the network had 348 nodes and was predominantly characterized by one major cluster comprising 283 stocks and the rest scattered among 10 substantially small clusters. The optimal partition shows that the returns from the 283 stocks were overall positively correlated while there were 65 stocks in the network whose returns had different patterns and therefore created 10 separate clusters. The structure of the network goes through a major change in 2020, where a single dominant cluster emerges, consisting of over 97\% of the stocks (298 out of the total 306 stocks), a change possibly attributable to the market’s reaction to the global pandemic conditions at the time. The year 2021 saw a reversion to a structure somewhat reminiscent of the 2019 structure, showcasing 13 clusters with one predominant cluster housing 292 constituents. By 2023, the network's composition had changed into five clusters, with two primary clusters comprising 189 and 231 constituents, respectively. Except for the networks of the years 2008 and 2020, multiple major clusters exist in all networks distinguishing years of financial downturn (visible with distinct node colours in Fig. \ref{fig:s_and_p_500_viz_1}).
	
	In summary, the partitions obtained by Troika allow us to infer clusters from correlations which in turn highlight the temporal changes of a portfolio or market index over the years. Compared to the network of a portfolio with one cluster encompassing all the network (the year 2008) or almost all the network (the year 2020), the observed presence of multiple sizable negatively correlated clusters is interpretable as a balanced portfolio with diverse sectors which offer opportunities for hedging against sector-specific risks.
	
	\section{Discussion and Conclusion}
	\label{s:discuss}
	
	We proposed an approximation algorithm for the CP problem and demonstrated its performance and applicability. The comparative analysis on five datasets provided in Section \ref{s:results-cp} indicates the practical advantages of Troika in solving the CP problem, in comparison to two existing alternatives. Troika improves upon partitions from the Combo algorithm and returns solutions that are substantially closer to optimal. Note that there are some CP benchmark instances that are randomly generated and therefore are less particular in terms of network structure. The \textit{Random} dataset from \cite{sorensen2023cp} include such instances which are used in multiple studies \cite{du2022solving,jovanovic2023fixed}. On eight challenging and non-trivial instances from the Random dataset \cite{sorensen2023cp}, we observed that Troika outperforms Combo, but shows a limitation for achieving any time-quality tradeoff advantage over Gurobi IP on these random unstructured instances. However, Troika outperforms the state-of-the-art Gurobi IP solver in solution quality and/or solve time on most CP instances as shown in Figs.\ \ref{fig:abr}--\ref{fig:ba}.
	
	Unlike common heuristic algorithms that rely on local or greedy optimization approaches, Troika is an approximation optimization algorithm for the CP problem and returns partitions with a guaranteed proximity to global optimality. Note that the descriptive comparisons we provided are not all statistically significant because some performance differences between these methods are marginal. A future study can be aimed to rank alternative CP methods. Such a study can use a Friedman test \cite{friedman1937} followed by a post-hoc Li test \cite{li2008multiple} to determine the statistically significant performance differences among existing CP methods.
	
	Troika handles networks with up to 5000 edges providing close-to-optimal solutions within a reasonable amount of time on standard hardware. For large-scale challenging instances (which are not solvable within an hour according to \cite{sorensen2023cp}), Troika returns high quality solutions within 10 minutes. On a wide range of benchmark instances, Gurobi IP returns solutions of lower quality if operationalized with the same time limit and optimality gap tolerance as Troika. For most benchmark instances, we showed that Troika improves the lower quality partitions of Combo even if a split second of extra time is available. Another advantage of Troika over a heuristic CP method is that for high-quality partitions of Combo, Troika ensures that the partition satisfies the user-specified optimality gap tolerance. In certain cases, networks with more than 5000 edges can also be processed by Troika to obtain a guaranteed approximation of the optimal partition within a reasonable time. This was exemplified in the analysis of the ``lecturers" network instance which has over 300,000 edges and nearly 800 nodes. A solution for this large instance is approximated within 0.01 of the optimal in 248.65 seconds by Troika. Remarkably, Gurobi IP fails to converge or even reach Troika's approximate solution for the lecturers instance in 4 hours. 
	
	Despite the relative efficiency of Troika, achieving global optimality in the CP problem for certain network structures and larger networks remains a practical challenge that no exact or approximation method has solved to the best of our knowledge. For these networks, Troika offers flexibility by allowing the user to specify an optimality gap tolerance or a specific time limit as stopping criteria. This ensures that the algorithm returns a partition alongside the maximum optimality gap, when operating under constrained conditions.
	
	We made a connection between the CP problem and optimization-based community detection. Using the modularity objective function as an example to make this connection explicit, we provided a reduction that converts any modularity maximization instance into a CP instance. This reduction makes the Troika algorithm capable of approximating the maximum modularity of a network and finding the partition that satisfies a user-specified optimality gap tolerance. Comparing this secondary usage of Troika to eight algorithms that were deliberately designed for modularity maximization, we observed that Troika outperforms them in returning partitions with closer proximity to the globally maximum modularity partitions.
	
	We also demonstrated a real-world use case of the CP problem by deploying the Troika algorithm for analyzing the networks of correlations of returns among the S\&P 500 stocks. Partitions obtained by Troika reveal the network-level temporal changes for a major part of the US stock market over the analyzed period of 2000-2024. The results showed that Troika is useful for clustering networks of correlations. Specifically for portfolio networks, it uncovers temporal changes to the network structure, including the 2008 financial crisis and COVID-19 impacts on the clusters of positively correlated stocks within the S\&P 500 market index.

	From a practical perspective, Troika addresses a challenge in computational science by offering an effective method for CP filling a much-needed gap on approximating globally optimal solutions for small and mid-sized instances of practical relevance. In the future, exploring alternative lower bound heuristics will be crucial in developing CP approximation algorithms of higher efficacy. We hope this work facilitates future developments in network clustering and optimization.
	
	\section{Materials and Methods}
	\label{s:materials}
	
\begin{figure}[ht!]
		\centering
		\includegraphics[width=1\linewidth]{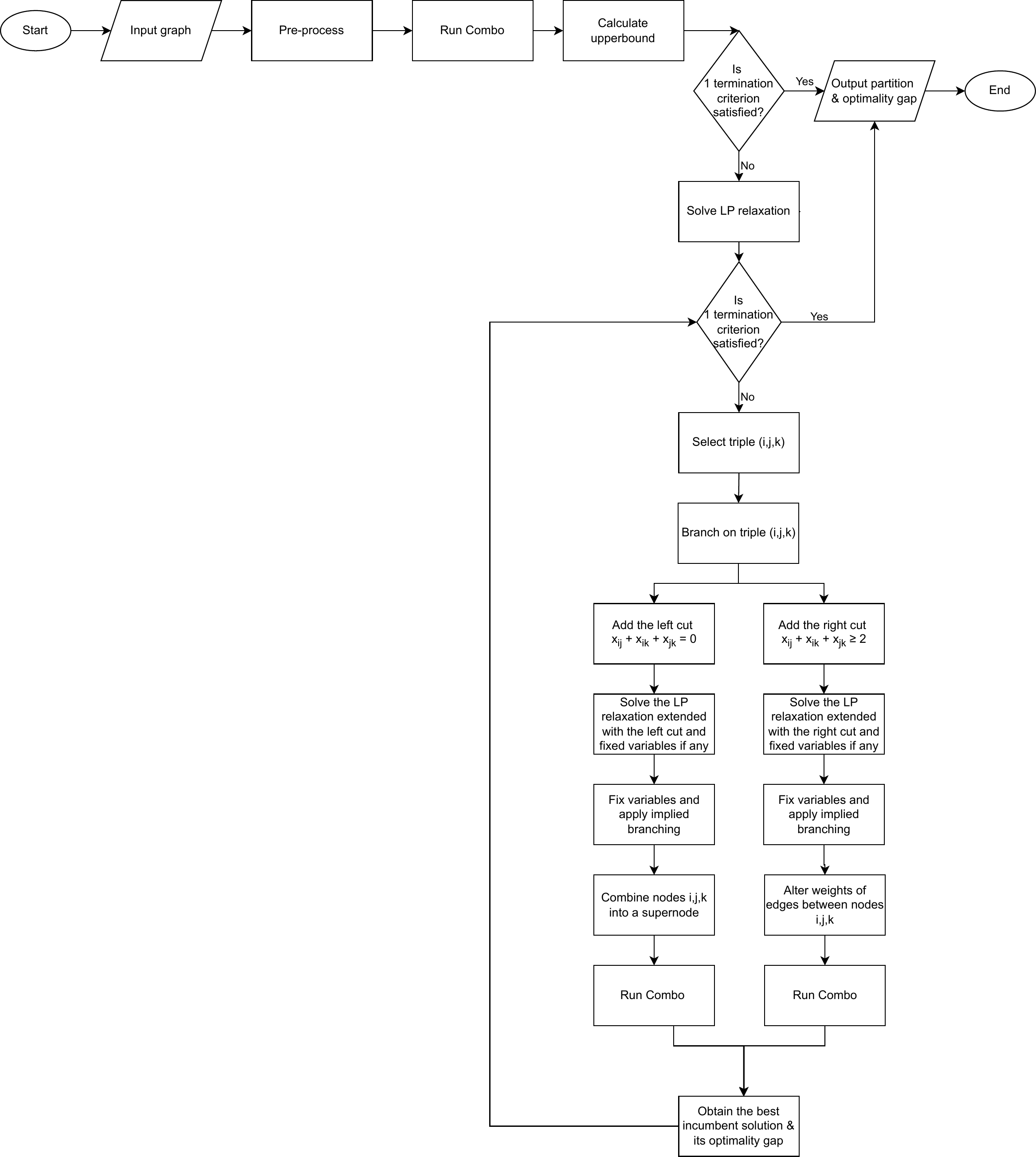}
		\caption{A flowchart of the key parts of the Troika algorithm}
		\label{fig:flowchart}
	\end{figure}

	This section provides the technical details of the Troika algorithm. These technical details are the building blocks of the Troika algorithm that allow it to approximate the optimal solution of the CP problem as demonstrated in Sections \ref{s:results-cp}--\ref{s:results-portfolio}. In Section \ref{ss:pre}, we explain two graph pre-processing steps that reduce the size of the graph input that Troika receives and builds optimization models for. In Section \ref{ss:variable}, a trick from integer programming is operationalized in Troika to increase its efficiency by fixing values of certain binary decision variables. In Section \ref{ss:implied}, another integer programming technique is operationalized in Troika which (1) generates additional cuts that strengthen the optimization model and (2) fixes the values of additional decision variables. Finally, Section \ref{ss:selection} discusses the design choices of Troika on how node triples are prioritized and selected to be branched on for an efficient exploration of the feasible space. A schematic representation of the key steps in the Troika algorithm is provided as a flowchart in Figure \ref{fig:flowchart}.
	
	\subsection{Graph pre-processing}
	\label{ss:pre}
	
	Troika leverages two graph pre-processing steps: pendant clique and node reduction, and component-wise processing for disconnected graphs.
	
	\subsubsection{Component-wise processing for disconnected graphs}
	If the input Graph $G$ is not fully connected, it is divided into its separate connected components. In the context of the CP problem, since only internal weights are considered in the objective function, each component's objective value can be optimized separately, with the collective partition forming the output. This straightforward decomposition of a problem substantially reduces the total number of variables involved, thus enhancing the performance of the algorithm in networks that have several components (disconnected graphs).
	
	\subsubsection{Pendant cliques and node reduction}
	Additionally, the Troika algorithm benefits from recognizing specific structural patterns, such as pendant nodes and cliques, to expedite feasible space exploration. In any optimal solution, each pendant node (i.e.\ node incident on precisely one edge) that has a positive degree, always belongs to the same cluster as its sole neighbour. We transform the original graph $G$ into a reduced graph $G'$ where each positive-degree ($d_i>0$) pendant node $i$ is replaced with a self-loop at the neighboring node with a weight of $d_i$. The weight of the self-loop reflects the contribution of the reduced positively weighted edge to the optimal objective function value. If the edge weight is negative, a separate cluster is created for the pendant node to be later appended to the output partition.
	
	\begin{figure}[!ht]
		\centering
		\subfloat[]{\includegraphics[width=0.4\textwidth]{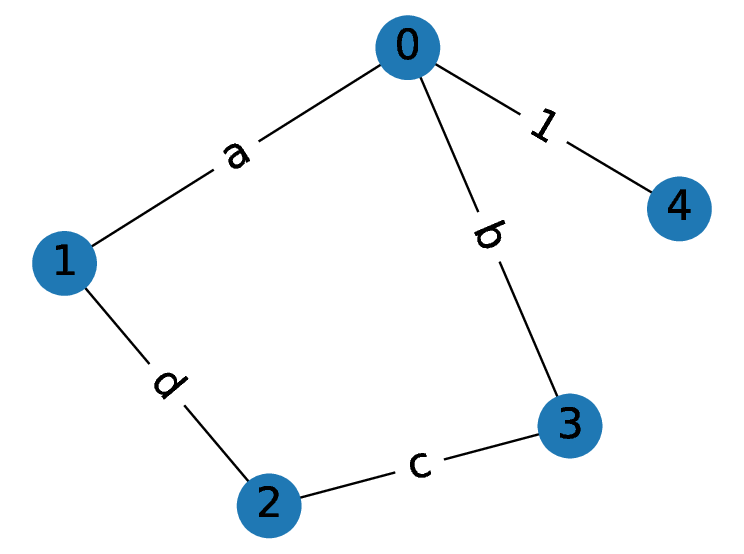}\label{fig:pendant_present}}
		\hfill
		\subfloat[]{\includegraphics[width=0.4\textwidth]{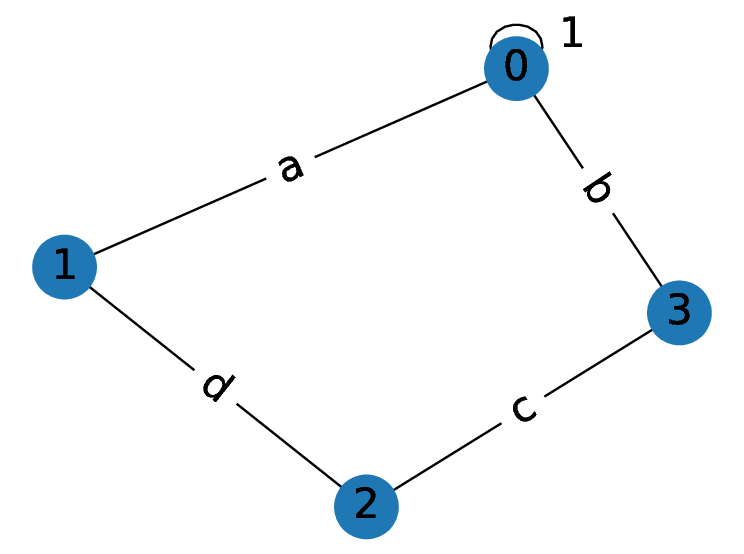}\label{fig:pendant_reduced}}
		\caption{(a) Pendant node 4 is positively connected to its sole neighbor node 0. (b) After reduction, node 4 and its edge are replaced with a self-loop at node 0 that has the same weight as the edge (0,4).}
		\label{fig:pendant_nodes}
	\end{figure}
	
	Cliques are defined as complete sub-graphs with all internal edges bearing positive weights. Considering a node to be a 1-clique, the pendant node reduction idea can be generalized to reductions for pendant cliques of arbitrary size $s$. An $s$-clique is pendant if all its nodes are incident on $s-1$ edges, except one node that is incident on $s$ edges. The exceptional node is called a \textit{connector}. Pendant cliques can be replaced by a self-loop on the connector node whose weight accounts for the contribution of the clique to the optimal objective function value. So, each pendant cliques (of any size) can be condensed into a self-loop on the corresponding connector node, ensuring the allocation of all nodes of the clique to the same cluster. This pre-processing step may substantially reduce the number of variables and constraints in the optimization models within the Troika algorithm. An alternative and more general pre-processing approach for CP is discussed in \cite{belyi2022network}.
	
	\begin{figure}[ht]
		\centering
		\subfloat[]{\includegraphics[width=0.4\textwidth]{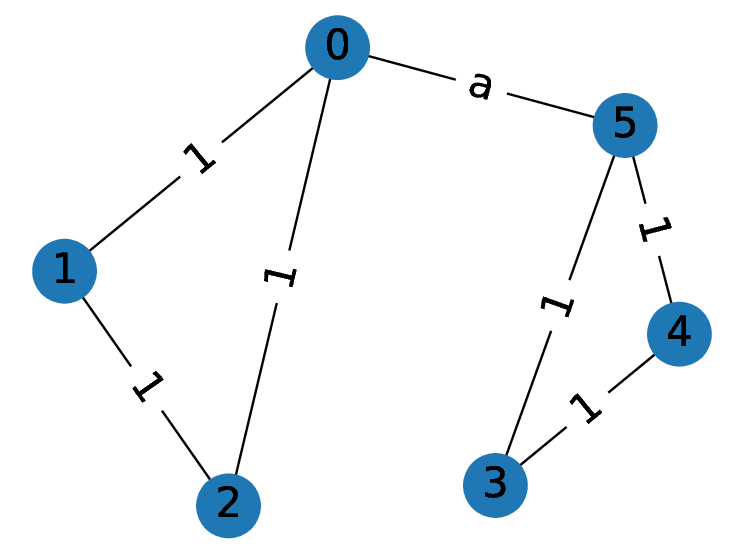}\label{fig:clique_present}}
		\hfill
		\subfloat[]{\includegraphics[width=0.4\textwidth]{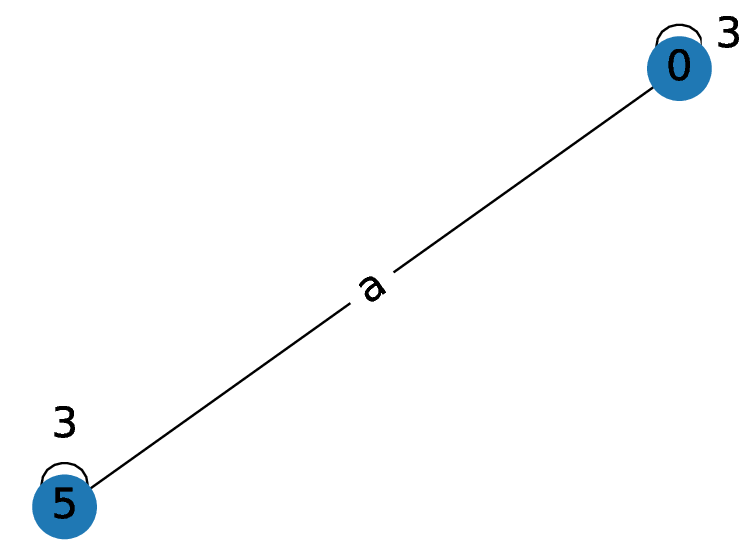}\label{fig:clique_reduced}}
		\caption{(a) Input graph $G$ with two distinct positive cliques: the first comprising nodes 0, 1, and 2, and the second consisting of nodes 3, 4, and 5. (b) The post-reduction graph $G'$, where each of the two cliques is replaced with a self-loop at their corresponding two connector nodes 0 and 5. Weights of each self-loop equals the sum of positive weights of the clique that it has replaced.}
		\label{fig:cliques}
	\end{figure}
	
	\subsection{Variable fixing}
	\label{ss:variable}
	
	Troika utilizes a variable fixing technique to solve the IP faster. Variable fixing can be used to determine the definitive value of certain variables at a point in the branch and cut process and for all subsequent LPs. Specifically, a binary variable $x_{ij}$ can be fixed to either zero or one when its reduced cost surpasses a certain threshold in the current LP's optimal solution. Suppose $x_{ij}$ is set to zero with a reduced cost of $c_{ij}$, in an optimal LP solution, where the optimal objective function value is denoted as $z^*_{LP}$. It is observed that $c_{ij}$ takes a negative value when $x_{ij}$ is at its lower bound (i.e., zero). Let $LB$ represent the objective value of the incumbent solution within the B\&B search tree. It is deduced that any feasible solution with $x_{ij} = 1$ will have an objective value not exceeding $z^*_{LP} + c_{ij}$. Consequently, $x_{ij}$ is fixed to zero if $z^*_{LP} + c_{ij} \leq LB$. Conversely, $c_{ij}$ will be positive if $x_{ij}$ is at its upper bound (i.e., one), prompting us to fix $x_{ij}$ to one if $z^*_{LP} - c_{ij} \leq LB$.
	
	Upon applying these conditions to fix variables, Troika proceeds to determine the states of additional variables through logical deductions. For instance, if nodes $i$ and $j$ belong to the same cluster ($x_{ij}$ fixed to zero) and nodes $k$ and $j$ do not share the same cluster ($x_{jk}$ fixed to one), it logically follows that nodes $i$ and $k$ must be in different clusters, leading to $x_{ik}$ being fixed to one. This logical implication is formalized as follows:
	\begin{equation}
		x_{ij} = 0 \land x_{jk} = 1 \rightarrow x_{ik} = 1
	\end{equation}

	\subsection{Implied branching and fixing}
	\label{ss:implied}
	
	We further explore the synergies between branching and variable fixing to increase the convergence speed of the algorithm. By drawing logical conclusions from the states of fixed variables and the already established branching cuts, further variables can be fixed and additional cuts can be introduced to enhance the separation in both the right and left branches of the B\&B tree \cite{aref2022bayan}.
	
	Consider a scenario in the right branch where the constraint $x_{ij} + x_{jk} + x_{ik} \geq 2$, pertaining to the triple $(i, j, k)$, has been added to the LP formulation. Suppose there exists a fixed variable related to one of these nodes, for instance, $x_{ip} = 0$. This precondition enables the introduction of a novel cut into the LP model as follows:
	\begin{equation}
		x_{ij} + x_{jk} + x_{ik} \geq 2 \land x_{ip} = 0 \rightarrow x_{jk} + x_{jp} + x_{kp} \geq 2.
	\end{equation}
	
	Similarly, in the context of a left branch, consider the constraint $x_{ij} + x_{jk} + x_{ik} = 0$ has been added to the LP model. If a variable $x_{ip}$, associated with one of the nodes in the triple $(i, j, k)$, is fixed, two more variables can be fixed as illustrated below:
	\begin{equation}
		x_{ij} + x_{jk} + x_{ik} = 0 \land x_{ip} = 0 \rightarrow x_{jp} = 0, \; x_{kp} = 0,
	\end{equation}
	\begin{equation}
		x_{ij} + x_{jk} + x_{ik} = 0 \land x_{ip} = 1 \rightarrow x_{jp} = 1, \; x_{kp} = 1.
	\end{equation}
	
	The integration of logical inferences with variable states serves a dual purpose: it not only simplifies the process of fixing variables but also supports the creation of strategic cuts. This dual functionality considerably accelerates the Troika algorithm.

	\subsection{Triple selection for branching}
	\label{ss:selection}
	
	Troika attempts to select the best triple for branching, facilitating an earlier detection of infeasibility, and enabling more variable fixing opportunities. 
	
	During each search iteration, the algorithm first identifies all triples that violate the two constraints in Eq.~\eqref{con:left_cut} and Eq.~\eqref{con:right_cut}. Subsequently, it prioritizes certain triples for further evaluation based on their respective edge weights. Specifically, as defined in Section~\ref{ss:b&c}, we introduced $\mathcal{T}$ as the union of the subsets $T^k_+$, $T^j_+$, and $T^i_+$, over which the transitivity constraints are defined. Moreover, $\mathcal{T}$ can be partitioned to $\mathcal{T} = T_3 \cup T_2 \cup T_1$ where
	
	\begin{itemize}
		\item $T_3$ denotes the triples that precisely have three strictly positive edge weights,
		\item $T_2$ denotes the triples that precisely have two strictly positive edge weights, and
		\item $T_1$ denotes the triples that precisely have one strictly positive edge weight.
	\end{itemize}

	The Troika algorithm adopts a structured approach for triple selection, starting with the triples in $T_3$. It only proceeds to select triples from $T_2$ after all triples in $T_3$ have been utilized. Similarly, selection from $T_1$ commences only after all triples in $T_2$ are exhausted. The selection of the best triple from the chosen subset is then guided by three key B\&B-node-specific factors: (1) the greater overlap between its nodes with those in the triples already used for branching, (2) the count of its associated fixed variables, and (3) the absolute degree of its nodes. A binary indicator, $\beta_i$, is assigned the value one if node $i$ is included in any branching triples of the parent nodes. The quantity of fixed variables linked to node $i$ is represented as $f_i$. For every node $i$, possessing a degree $d_i$, we compute a score $s_i$ according to Eq.\ \eqref{eq:score}, incorporating $\beta_i$, $f_i$, and $d_i$ to address the three outlined criteria. 
	
	\begin{equation}
		\label{eq:score}
		s_i = 1 - e^{-f_i} + \beta_i + \frac{|d_i|}{n - 1}
	\end{equation}

	The collective score for a triple $(i, j, k)$ is calculated as the sum of $s_i$, $s_j$, and $s_k$. Selection of a triple for branching is then carried out using a roulette wheel selection mechanism based on the the calculated scores of the candidate triples.

	\section*{Acknowledgments}
	Troika algorithm is built on a foundation of earlier developments made accessible by other researchers. We are thankful to Alexander Belyi, Stanislav Sobolevsky, Alexander Kurbatski, Carlo Ratti, Riccardo Campari, Philipp Kats, Mahdi Mostajabdave, and Hriday Chheda for making their algorithms publicly accessible. We are also thankful for the helpful comments from to the anonymous reviewers.
	
	
	%
	%
	%

	
	\section*{Appendices}
	
	The following appendices provide additional results as well as references for accessing the network data.

	\paragraph*{Accessing the network data}
	
	

	Each of the 53 real networks used in Section \ref{s:results-community} was loaded from the network repository \href{https://networks.skewed.de/}{Netzschleuder} as simple unweighted and undirected graph $G$. Then, a CP problem was defined according to the weighted graph $G'$ whose edge weights are the entries of the modularity matrix of $G$. The data for all these real networks ($G$) are available in a FigShare data repository \cite{Aref2025dataTroika}. The same FigShare repository contains data on all synthetic networks (LFR and ABCD networks) used in Section \ref{s:results-community} as well as all BA instances used in Section \ref{ss:ba} and all portfolio networks used in Section \ref{s:results-portfolio}. Other networks used in our study were from \cite{sorensen2023cp}. S{\o}rensen and Letchford \cite{sorensen2023cp} have shared their CP instances and the optimal solutions that were available in a public GitHub repository \url{https://github.com/MMSorensen/CP-Lib}.

	
	
	
	\paragraph*{Additional numerical results}
	
	Tables \ref{table:abr_table}--\ref{tab:sp500} provide additional numerical results.
	
	\begin{table}[ht]
		\centering
		\begin{adjustbox}{max width=\textwidth}
			\begin{tabular}{ccccccccc}
				\hline
				Instance name & $n$ & $m$ & \makecell{ time \\ Troika} & \makecell{ time \\ Gurobi IP} & \makecell{ time \\ Combo } & \makecell{ $Q$ \\ Troika} & \makecell{ $Q$ \\ Gurobi IP} & \makecell{ $Q$ \\ Combo } \\
				\hline
				\rowcolor{mycolor} bridges & 108 & 5117 & 0.78 ± 0 & 10.64 ± 0.14 & 0.22 ± 0 & 3867 ± 0 & 3866 ± 0 & 3867 ± 0 \\
				\rowcolor{mycolor} cars & 33 & 528 & 0.02 ± 0 & 0.08 ± 0 & 0.01 ± 0 & 1501 ± 0 & 1501 ± 0 & 1501 ± 0 \\
				\rowcolor{mycolor} cetacea & 36 & 619 & 0.01 ± 0 & 0.05 ± 0 & 0 ± 0 & 967 ± 0 & 967 ± 0 & 967 ± 0 \\
				\rowcolor{mycolor} companies & 137 & 9316 & 4.36 ± 0 & 4.95 ± 0.01 & 4.33 ± 0 & 81802 ± 0 & 78548 ± 0 & 81802 ± 0 \\
				\rowcolor{mycolor} hayes-roth & 160 & 12720 & 22.73 ± 0.08 & 47.33 ± 0.68 & 0.24 ± 0 & 2797 ± 0 & 2733 ± 0 & 2797 ± 0 \\
				\rowcolor{mycolor} lecturers & 797 & 306915 & 248.65 ± 0.62 & 600.67 ± 0.02 & 11.17 ± 0.01 & 14306 ± 0 & 1189 ± 0 & 14298 ± 0 \\
				\rowcolor{mycolor} lung-cancer & 32 & 483 & 0.01 ± 0 & 0.06 ± 0 & 0.01 ± 0 & 3472 ± 0 & 3472 ± 0 & 3472 ± 0 \\
				\rowcolor{mycolor} lymphography & 148 & 9101 & 3.95 ± 0 & 113.57 ± 1.52 & 2.89 ± 0 & 19174 ± 0 & 19170 ± 0 & 19174 ± 0 \\
				\rowcolor{mycolor} micro & 40 & 660 & 0.02 ± 0 & 0.1 ± 0 & 0.01 ± 0 & 966 ± 0 & 966 ± 0 & 966 ± 0 \\
				primary-tumor & 339 & 54949 & 178.32 ± 0.99 & 2.88 ± 0.15 & 176.15 ± 0.11 & 323614 ± 0 & 322867 ± 0 & 323614 ± 0 \\
				\rowcolor{mycolor} soup & 209 & 21736 & 8.66 ± 0.06 & 48.83 ± 0.52 & 0.27 ± 0 & 4618 ± 0 & 4425 ± 0 & 4618 ± 0 \\
				\rowcolor{mycolor} soybean-21 & 47 & 1081 & 0.02 ± 0 & 0.38 ± 0.02 & 0.02 ± 0 & 3041 ± 0 & 3041 ± 0 & 3041 ± 0 \\
				soybean-35 & 47 & 1081 & 0.07 ± 0 & 0.01 ± 0 & 0.07 ± 0 & 14613 ± 0 & 14613 ± 0 & 14613 ± 0 \\
				\rowcolor{mycolor} soybean-large & 307 & 45483 & 93.15 ± 0.03 & 603.78 ± 0.12 & 77.88 ± 0.1 & 316469 ± 0 & 293892 ± 0 & 316469 ± 0 \\
				\rowcolor{mycolor} sponge & 76 & 2781 & 0.29 ± 0 & 0.75 ± 0.03 & 0.21 ± 0 & 25677 ± 0 & 25677 ± 0 & 25677 ± 0 \\
				\rowcolor{mycolor} ta-evaluation & 151 & 11325 & 0.5 ± 0 & 1.32 ± 0.07 & 0.09 ± 0 & 1108 ± 0 & 1108 ± 0 & 1108 ± 0 \\
				\rowcolor{mycolor} uno & 54 & 1431 & 0.03 ± 0 & 0.37 ± 0.01 & 0.02 ± 0 & 798 ± 0 & 798 ± 0 & 798 ± 0 \\
				\rowcolor{mycolor} uno\_1a & 158 & 12403 & 1.76 ± 0.02 & 5.5 ± 0.09 & 1.21 ± 0.01 & 12197 ± 0 & 12197 ± 0 & 12197 ± 0 \\
				\rowcolor{mycolor} uno\_1b & 139 & 9591 & 1.29 ± 0 & 4.07 ± 0.12 & 0.95 ± 0 & 11775 ± 0 & 11775 ± 0 & 11775 ± 0 \\
				\rowcolor{mycolor} uno\_2a & 158 & 12403 & 4.55 ± 0 & 7.82 ± 0.04 & 3.9 ± 0.02 & 72820 ± 0 & 72820 ± 0 & 72820 ± 0 \\
				\rowcolor{mycolor} uno\_2b & 145 & 10440 & 3.93 ± 0 & 6.88 ± 0.11 & 3.62 ± 0 & 71818 ± 0 & 71818 ± 0 & 71818 ± 0 \\
				\rowcolor{mycolor} uno\_3a & 158 & 12403 & 6.15 ± 0 & 9 ± 0.24 & 6.02 ± 0.02 & 73068 ± 0 & 73068 ± 0 & 73068 ± 0 \\
				uno\_3b & 147 & 10731 & 5.68 ± 0 & 0.19 ± 0.01 & 5.64 ± 0.01 & 72629 ± 0 & 72629 ± 0 & 72629 ± 0 \\
				\rowcolor{mycolor} wildcats & 30 & 381 & 0.01 ± 0 & 0.04 ± 0 & 0.01 ± 0 & 1304 ± 0 & 1304 ± 0 & 1304 ± 0 \\
				\rowcolor{mycolor} workers & 34 & 561 & 0.01 ± 0 & 0.07 ± 0 & 0.01 ± 0 & 964 ± 0 & 964 ± 0 & 964 ± 0 \\
				\rowcolor{mycolor} zoo & 101 & 4263 & 0.82 ± 0 & 1.9 ± 0.01 & 0.66 ± 0 & 16948 ± 0 & 16948 ± 0 & 16948 ± 0 \\
				\hline
			\end{tabular}
		\end{adjustbox}
		\caption{Comparison of Troika, Gurobi IP, and Combo on solve time (mean ± standard deviation) and objective value on returned partitions for ABR instance. Highlighted rows show instances where Troika outperformed Gurobi IP in solve time and returned a solution that was equal or better.}
		\label{table:abr_table}
	\end{table}
	
	\begin{table}[!ht]
		\caption{Median EOS and mean EOS for Troika and eight modularity-based algorithms based on 100 real and synthetic networks}
		\centering
		\begin{tabular}{lll}
			\hline
			Algorithm & Median EOS & Mean EOS \\ \hline
			Troika & 0 & 0.000311 \\
			Belief & 0.065864 & 0.299997 \\
			CNM & 0.014007 & 0.035122 \\
			Combo & 0 & 0.001762 \\
			EdMot & 0.026774 & 0.18119 \\
			GNN & 0 & 0.007114 \\
			Leiden & 0 & 0.00635 \\
			Louvain & 0.000331 & 0.01089 \\
			Paris & 0.135471 & 0.257604 \\ \hline
		\end{tabular}
		\label{tab:success}
	\end{table}

	\begin{table}[!ht]
		\centering
		\caption{Details of the optimal partition obtained by the Troika algorithm for the S\&P 500 networks}
		\label{tab:sp500}
		\begin{tabular}{lp{3cm}p{2cm}p{2cm}p{3cm}p{3cm}}
			\hline
			Year & Order of the network ($n$) & Number of clusters & Average cluster size & Largest cluster \\ \hline
			2000 & 149 & 11 & 13.54 & 96 \\ 
			2001 & 214 & 8 & 26.75 & 152 \\ 
			2002 & 230 & 11 & 20.9 & 209 \\ 
			2003 & 233 & 6 & 38.83 & 226 \\ 
			2004 & 204 & 4 & 51.0 & 151 \\ 
			2005 & 196 & 12 & 16.33 & 138 \\ 
			2006 & 244 & 9 & 27.11 & 201 \\ 
			2007 & 143 & 5 & 28.6 & 69 \\ 
			2008 & 294 & 1 & 294.0 & 294 \\ 
			2009 & 303 & 6 & 50.5 & 292 \\ 
			2010 & 323 & 7 & 46.14 & 241 \\ 
			2011 & 255 & 7 & 36.42 & 190 \\ 
			2012 & 239 & 8 & 29.87 & 153 \\ 
			2013 & 279 & 13 & 21.46 & 242 \\ 
			2014 & 300 & 13 & 23.07 & 220 \\ 
			2015 & 219 & 7 & 31.28 & 128 \\ 
			2016 & 374 & 9 & 41.55 & 341 \\ 
			2017 & 239 & 14 & 17.07 & 195 \\ 
			2018 & 259 & 6 & 43.16 & 131 \\ 
			2019 & 348 & 11 & 31.63 & 283 \\ 
			2020 & 306 & 5 & 61.2 & 298 \\ 
			2021 & 362 & 13 & 27.84 & 292 \\ 
			2022 & 323 & 6 & 53.83 & 274 \\ 
			2023 & 430 & 5 & 86.0 & 231 \\ 
			2024 & 361 & 24 & 15.04 & 270 \\ \hline
		\end{tabular}
	\end{table}

	\FloatBarrier
	
	

\end{document}